\documentclass[11pt]{article}
\pdfoutput=1
\usepackage{amsmath,amssymb,epsfig,amsfonts}
\usepackage{graphicx,subfigure}
\usepackage[usenames, dvipsnames]{color}
\usepackage[pagebackref,hidelinks]{hyperref}
\usepackage{cite}
\usepackage{verbatim} 
\usepackage{float}
\restylefloat{table}
\usepackage[all]{xy}
\usepackage{pdflscape}
\usepackage{tikz}
\usetikzlibrary{shapes.geometric}
\usepackage{array}
\usepackage{youngtab}
\usepackage{multirow}
\usepackage{color}
\usepackage{ulem}
\usepackage{tikz-cd}

\DeclareFontFamily{U}{wncy}{}
    \DeclareFontShape{U}{wncy}{m}{n}{<->wncyr10}{}
    \DeclareSymbolFont{mcy}{U}{wncy}{m}{n}
    \DeclareMathSymbol{\Sh}{\mathord}{mcy}{"58} 

\makeatletter

\renewcommand*{\backref}[1]{}
\renewcommand*{\backrefalt}[4]{({%
    \ifcase #1 Not cited.%
          \or Page~#2.%
          \else Pages #2.%
    \fi%
    })}

\def\bbP{{\mathbb{P}}}
\def\bbZ{{\mathbb{Z}}}

\def\bbF{{\mathbb{F}}}

\def\fku{{\mathfrak{u}}}

\def\fkg{{\mathfrak{g}}}

\makeatother
\hypersetup{
    pdftitle={Phases of 5d SCFTs from M-/F-theory on Non-Flat Fibrations},
    pdfauthor={Fabio Apruzzi, Ling Lin, Christoph Mayrhofer},
    pdfsubject={SCFTs, F-theory, M-theory, non-flat fibration}
}

\begin{document}

\baselineskip=18pt  
\numberwithin{equation}{section}  



\vspace*{-2cm} 
\begin{flushright}
{\tt UPR-1295-T}\\
\end{flushright}

\vspace*{0.8cm} 
\begin{center}
 {\LARGE  Phases of 5d SCFTs from M-/F-theory on Non-Flat Fibrations}

 \vspace*{1.8cm}
 {Fabio Apruzzi$^{1}$, Ling Lin$^2$, Christoph Mayrhofer}

 \vspace*{1.2cm} 

{\it $^1$Mathematical Institute, University of Oxford, Woodstock Road, \\
Oxford, OX2 6GG, United Kingdom}

\bigskip
{\it $^2$Department of Physics and Astronomy, University of Pennsylvania,  \\
 Philadelphia, PA 19104-6396, USA}

 \bigskip
  
 {{\tt Fabio.Apruzzi@maths.ox.ac.uk}$\,,\quad$ {\tt lling@physics.upenn.edu}$\,,\quad$ {\tt christoph.k.mayrhofer@gmail.com}}

%
%
%
\vspace*{0.8cm}
\end{center}
\vspace*{.5cm}
%
\noindent
We initiate the systematic investigation of non-flat resolutions of non-minimal singularities in elliptically fibered Calabi--Yau threefolds. 
Compactification of M-theory on these geometries provides an alternative approach to studying phases of five-dimensional superconformal field theories (5d SCFTs).
We argue that such resolutions capture non-trivial holonomies in the circle reduction of the 6d conformal matter theory that is the F-theory interpretation of the singular fibration.
As these holonomies become mass deformations in the 5d theory, non-flat resolutions furnish a novel method in the attempt to classify 5d SCFTs through 6d SCFTs on a circle.
A particularly pleasant aspect of this proposal is the explicit embedding of the 5d SCFT's enhanced flavor group inside that of the parent 6d SCFT, which can be read off from the geometry.
We demonstrate these features in toric examples which realize 5d theories up to rank four.


\newpage

\tableofcontents
\newpage

\section{Introduction}

One of the remarkable achievements of string theory is to provide evidence for the existence of higher dimensional super(symmetric-)conformal field theories (SCFTs). 
Such theories are always strongly coupled and lack a canonical Lagrangian description, thus making them difficult to approach using traditional quantum field theory techniques.  
Compared to these methods, the crucial advantage of string theory comes from the geometrization of supersymmetric gauge dynamics---including non-perturbative effects---which via dualities can be described by various brane constructions.
Perhaps one of the most recent success stories in this context is the classification of 6d ${\cal N} = (1,0)$ SCFTs via compactifications of string and M-/F-theory \cite{Witten:1995ex, Witten:1995zh,
Strominger:1995ac, Seiberg:1996qx, WittenSmall,
Ganor:1996mu,Morrison:1996pp,Seiberg:1996vs, Bershadsky:1996nu,
Brunner:1997gf, Blum:1997fw, Aspinwall:1997ye, Intriligator:1997dh, Hanany:1997gh,Heckman:2013pva, Gaiotto:2014lca, DelZotto:2014hpa,
DelZotto:2014fia, Heckman:2015bfa, Bhardwaj:2015xxa,
Apruzzi:2013yva}.

Similarly, 5d ${\cal N}=1$ SCFTs have been constructed in many ways, e.g., as the world volume theory of D4 branes probing a D8/O8$^-$ stack \cite{Seiberg:1996bd, Brandhuber:1999np, Bergman:2012kr}, using webs of $(p,q)$-five-branes in type IIB \cite{Aharony:1997ju,Aharony:1997bh,DeWolfe:1999hj,Bergman:2015dpa, Zafrir:2015rga, Zafrir:2015ftn, Ohmori:2015tka, Hayashi:2017btw,Hayashi:2018bkd,Hayashi:2018lyv}, or via holography \cite{Ferrara:1998gv,Brandhuber:1999np,Bergman:2012kr,Bergman:2013koa,Passias:2018swc,Passias:2012vp,Jafferis:2012iv,Gutperle:2018axv,DHoker:2016ujz,DHoker:2016ysh,DHoker:2017mds,Apruzzi:2014qva,Kim:2015hya,Kim:2016rhs,Lozano:2012au,Lozano:2013oma,Gutperle:2017tjo,Kaidi:2017bmd,Gutperle:2018wuk,Bergman:2018hin,Fluder:2018chf,Kaidi:2018zkx,DHoker:2017zwj,Gutperle:2018vdd,Apruzzi:2018cvq,Hong:2018amk,Malek:2018zcz}. 
An alternative approach, which will also be the focus of this work, utilizes geometric engineering via M-theory on singular Calabi--Yau threefolds \cite{Morrison:1996xf, Intriligator:1997pq, Douglas:1996xp,DelZotto:2017pti, Xie:2017pfl,Jefferson:2017ahm, Jefferson:2018irk}. At last, a complementary method to determine the gauge theory phases of 5d SCFTs has be given in \cite{Closset:2018bjz}. It consists of studying the full spectrum in the context of M-theory/IIA string theory duality, by finding the circle fibrations of the M-theory Calabi--Yau threefolds and analyzing the reduction on that $S^1$. 

Renormalization group (RG) flows triggered by mass deformations connect different 5d $\mathcal N=1$ supersymmetric gauge theories. In the moduli space of each of these different gauge theories a 5d UV fixed point can be present. 
On the Coulomb branch of such gauge theories, the dynamics is described by a supersymmetric quantity called prepotential, $\cal F$, which is a real-valued function in the Coulomb branch parameters.
Its derivatives encode the metric on the Coulomb branch, the related kinetic terms, as well as the tension of non-perturbative objects such as monopole strings.
Traditionally, one can infer the existence of a 5d conformal fixed point as well as the transitions between different phases of the theory by positivity conditions of derivatives of $\cal F$ \cite{Seiberg:1996bd}.
Using geometric engineering, these conditions have been translated into properties of Calabi--Yau threefolds \cite{Morrison:1996xf, Intriligator:1997pq,DelZotto:2017pti,Xie:2017pfl} and subsequently refined in \cite{Jefferson:2017ahm, Jefferson:2018irk}, setting up the stage for a classification program which was recently successfully applied to 5d SCFTs of rank 1 and 2 \cite{Jefferson:2017ahm,Jefferson:2018irk,Bhardwaj:2018yhy}.

An observation from the rank 1 and 2 classification, which is also backed up by other, higher rank examples constructed in the literature, is that all 5d SCFTs seem to have ``parent'' 6d SCFT.
To be more precise, the conjecture is that through mass deformations and RG flows, any 5d SCFT is connected to a so-called ``5d Kaluza--Klein (KK) theory'' \cite{Jefferson:2018irk, DelZotto:2017pti}, which is an $S^1$ reduction of a 6d SCFT without any non-trivial holonomies.
These theories do not have an honest 5d fixed point, because morally speaking, the degrees of freedom in the UV reassemble into those of a 6d SCFT (hence the name).
However, by mass-deforming a 5d KK-theory, which from the circle reduction perspective amounts to turning on non-trivial holonomies, one can now flow to a different 5d theory that has a genuine 5d UV fixed point.

Given that all 6d SCFTs are classified by F-theory, one way to test the above conjecture would be, at least in principle, to dimensionally reduce all F-theory constructions and study the possible 5d theories obtainable by mass deformation and RG flow.
From a field theoretic perspective, this is anything but an easy task.
The difficulty comes from our incomplete understanding of strongly coupled dynamics, in particular what the possible mass deformations of a given 6d SCFT on an $S^1$ are.

From a stringy perspective, however, the geometric construction allows us to relate 6d and 5d theories via the duality between F- and M-theory \cite{Vafa:1996xn,Morrison:1996na,Morrison:1996pp}:
\begin{align}\label{eq:M-/F-theory_duality}
\begin{split}
  & \text{\it M-theory on elliptically fibered Calabi--Yau threefold } Y_3 \quad \cong \quad \text{\it F-theory on } Y_3 \times S^1  .
\end{split}
\end{align}
For weakly coupled 6d theories, the corresponding threefold has minimal singularities according to the Kodaira classification.
Under the duality, (partial) blow-up resolutions of these singularities correspond to turning on holonomies of 6d gauge fields along the $S^1$ when descending to 5d, which pushes the 5d theory onto its Coulomb branch.

On the other hand, 6d SCFTs are known to arise from non-minimal singularities of the elliptic fibration \cite{Bershadsky:1996nu, DelZotto:2014hpa} over codimension\footnote{Here and in what follows codimension refers to the codimension of the base of the elliptic fibration.} two loci, i.e., points $p$ in the base.
In terms of a Weierstrass model of $\pi: Y_3 \rightarrow B_2$,
\begin{equation}
y^2=x^3 + f \, x + g \, ,
\end{equation}
with discriminant $\delta=27g^2+4f^3$, such singularities are characterized by the vanishing orders
\begin{equation}
\text{ord}(f|_p, g|_p, \delta|_p) \geq (4,6,12)
\end{equation}
of the Weierstrass functions at $p$.
To make sense of such singularities, one typically blows up such points $p$ in the base into a collection of rational curves $\Sigma_i$, until the resulting total space has only minimal singularities.
Physically, this corresponds to pushing the 6d SCFT onto its weakly coupled tensor branch.
Indeed, the restriction coming from the geometry on how such base blow-ups are compatible with the Calabi--Yau condition on $Y_3$ is the basis for the 6d classification \cite{Heckman:2015bfa, Bhardwaj:2015xxa}.

Given the better handle---both physically and geometrically---of the 6d theory on the weakly coupled tensor branch, one approach to study the relationship between 6d and 5d SCFTs is to first reduce the 6d tensor branch theory on an $S^1$.
This yields a weakly coupled 5d gauge theory whose UV limit is an aforementioned 5d KK-theory.
Then one can mass deform this gauge theory, leading to phases which can have 5d UV fixed points. 
The geometric counterpart of this process is to consider M-theory on the base-blown-up threefold---yielding the weakly coupled phase of the KK-theory---and then consider geometric transitions to obtain suitable geometries supporting 5d SCFTs.
Indeed, there has been a lot of recent activity along these lines \cite{DelZotto:2017pti,Jefferson:2017ahm, Jefferson:2018irk, Bhardwaj:2018yhy}.

The classification of 6d SCFTs revealed the existence of so-called \textit{conformal matter} (which are 6d SCFTs by themselves) as building blocks for the generalized quiver structure of 6d SCFTs \cite{DelZotto:2014fia}.
Such theories are constructed in F-theory via a collision of two non-compact divisors $W_{1,2}$ carrying ADE groups $G_{1,2}$ at a smooth point $p$ in the F-theory base $B_2$.
If the fiber singularity at the collision point $p$ was of minimal type, this would just correspond to ordinary charged matter.
In that analogy, one can think of the strongly coupled sector at the non-minimal singularity over $p$ as a type of generalized matter charged under the flavor symmetry $G_{1,2}$.

For the circle reduction of conformal matter theories, we propose an alternative geometric procedure to analyze the resulting 5d theories:
Instead of blowing up the base to reach a fibration with only Kodaira fibers, as was done in \cite{DelZotto:2014fia}, one can also resolve the non-minimal singularities of the total space $Y_3$ via fiber blow-ups which do not change the base $B_2$.
Because of the severity of the singularity, such a resolution introduces, in addition to curves, also surface components into the fiber.
The resulting fibration is thus \textit{non-flat}, that is, it does not have equi-dimensional fibers.

Being divisors in a Calabi--Yau threefold, these compact surface components immediately give rise to gauged $\fku(1)$s in the M-theory compactification.
Since they are typically ruled surfaces, they can be blown-down to a curve, which enhances the $\fku(1)$s to a non-abelian algebra \cite{Morrison:1996xf, Intriligator:1997pq}.
The basic field content of these gauge theories come from the spectrum of M2-branes wrapping fibral curves of the ruling in these surfaces.
The careful analysis of the K\"ahler cone of the resolved Calabi--Yau matches the 5d prepotential analysis which describes the Coulomb branches of the 5d gauge theory phases.
Furthermore, since by construction the surfaces can be blown down to a point (namely the fiber singularity), the corresponding gauge theory also has a strong coupling limit.

In fact, different blow-up resolutions of the same non-minimal singularity will in general lead to different geometries for the surface components, thus yielding different weakly coupled gauge theories and SCFT limits.
One key point of the fiber resolution picture, however, is that we can easily identify the ``parent'' 6d conformal matter theory from which these different 5d phases originate:
It is the F-theory compactification defined on the singular elliptic threefold.

Compared to the local analysis, where the focus is just on the geometry of the compact surfaces, our approach provides another advantage, namely a geometric way to identify the flavor symmetry in the strong coupling limit. Flavor symmetry enhancements in 5d have already been successfully studied with other methods, such as analyzing the spectrum of operators charged under the instantonic $U(1)$ symmetry and the superconformal index \cite{Bergman:2013aca,Zafrir:2014ywa,Mitev:2014jza,Hwang:2014uwa,Tachikawa:2015mha,Yonekura:2015ksa,Zafrir:2015uaa,Bergman:2016avc}, or exploring the Higgs branch at infinite coupling by compactifying on a $T^2$ to a 3d $\mathcal N=4$ theory and constructing the Coulomb branch of the mirror dual Lagrangian theory \cite{Ferlito:2017xdq, Cabrera:2018jxt}.

A direct method for identifying flavor symmetries from the geometry has been previously presented in \cite{Xie:2017pfl}.
There, the authors conjectured that any canonical singularity in a non-compact Calabi--Yau threefold defines a 5d ${\cal N} = 1$ SCFT.
Using toric examples, it was shown that resolutions of these singularities encode global symmetries, both at weak and strong coupling, in terms of collapsing non-compact divisors.
Our approach is similar in practice, but conceptually it identifies these collapsing divisors as the exceptional divisors that resolve the minimal singularities over the codimension one loci $W_i$, thus revealing the higher dimensional origin of the 5d global symmetry.

This identification is possible because we not only perform the fiber resolution at the non-minimal point, but also over $W_i$.
Concretely, it allows us to track how the generic Kodaira fibers of $\hat{G}_i$ (the affinized version of $G_i$) split and become curves in the non-flat surfaces.
Since the generic fiber is homologous to the sum of its split products at a special fiber, 
it is not hard to see that, by blowing down the non-flat surfaces (either to a curve or a point), some of the curve components over $W_i$ may be forced to shrink too, thus enhancing the singularity over $W_i$ to $H_i \subset \hat{G}_i$.
In this way, the global symmetry $H_i$ of the 5d theory appears as a subgroup of the flavor symmetry of the 6d parent theory (affinized by the KK-$U(1)$ upon circle reduction).

In this paper, we will exemplify our proposal by considering non-flat resolutions of singularities which in 6d give rise to conformal matter with $(E_8,G)$ flavor symmetry, where $G \in \{\emptyset,SU(2), SU(3)\}$.
Their circle reductions yield 5d theories of rank 1, 2 and 4, respectively.
For simplicity, we will restrict ourselves to constructions based on so-called ``tops'' \cite{Candelas:1996su, Bouchard:2003bu}, where the elliptic fibration $Y_3$ is resolved through a hypersurface embedding into a (semi-)toric ambient space ${\cal X}_4$.
As we will demonstrate, our method to identify the global symmetries agrees with known results.

The paper is organized as follows.
In section \ref{sec:5d_gauge_theories}, we briefly review the gauge dynamics of 5d SCFTs and their construction via geometric engineering in M-theory.
Section \ref{sec:M-/F-duality_SCFTs} then makes connection to 6d theories via M-/F-theory duality, and discusses the physical difference of base and fiber blow-ups.
This motivates the study of non-flat resolutions of non-minimal elliptic singularities, which is detailed in section \ref{sec:non-flat_general}.
Here, we also explain the relationship between flavor symmetries in 6d and 5d, and how the non-flat geometry makes them manifest.
To demonstrate these ideas, we then turn to an explicit construction of rank one theories in section \ref{sec:rank_one}, where we analyze in detail all phases that are torically available.
Because the higher rank theories have a more complex RG-flow network, we will leave a full classification of their non-flat resolutions for future works.
Instead, section \ref{sec:rank_two} will contain isolated cases of rank two and four theories, where we focus on the appearance of 5d quiver theories and dualities between them.
Finally, after a summary, section \ref{sec:summary} will touch upon some aspects and open questions which we believe are worthwhile pursuing in the future.

\paragraph{Note Added:} In the final stages of preparing this manuscript, the work \cite{Bhardwaj:2018vuu} appeared which extends the results of \cite{Bhardwaj:2018yhy}.
While the motivation there is similar to ours, namely to study the Coulomb branch of 5d SCFTs from a parent 6d SCFT via M-/F-theory duality, the practical approaches are different.
The starting point of \cite{Bhardwaj:2018vuu} is the 6d tensor branch, that is, the elliptic threefold after blow-ups in the base.
The authors argue that they can reach the geometries (which cannot have a flat fibration) that describe the 5d gauge theory phases with UV fixed point in their Coulomb branch. 
This is done by a complicated series of flop transitions.
In contrast to this, for a large class of examples our approach gives directly these geometries, which manifestly describe the 5d gauge theories.
Our methods also allow us to study the existence of a 5d fixed point in the moduli space of these theories, as well as the enhanced flavor symmetries.

\section{5d Supersymmetric Gauge Theories}\label{sec:5d_gauge_theories}

In this section, we collect some basic facts about five dimensional supersymmetric gauge theories.
For a recent overview see, e.g., \cite{Jefferson:2017ahm,Jefferson:2018irk} and references therein.
Throughout this paper, unless otherwise stated, we will adopt the notation of using letters in lower case Fraktur ($\fkg$, $\mathfrak{su}$, ...) for gauge symmetries and normal upper case letters ($G$, $SU$, ...) for global and flavor symmetries.

The field content of a 5d gauge theory is given by vector multiplets $\mathcal A$ and hypermultiplets $\mathbf h$.
A vector multiplet consists of a 5d gauge potential $A_{\mu}$ transforming in the adjoint representation of $\mathfrak{g}$, a collection of real scalars $\phi^i$, $i=1, \ldots, \text{ rank}(\mathfrak g)$, and an adjoint fermion $\lambda$.
Hypermultiplets have two complex scalars (and a fermion $\psi$) transforming in an irreducible representation and its conjugate, $\mathbf R_{\mathfrak g} \oplus \mathbf{R}^c_{\mathfrak g}$, respectively.
If $\mathbf{R}_{\mathfrak g}$ is real or pseudo-real (i.e., the conjugate representation is the same), the matter comes in half-hypermultiplets transforming in $\mathbf R_{\mathfrak g}$. 
To summarize we have the following multiplets:
\begin{align}
\begin{split}
  &\mathcal A = (A_{\mu} , \phi^i, \lambda), \quad A_{\mu}, \lambda \in \mathbf{Adj}_{\mathfrak g}, \quad i=1, \ldots, \text{rank}(\mathfrak g) \, ,\\
  & \mathbf h = ( h \oplus h^c , \psi), \quad h \in \mathbf R_{\mathfrak g}, \quad h^c \in \mathbf R_{\mathfrak g}^c, \quad \psi \in \mathbf {\mathbf R}_{\mathfrak g} \oplus {\mathbf R}_{\mathfrak g}^c \, ,
\end{split}
\end{align}

In general, besides the gauge symmetry and the 5d Lorentz symmetry, we have a global $SU(2)_R$ R-symmetry rotating the two scalars $(h, h^c)$, and a flavor symmetry $G_f$ acting on the hypermultiplets.
Moreover, there is a global topological $U(1)_T$ symmetry associated with the current
\begin{equation}
J_{T}= \frac{1}{8\pi^2}  \star \text{Tr} (F \wedge F).
\end{equation} 
This symmetry acts on gauge instantons and can provide a further enhancement of the flavor symmetry at the UV fixed point.

5d gauge theories have a Coulomb branch, which is parametrized by the vacuum expectation values (vevs) of the scalar $\phi^i$ in the vector multiplets. 
At a generic point of this moduli space, when all the scalars have non-trivial vevs, the gauge symmetry is broken to $\fku(1)^{\text{rank}(\mathfrak g)}$.
Of course, there can be regions of the Coulomb branch where some of the vevs are trivial, in which case a non-abelian subalgebra of $\fkg$ is preserved. 
The Higgs branch on the other hand is parametrized by vevs of the complex scalars in the hypermultiplets.

Gauge theories in 5d can have a discrete theta angle which signals the sign flip of the partition function when integrated over spaces of different topology.
It is defined as an element of $\pi_4(G)$, where $G$ is the simply connected Lie group associated to $\fkg$.
For instance, we will see shortly that some of the rank one theories with $\fkg = \mathfrak{su}(2)$ can have a non-trivial discrete theta angle, as $\pi_4(SU(2)) \cong \mathbb Z_2$.

There exists a non-renormalizable effective Lagrangian on the Coulomb branch, which can be written as
\begin{equation}
\mathcal L_\text{eff}= G_{ij} \, d\phi^i \wedge \star d \phi^j + G_{ij} \, F^i \wedge \star F^j + \frac{c_{ijk}}{24 \pi^2} \, A^i \wedge F^j \wedge F^k + \ldots
\end{equation}
The couplings that appear, namely the metric $G_{ij}$ of the Coulomb moduli space and the Chern--Simons couplings $c_{ijk}$ can be derived from a quantity called the prepotential, $\mathcal F$:
\begin{align}
  \begin{split}
    & G_{ij} = \frac{\partial^2\mathcal F}{\partial \phi^i \partial \phi^j} \, ,\\
    & c_{ijk} = \frac{\partial^3\mathcal F}{\partial \phi^i \partial \phi^j\partial \phi^k} \, .
  \end{split}
\end{align}
Moreover, it is important to notice that $\partial_i \partial _j \partial_k \mathcal F= c_{ijk}$ are constant integers, $c_{ijk} \in \mathbb Z$.
This is because, even though the Chern--Simons term is not gauge invariant, its integral should be well defined modulo $2\pi i \, \bbZ$. 
The prepotential receives in general a classical contribution which reads
\begin{equation} \label{eq:prepcl}
\delta \mathcal{F}_{\rm classical}=\frac{1}{2}\,m_0 h_{ij} \phi^i \phi^j + \frac{ \kappa}{6} \, d_{ijk} \phi^i \phi^j \phi^k \, ,
\end{equation}
where $h_{ij}$ is the Cartan matrix of the Lie algebra $\mathfrak g$ and $d_{ijk}= \frac{1}{2} {\rm tr}_{\rm fund}\left( T_a (T_b T_c + T_c T_b)\right)$. 
The quantum contribution can be computed by evaluating the one-loop $AAA$ amplitude with fermions charged under the gauge symmetry that run in the loop \cite{Witten:1996qb}:
\begin{equation} \label{eq:prepq}
\mathcal \delta F_{\rm quantum}= \frac{1}{12} \left( \sum_{\alpha \in \mathbf L} |\alpha \cdot \phi|^3 - \sum_f \sum_{ w_f \in \mathbf W_f} |w \cdot \phi + m_f|^3\right) \, .
\end{equation} 
Here, $\alpha$ are the roots in the root lattice $\mathbf L$, $\phi \equiv \langle \phi \rangle$ is the vev that parametrize the Coulomb branch, $f$ is the number of hypermultiplets in a certain representation $\mathbf R_f$ (and its conjugate) of $\mathfrak g$, $\mathbf W_f$ are the weights of ${\bf R}_f$, and $\cdot$ is the bilinear product on the root lattice constructed via $h_{ij}$.
Gauge invariance and quantization of the integral of the Chern--Simons term implies that the quantum contribution is exact at one-loop (by dimensional analysis higher loops will violate the condition $c_{ijk} \in \mathbb Z$).
Furthermore, it is also important to highlight the fact that the fermions in the matter hypermultiplets contribute opposite sign with respect to the fermions in the vector multiplet. 
The total prepotential is given by
\begin{equation} \label{eq:preptot}
\mathcal F =\delta \mathcal{F}_{\rm classical} + \mathcal \delta F_{\rm quantum}.
\end{equation}

The Coulomb branch moduli space is known to have a wedge structure, and the Weyl chambers are defined by $\alpha \cdot \phi$ being positive (or negative) everywhere in a connected region; $\alpha \cdot \phi=0$ are the boundaries of the chambers.
On the other hand, it is still possible for $w \cdot \phi + m_f$ to change sign in a single Weyl chamber, which means that, because of the absolute values in \eqref{eq:prepq}, there can be different prepotentials within a single Weyl chamber.

The first derivative of the prepotential describes the tension of non-perturbative objects of the theory.
The BPS object in 5d gauge theories are monopole strings and electric particles (under compactification on a circle to four dimensions, these become magnetic monopoles and electric particles), whose central charges are given by
\begin{equation}
Z_e=\sum_{i} n_e^i \phi^i + m_0 n_I \, ,\qquad \qquad  Z_m=\sum_{i} n_m^i T(\phi)_i \, ,
\end{equation}
where $n_e^i, n_m^i$ are integral electric and magnetic charges.
Moreover, the instanton charge is defined as
\begin{equation}
n_I\equiv \frac{1}{8\pi^2}\int_{S^4} {\rm Tr}(F \wedge F),
\end{equation}  
where ${\rm Tr}\equiv {\rm tr}_{\rho}/{\rm Index}(\rho)$ is the normalized trace with respect any representation $\rho$ of $\fkg$. Finally, $T(\phi)_i$ is the tension of the monopole strings given by
\begin{equation}
T(\phi)_i\equiv \partial_i \mathcal F.
\end{equation}
These tensions can be thought of as a dual coordinate base on the Coulomb branch with respect to $\phi^i$.

Lastly, the first and second derivatives of the prepotential must be continuous all over the Coulomb branch, whereas $c_{ijk}$ can jump in integer values.
In particular, such jumps at the boundaries of the sub-wedges signals the presence of charged matter, which become massless at these boundaries.

\subsection{5d fixed points}

For unitarity and consistency of the theory, the gauge kinetic term must be positive which implies that 
\begin{equation} \label{eq:posconst}
G_{ij}=\partial_i \partial_j \mathcal F
\end{equation}
is positive semi-definite.
In general, the metric takes the following form,
\begin{equation} \label{eq:KMSmetric}
G_{ij} \sim m_0 h_{ij} +\kappa \, d_{ijk}\phi^k + c_{\text{quantum}, \, ijk}\phi^k,
\end{equation}
and when the quantum contribution $c_{\text{quantum}, \, ijk}$ is positive in the Weyl chamber, it is possible to have a scale invariant fixed point where the classical mass and inverse coupling are set to zero, $m_0= g_\text{classical}^{-2}=0$.
This is not possible when the quantum contribution is negative due to the positivity constraint \eqref{eq:posconst}.
So, the existence of a fixed point depends on the positivity of the second derivative of the quantum prepotential $\delta \mathcal{F}_\text{quantum}$, or, in other words, whether the prepotential is a convex function throughout the Weyl chamber. 
Moreover, from \eqref{eq:posconst}, we can see that charged matter hypermultiplets give concave contribution to $\mathcal F$, which means that there is an upper bound on the matter content.
However, the positivity of the second derivative is not the only constraint.
For instance one should also verify that the tensions of the monopole strings stay positive throughout the chamber,
\begin{equation}
T(\phi)_i >0, \qquad  \forall i = 1, \ldots, \text{rank}({\fkg}).
\end{equation}
Sometimes, it can happen that the metric is not positive in some region of the Coulomb branch.
However, before the metric becomes negative definite, the tension of some BPS object will change sign.
This signals a break-down of the effective description, since the vanishing tension of a BPS object indicates non-perturbative physics at that point.
In fact, in these cases there is usually an alternative effective description which is perturbatively valid at that point \cite{Jefferson:2017ahm}, which fully describes the dynamic of the Coulomb branch.
These more refined analysis allows also for quiver gauge theories, \cite{Bergman:2012kr}, and it will become more clear in the geometric description.

Finally, we note that many different gauge/quiver theories can have the same fixed point in the UV.
A 5d SCFT, if it exist in the moduli space of a gauge/quiver theory, is then believed to be characterized by their rank, which is the same as the Coulomb branch dimension, and by the enhanced flavor symmetry group.

\subsection[\texorpdfstring{$\mathfrak{su}(2)$}{su(2)} gauge theory with matter]{$\mathbf{\mathfrak{su}(2)}$ gauge theory with matter}\label{sec:review_su2_in_5d}

Let us discuss the simple example of rank one theories, i.e., $\fkg = \mathfrak{sp}(1) \cong \mathfrak{su}(2)$ gauge theory with matter. 
The Coulomb branch is characterized by $\phi^1\geq0$. 
The gauge symmetry is broken to the Cartan $\fku(1)$ when $\phi^1>0$. 
If we consider a theory with $N_f$ hypermultiplets in the fundamental of $\mathfrak{su}(2)$, the classical flavor symmetry is $SO(2N_f) \times U(1)_T$, where $U(1)_T$ is the topological $U(1)$.
At the boundary of the sub-wedge regions in the Weyl camber, where all masses of the flavors vanish, $m_f=0$, the quantum prepotential reads 
\begin{equation}
\mathcal F= \frac{1}{12} (16- 2N_f) \phi^3.
\end{equation}
It is clear from this that, in order to have a non-trivial fixed point, we need $N_f \leq 7$, whereas for $N_f=8$ there is no singularity in the moduli space, since the metric \eqref{eq:KMSmetric} in the limit $m_0 \rightarrow 0$ is trivial. For this reason, the theory with $N_f=8$ has no 5d fixed point, but it has a 6d UV completion, i.e., the E-string theory.

For $N_f\leq 7$, the strong coupling limit $g_{\rm classical}^{-2} \rightarrow 0$ enhances the flavor symmetry to $E_{{n_f}+1}$ \cite{Seiberg:1996bd}, where 
\begin{align}
  \begin{split}
    & E_1= SU(2) \, ,\\
    & E_2= SU(2) \times U(1) \, ,\\
    & E_3 = SU(3) \times SU(2) \, ,\\
    & E_4 = SU(5) \, ,\\
    & E_5 = SO(10) \,,
  \end{split}
\end{align}
and $E_6, E_7, E_8$ are the exceptional Lie groups. Moreover there are two outliers where the flavor symmetry at infinite coupling (conformal point) is $E_0= \emptyset$ and $\widetilde E_1=U(1)$. The $E_1$ and $\widetilde E_1$ gauge theories are distinguished by a different discrete theta angle for the gauge group, which is $SU(2)_0$ for $E_1$ and $SU(2)_{\pi}$ for $\widetilde E_1$.

As we will see later, we can reproduce this result also geometrically, including the two outliers which are difficult to track field theoretically.
Because we limit ourselves to toric constructions, our geometric examples do not cover cases $N_f \geq 3$.
However, as we will emphasize in section \ref{sec:non-flat_general}, we expect this limitation to be lifted for general, non-toric models.

\subsection{5d gauge theories and fixed points from geometry: brief summary}\label{sec:5d_gauge_theory_geometry}

Here we briefly list the fundamental ingredients to understand 5d gauge theories and their Coulomb branch phases from M-theory on singular Calabi--Yau threefold and their resolutions, following the work \cite{Intriligator:1997pq}.

We study M-theory on singular Calabi--Yau threefolds $Y$, whose blow-up resolution $\hat{Y}$ contains a bouquet of complex surfaces $S_i$.
Though the precise geometry of the $S_i$ depend on the resolution procedure, these surfaces are generically ruled, i.e., carry a fibration structure $S_i \rightarrow \beta_i$ with the generic fiber $\epsilon_i$ being a $\bbP^1$.
The collection of surfaces is usually arranged as the Dynkin diagrams of the associated gauge theory. 
When the surfaces intersect along (multi-)sections $\gamma_{ij}$ of the rulings on both $S_i$ and $S_j$, one can shrink the surfaces to a curve by blowing down the fibers $\epsilon_i$ of the rulings.\footnote{For a visualization of these configurations, we again refer the reader to \cite{Intriligator:1997pq}.}
This produces a curve-worth of singularities inside $\hat{Y}$, which realizes a gauge theory as follows:
\begin{itemize}
  \item Vector multiplets: 
    \begin{itemize}
      \item The simple roots of the adjoint weights are given by the M2-brane states wrapping the fiber curves $\epsilon_i$.
      \item The uncharged weights (i.e., those spanning the Cartan subalgebra) arise from KK-reduction of the M-theory $C_3$-form along the harmonic $(1,1)$-form Poincar\'{e}-dual to the divisor classes $[S_i] \in H_4(\bbZ, \hat{Y})$.
    \end{itemize}
  
  \item Fundamental hypermultiplets: These are given by M2-branes wrapping special fibers $\pm[\sigma_i]$ of the ruled surfaces $S_i$ with intersection numbers
    \begin{equation}
      S_j \cdot \sigma_i = \begin{cases}
        \pm 1,& \quad i=j, \\
        \mp 1, &\quad i=j+1,\\
        0, &\quad i\neq j,j+1.
      \end{cases}
    \end{equation}
  
  \item Adjoint hypermultiplets: These arise from the moduli of M2-branes wrapping $\epsilon_i$. Since $\epsilon_i$ is a fiber, its moduli space is precisely the base $\beta_i$ with genus $g$, and hence there are $g$ adjoint hypermultiplets \cite{Witten:1996qb}.

  \item Hypermultiplets with more exotic representations: If the base $\beta_i$ of intersecting surfaces are of different genera, M2-branes on special fiber components $\sigma$ can carry weights of (anti-)symmetric representations. 
\end{itemize}
Because we are shrinking the fibers, all the M2-brane states listed above will become massless, and hence give rise to well-defined gauge dynamics.
The precise gauge algebra depends on the topology of the bases $\beta_i$ and the gluing curves $\gamma_{ij}$.
For example, to realize an $\mathfrak{su}(n)$ gauge symmetry, $n-1$ surfaces, all ruled over a genus $g$ curve, must intersect along a chain such that $\gamma_{i,i+1} = S_i \cap S_{i+1}$ is a section of the rulings on $S_i$ and $S_{i+1}$.

To go to strong coupling, all surfaces must be further blown down to a point,
At this stage, also M2-branes states wrapping the bases $\beta_i$ will become massless, signaling additional light degrees of freedom in the spectrum.
In fact, one can dually see massless excitations of monopole strings, which come from M5-branes wrapping the surfaces that are tensionless in the singular limit.
These light states indicate a break down of perturbative physics.

Let us now relate the Coulomb branch scalars to geometric quantities.
The Coulomb branch is identified with the negative K\"ahler cone of $\hat{Y}$, with an the extra condition imposed by the shrinkability of curves in the singular limit $\hat{Y} \rightarrow Y$:
\begin{equation}
-\mathcal K(\hat{Y}/Y)= \left\{ S= \sum \phi^i S_i \; |\; -S\cdot \sigma >0 \quad \forall \sigma \in H_2(\hat{Y}) \text{ mapping to points in } Y \right\}.
\end{equation}
The prepotential is computed geometrically by
\begin{equation} \label{eq:geomprep}
\mathcal F = \frac{1}{6}c_{ijk}\phi^i \phi^j \phi^k \qquad \text{with} \qquad c_{ijk}=S_i \cdot S_j \cdot S_k \, .
\end{equation}
There can be different blow-up resolutions which lead to the same singular $Y$.
Physically, these resolutions correspond to different phases of the theory which are dual to each other.
These phases are all related by a sequence of flop transition, i.e., blowing down curves and blowing up new ones somewhere else.

Lastly, a 5d gauge theory has one further discrete label.
E.g., $\mathfrak{su}(n)$ theories with $n\geq 3$ are labelled by Chern--Simons level $\kappa$, which one can in principle determine this geometrically by comparing the prepotential \eqref{eq:geomprep} with the field theory data \eqref{eq:prepcl} -- \eqref{eq:preptot}.
For $\mathfrak{su}(2)$ and $\mathfrak{sp}(n)$ theories, there is a discrete theta angle $\theta = 0$ or $\pi$.

\section{SCFTs in M-/F-theory Duality}\label{sec:M-/F-duality_SCFTs}

In this section, we will briefly review the F-theoretic description of 6d conformal matter theories.
We then discuss the state of the art methods to relate these to 5d SCFTs via the duality between M- and F-theory  and motivate the study of non-flat resolutions.

\subsection{F-theory on elliptic threefolds with non-minimal singularities}

While the classification of 5d SCFTs is an open problem, the situation in 6d is much more pleasant:
there, all ${\cal N} = (1,0)$ SCFTs\footnote{Strictly speaking, the classification is for all 6d ${\cal N} = (1,0)$ SCFTs with an stringy embedding.} have been classified using the language of F-theory \cite{Heckman:2015bfa,Bhardwaj:2015xxa} (for a recent review, see \cite{Heckman:2018jxk}).
That is, any 6d SCFT can be described geometrically by a singular elliptically fibered Calabi--Yau threefold $\pi: Y_3 \rightarrow B_2$.

The physics of an F-theory compactification is determined by a Weierstrass model for $Y_3$,
\begin{align}\label{eq:weierstrass_model_generic}
  y^2 = x^3 + f\,x + g \, , \quad f \sim K^{-4}_{B_2} \, , \, g\sim K^{-6}_{B_2} \, ,
\end{align}
where $K_{B_2}$ is the canonical bundle of the base $B_2$.
The degeneration locus of the elliptic fiber, given by the vanishing of the discriminant,
\begin{align}
  \{ \delta := 4\,f^3 + 27\,g^2 = 0 \} \subset B_2 \, ,
\end{align}
corresponds to the location of 7-branes in the non-perturbative type IIB interpretation of F-theory \cite{Vafa:1996xn}.
The worldvolume dynamics of these branes gives rise to 6d ${\cal N} = (1,0)$ Yang--Mills theories, whose gauge algebras $\fkg_i$ are of the same ADE-type as the Kodaira singularities determined by the vanishing orders of $(f,g,\delta)$ as well as monodromy effects along irreducible components $W_i$ of $\delta$ \cite{Morrison:1996na, Morrison:1996pp, Bershadsky:1996nh, Aspinwall:2000kf, Grassi:2000we} (see also \cite{Weigand:2018rez,Cvetic:2018bni}).

Furthermore, at intersection points $p = W_i \cap W_j$, the fiber singularity in general enhances, indicated by a higher vanishing order of $(f,g,\delta)$.
For minimal singularities, i.e., ord$(f|_p , g|_p , \delta|_p) < (4,6,12)$, the corresponding ADE algebra $\mathfrak{h}_p \supset \fkg_i \oplus \fkg_j$ indicates the presence of matter states in representations ${\bf R}_k$ according to the ``Katz--Vafa'' rule \cite{Katz:1996xe,Bershadsky:1996nh}:
\begin{align}
  \begin{split}
    & \mathfrak{h} \supset \fkg_i \oplus \fkg_j \, , \\
    & {\bf adj}_\mathfrak{h} \longrightarrow {\bf adj}_{\fkg_i} \oplus {\bf adj}_{\fkg_j} \oplus \bigoplus_k {\bf R}_k \, .
  \end{split}
\end{align}
However, if the singularity types on $\Sigma_{i,j}$ are too severe, their collision will lead to a so-called non-minimal singularity with ord$(f|_p,g|_p,\delta|_p) \geq (4,6,12)$.
For such singularities, there is no associated ADE-algebra, and hence no conventional matter states.

To make sense of them in F-theory compactifications to 6d, one can blow-up the base $B_2$ at the point $p$.
This procedure introduces a collection of rational curves $\Sigma_i \subset \tilde{B}_2$ in the blown-up base, over which the (pulled-back) elliptic fibration $\tilde{\pi}: \tilde{Y}_3 \rightarrow \tilde{B}_2$ only has minimal singularities, i.e., ordinary gauge algebras $\fkg_i$ and matter representations.

Physically, the blow-up curves $\Sigma_i$ support 6d tensor multiplets dual to BPS strings, the latter arising from D3-branes wrapping $\Sigma_i$ in the IIB picture.
The volume of $\Sigma_i$ correspond to the vacuum expectation value of the scalar in the tensor multiplet, and parametrize the so-called tensor branch of the 6d theory.
Furthermore, if $\Sigma_i$ carries singular fibers, i.e., supports a gauge theory, its squared gauge coupling is proportional to vol$(\Sigma_i)^{-1}$.

Blowing down these curves, thus reproducing the non-minimal singularity, we immediately see that the gauge coupling becomes formally infinite.
In addition, even if $\Sigma_i$ carries no gauge symmetry, the strings from wrapped D3-branes still become tensionless, such that their excitations give rise to infinitely many light degrees of freedom.
Both observations indicate a strongly coupled sector without a Lagrangian description.
To obtain an honest SCFT, the F-theory base $B_2$ has to decompactify in order to decouple gravity.
In this limit, all the irreducible components $W_i$ of the discriminant decompatify as well.
Since the gauge coupling is inversely proportional to their volume, we see that the gauge symmetries on $W_i$ become non-dynamical, i.e., global, or flavor symmetries of the SCFT.

Some 6d SCFTs can also arise solely from a singular point $p$ of the base, without colliding codimension one singularities \cite{DelZotto:2014fia}.
These SCFTs typically have no global symmetries.
In addition, it is not entirely clear how the concept of fiber resolutions works in this context (see, however, \cite{Anderson:2018heq} for recent studies of such examples).
For the purpose of this paper, we will therefore assume that $p\in B_2$ is smooth, in which case the non-minimal singularity in the fiber over $p$ has to come from the collision of two non-compact divisors $W_{1,2}$ with ADE fibers.
The associated Lie groups $G_{1,2}$ constitute then (a subgroup of) the global symmetry of the SCFT at $p$.
These theories are called \textit{6d conformal matter theories} and are important building blocks of 6d SCFTs.


\subsection{Circle reduction of conformal matter theories and fiber resolutions}\label{sec:M-/F-duality}

With the classification of 6d SCFTs on one hand, and the motivation to better understand 5d SCFTs on the other, a natural object to study is the circle reduction from 6d to 5d.
This is particularly appealing in the spirit of the M-/F-theory duality, as it connects the 6d classification directly to geometric engineering of 5d theories via M-theory.

Due the complications arising from the strong coupling nature of SCFTs, the practical way found in the literature is to first push the 6d F-theory model onto the tensor branch by blowing up the base into $\tilde{B}_2$, and to then study the 5d theory obtained by compactifying M-theory on $\tilde{Y}_3 \rightarrow \tilde{B}_2$ \cite{DelZotto:2017pti, Bhardwaj:2018yhy} (see also \cite{Jefferson:2017ahm, Jefferson:2018irk}).
This applies in particular to 6d conformal matter theories, which will be the only type of 6d SCFTs we consider for the rest of the paper.

After moving onto the tensor branch of the 6d theory, i.e., when we consider the threefold $\tilde{Y}_3$, the M-theory compactification sees a collection of compact surfaces in $\tilde{Y}_3$, which in general are blown-down to curves.
Indeed, the elliptic fibration $\tilde{Y}_3$ will generically have ADE-singularities of type $\fkg_i$ over the blow-up curves $\Sigma_i$.
Blowing up these singularities will yield rank$(\fkg_i)$ compact divisors $S_{i;a}$, $a = 1, ..., \, \text{rank}(\fkg_i)$ that are ruled surfaces over $\Sigma_i$, whose $\bbP^1$ fibers intersect in the Dynkin diagram of $\fkg_i$.
Following the discussion in section \ref{sec:5d_gauge_theory_geometry}, we conclude the presence of a 5d gauge theory with gauge algebra $\fkg_i$, when we shrink the fibers of all $S_{i;a}$.

On $\tilde{Y}_3$, there is another divisor $S_{i;0}$ ruled over $\Sigma_i$, whose fiber is the \textit{affine} component of the Kodaira fiber over $\Sigma_i$.
In 5d, one is allowed to shrink also the affine node, as long as at least one of the other fiber components remains at finite size.\footnote{Otherwise, the generic elliptic fiber of $\tilde{Y}_3$ would shrink to zero size.
Of course, we know that this limit corresponds to the 6d F-theory limit of M-theory on $\tilde{Y}_3$.}
The enhanced 5d gauge symmetry is then a sub-algebra $\mathfrak{h}_i$ of the ``affinized'' version $\hat{\fkg}_i$ of $\fkg_i$.
In a certain sense, one can view $\hat{\fkg}$ as augmenting the gauge symmetry $\fkg$ by the KK-$U(1)$ from the circle reduction.
However, we cannot enhance this full symmetry without simultaneously unwinding the $S^1$ and ending up in 6d.

Nevertheless, given that a further blow-down of $\Sigma_i$ in the base reduces the surfaces with blown-down fibers to a point, one might be tempted to say that this is the 5d SCFT limit of the $\mathfrak{h}_i$ gauge theory.
However, as pointed out in \cite{DelZotto:2017pti, Jefferson:2018irk}, this limit is a so-called 5d KK-theory, as the light degrees of freedom are all accompanied by a tower of massive KK-states.
To obtain theories with non-trivial 5d UV fix points, one has to mass deform the KK-theory.

Within the duality \eqref{eq:M-/F-theory_duality} between M- and F-theory, such mass deformations correspond to turning on non-trivial holonomies of the flavor symmetry along the $S^1$ on which we reduce the 6d F-theory.
Locally, these parameters change the volumes of special fibers inside the ruled surfaces $S_{i;a}$ supporting the gauge symmetry.
These special fibers give rise to charged matter states via wrapped M2-branes.
By tuning such volume parameters to formally change sign, the geometry undergoes a flop transition, whereby a holomorphic curve shrinks to zero size, and another one gets blown-up.
In extreme cases, when the flopped curves are generic fibers of a fibered surface, the surface will have to shrink to a curve of singularities before blown up again into a different surface.
Strictly speaking, such a process is a geometric transition rather than a flop.

The resulting surfaces $S_k$ will in general no longer be compatible with an elliptic fibration.
For example, in the cases of 5d rank one theories, the KK theory is obtained via compactification of the tensor branch of the 6d E-string theory.
Geometrically, we have a single compact blow-up curve $\Sigma_i \cong \bbP^1 \subset \tilde{B}_2$ over which the elliptic fibration $\tilde{Y}_3$ is generically smooth.
The resulting elliptic surface $S_{i;1} = \tilde{\pi}^{-1}(\Sigma_i) \cong \text{dP}_9$ has many curves (in fact, infinitely many) in form of sections, which upon (successive) flopping produces the other del Pezzo surfaces $S \cong \text{dP}_{n \leq 8}$.
Since these surfaces have no elliptic fibration structure, the full threefold cannot possibly be elliptically fibered---at least not in the ordinary way.
Thus, it seems that the geometry after such a transition is only distantly related to the original F-theory geometry one started with.

However, from the perspective of M-/F-theory duality, non-trivial $S^1$ holonomies of a gauge field supported on a base divisor $W \subset \{\delta =0 \}$ precisely correspond to a blow-up resolution of the fiber singularities over $W$.
For F-theory compactifications with only minimal singularities, i.e., only ordinary matter representations, such holonomies will generically make all matter states charged under the gauge algebra over $W$ massive.
Geometrically, this means that all singularities in codimension two in the base are also resolved through the appearance of additional curves in the fiber.\footnote{For our purposes, we assume that there are no terminal singularities in the vicinity of the non-minimal singularities. In general, their physical significance has been discussed in \cite{Arras:2016evy,Grassi:2018rva}.}
Furthermore, it is well-known that there can different phases of the fiber resolution related by flop transitions, which differ by the fiber structures in codimension two in the base.

With this in mind, we therefore ask if a \textit{fiber-resolution} of non-minimal singularities can teach us something non-trivial about the associated 5d SCFTs.\footnote{See \cite{Ganor:1996gu,Klemm:1996hh,Ganor:1996pc} for earlier works on circle compactifications of the E-string in this spirit.}
Following the above intuition, we further expect the resolution of codimension one fibers over $W_{1,2} \subset B_2$ to give rise to a non-trivial mass deformation of the 5d KK-theory, which is obtained via $S^1$-reduction of the 6d SCFT living at the intersection $p = W_1 \cap W_2$.

Given that an M-theory realization of 5d gauge theory requires (possibly collapsed) compact surfaces in the geometry, it might seem puzzling that a fiber resolution of an elliptic fibration alone could produce such a setting.
However, because of the non-minimal singularity type, the fiber resolution actually introduces surface components.
As such, the resulting total space is what is called a \textit{non-flat} fibration.







\section{Non-flat Resolutions of Non-Minimal Singularities}\label{sec:non-flat_general}

Resolutions of minimal singularities of elliptic fibrations always introduces rational curves into the fiber.
As such, the fiber of the resolved fibration $\hat{\pi} : \hat{Y}_3 \rightarrow B_2$ always has (complex) dimension one.
In algebraic geometry, such a fibration is called \textit{flat}.
Conversely, given a fibration where the dimension of the fiber jumps from a generic to a special fiber, one typically refers to it as being \textit{non-flat}.
For an elliptic fibration $\pi: Y_3 \rightarrow B_2$ with a non-minimal singularity over a \textit{smooth} point $p \in B_2$, we conjecture that a full resolution $\hat{Y}_3$ of $Y_3$ without changing the base $B_2$ will generically introduce surface components into $\hat{\pi}^{-1}(p)$, i.e., $\hat{\pi} : \hat{Y}_3 \rightarrow B_2$ is non-flat.

Although we do not know of a strict mathematical proof of this statement, we point out that this phenomenon has been observed frequently in the F-theory literature \cite{Candelas:2000nc,Denef:2005mm,Braun:2011ux,Mayrhofer:2012zy,Lawrie:2012gg,Braun:2013nqa,Cvetic:2013uta,Borchmann:2013hta,Lin:2014qga,Anderson:2016ler,Anderson:2017aux,Buchmuller:2017wpe,Dierigl:2018nlv,Achmed-Zade:2018idx,Tian:2018icz}.\footnote{Non-flat fibers occur frequently in elliptic fourfolds, which were mostly studied for phenomenological purposes. However, there, surface components in non-flat fibers in codimension three do not define divisors in a fourfold, and thus appear on different footings as in threefolds.
Consequently, we lack a good understanding of the resulting 4d/3d physics of F-/M-theory.}
More importantly, however, we expect this phenomenon from the M-/F-theory duality!
As discussed in the previous section, a fiber resolution corresponds to turning on holonomies along the $S^1$, which in turn generates a mass deformation of the 5d KK-theory.
For generic deformations, this will lead to a weakly coupled 5d gauge theory, which in M-theory has to be realized on compact surfaces.
Without changing the base or the codimension one structure of the fibration, the only ``place'' these surfaces $S_k$ can appear is therefore in the fiber over $p$.

As anticipated in subsection \ref{sec:5d_gauge_theory_geometry}, these surface components are generically ruled, and have reducible special fibers.
Moreover, the surfaces will generically intersect each other along curves.
In addition to the surface components $S_k$, the fiber $\hat{\pi}^{-1}(p)$ also contains other curves which lie completely outside the surfaces, with some intersecting the surfaces only in points.

Focusing just on the local geometry of the intersecting surfaces, one can apply the criteria from \cite{Morrison:1996xf,Intriligator:1997pq,Jefferson:2017ahm,Jefferson:2018irk} to analyze the resulting M-theory gauge theory (see also section \ref{sec:5d_gauge_theories}).
In addition, one will find curves $\Gamma$ inside the surfaces which have zero intersection number with the canonical divisors of the surfaces:
\begin{align}
  0 = \Gamma \cdot_{S_k} K_{S_k} = \Gamma \cdot_{\hat{Y}_3} S_k \, .
\end{align}
Because of this, the volume of these curves are not controlled by the K\"ahler parameters dual to the compact surfaces.
From the gauge theory perspective, it means that the masses of M2-brane states on these curves are not set by the Coulomb branch parameters, and hence correspond to external parameters of the gauge theory.

However, recall that the non-minimal singularity was a result of the collision of two ADE singularities $G_{1,2}$ along $W_{1,2}$.
Therefore, the fiber components over $W_i$, which at a generic point $q \in W_i$ form nodes of the affine Dynkin diagram of $G_i$, must also appear within the non-flat fiber $\hat\pi^{-1}(p)$.
Let us denote the (non-compact) exceptional resolution divisors over $W_i$ by $A_{i;a}$, $a = 1, ..., \text{rank}(G_i)$, and their fibers by $\bbP^1_{i;a}$.
If the singularity type at $p$ was minimal, then it is a well-known story that some of the $\bbP^1_{i;a}$ would split over $p$, giving rise to matter charged under $G_i$.\footnote{For example, in a collision of $G_1 = SU(N)$ with $G_2 = ``SU(1)"$, the latter corresponding to an $I_1$ fiber over $W_2$, one of the $N$ fibers over $W_1$ that corresponded to a simple root of $SU(N)$ would split into two curves carrying the weights of an (anti-)fundamental state.}
The same situation now also occurs within the non-flat fiber, except for the appearance of the surfaces $S_k \subset \hat\pi^{-1}(p)$ \cite{Lawrie:2012gg}.

In particular, the curves $\Gamma$ contained inside some $S_k$ with $\Gamma \cdot S_k = 0$, i.e., those whose volume correspond to the mass parameters of the gauge theory defined on $S_k$, can be identified with (split) components of the codimension one fibers.
We will argue in the following that this fact allows us to read off the global symmetries of the 5d gauge theory and its SCFT limit.

\subsection{Global symmetry from the global resolution}\label{sec:global_symmetry_general}

It was anticipated in \cite{Xie:2017pfl} via toric models that one can determine the global symmetries of a 5d SCFT explicitly from the degeneration of non-compact divisors of the Calabi--Yau threefold $\hat{Y}$.
As we will describe now, these divisors are naturally identified with the exceptional divisors which resolve the ADE singularities of the elliptic fibration in codimension one in the base.
Within the M-/F-theory duality, we thus identify the global symmetry as a subgroup of the global symmetries that arise from circle reducing the 6d theory.

Let us spell out these ideas in more detail.
First, we can identify the rank of the global symmetry group $G_f$ with the rank of the matrix $D_l \cdot \Gamma_m$, where $\Gamma_m$ are all independent classes of curves inside the compact surfaces $S_k$.
$D_l$ on the other hand are all independent classes of non-compact divisors of $\hat{Y}$, which include the exceptional divisors $A_{i;a}$ and (multi-)sections.
These divisors are dual to Cartan $U(1)$s of the flavor symmetry in 6d and the KK-$U(1)$, from which we expect 5d global symmetries to originate.

\begin{figure}[ht]
  \centering
  \includegraphics[width = \hsize]{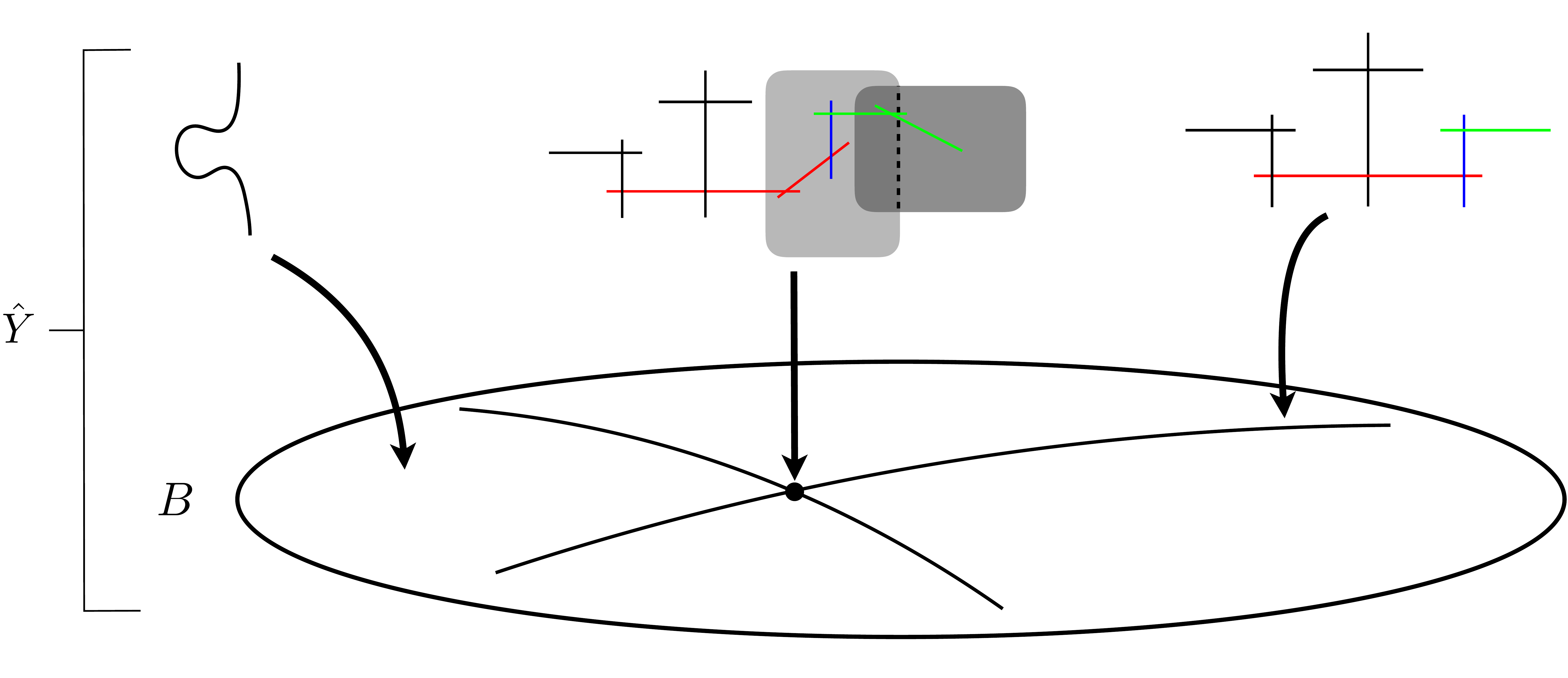}
  \caption{A cartoon illustrating the splitting of fibers after blowing up the non-minimal singularity over a point of the base.
  The gray squares represent the surface components of the non-flat fiber.
  In general, codimension one fibers can split into several curves which all (green) or only partly (red) lie inside the surfaces.
  They can also remain irreducible, and sit inside (blue) or outside (black) of any surface.
  By blowing down the surfaces to a point, the blue and green fibral curves are forced to shrink over the codimension one loci of the base.
  In the case depicted here, this would lead to an enhancement of an $SU(3)$ subgroup of the full $E_6$ flavor symmetry.
  }\label{fig:non-flat_fiber_cartoon}
\end{figure}

In order to determine the non-abelian part of $G_f$, we use the basic K\"ahler geometry fact that algebraic curves homologous to each other inside $\hat{Y}$ have the same volume.
Therefore, if one curve within a class shrinks to zero volume, then so do all the others.
Now, as we have pointed out above, some of the resolution curves over the codimension one loci $W_{1,2}$ may split into several curves at $p = W_1 \cap W_2$, of which one or some could be contained in a surface $S_k$.
In particular, it can also happen that a codimension one fibral curve remains unchanged over $p$, and simply sit as a whole inside one of the surfaces.
See figure \ref{fig:non-flat_fiber_cartoon} for illustrations of these cases.

When the elliptic fibration is fully resolved, all curves and surfaces are at finite size.
Then, the M-theory compactification is at a generic point on its Coulomb branch, where all charged states are massive.
The gauge symmetry here is just $\fku(1)^n$, where $n$ is the number of compact surfaces $S_k$.
If the surfaces $S_k$ are ruled, then by blowing down their generic fiber, the surfaces shrink to curves (which form singular loci of the total space), thus enhancing the gauge symmetry to a rank $n$ non-abelian algebra $\fkg$.
Because we are blowing down the generic fiber of $S_k$, any individual component of special fibers inside $S_k$ also shrinks.

In the fully blown-up case, i.e., where the singularities over $W_i$ are resolved, one can easily track how the generic fibers $\bbP^1_{i;a}$ splits into curves $\Gamma_l$ inside $\hat{\pi}^{-1}(p)$ (including possible multiplicities).
But because fibral curves of $A_{i;a}$ are homologous to each other, we have, in terms of homology classes,
\begin{align}\label{eq:homology_fiber_split}
  [\bbP^1_{i;a}] = \sum_l \lambda_{l} \, [\Gamma_{l}] \, .
\end{align}
The relative locations of the split components $\Gamma_l$ and the non-flat surfaces now allow for several possibilities.
\begin{enumerate}
  \item One of the $\Gamma_l$ is not contained in any non-flat surface $S_k$.
    In this case, $\bbP^1_{i;a}$ does no shrink regardless of the fate of $S_k$.
  \item All $\Gamma_l$ are inside the non-flat surfaces, and become fibers of the ruling on $S_k$.
    In this case, once we enhance the gauge symmetry to $\fkg$ by shrinking along the ruling, $\bbP^1_{i;a}$ shrinks everywhere over $W_i$.
  \item All $\Gamma_l$ are inside the non-flat surfaces, but some form a (multi-)section of the ruling on $S_k$.
    In this case, $\bbP^1_{i;a}$ will only shrink when we blow down $S_k$ to a point, i.e., go to the SCFT limit of the gauge theory.
\end{enumerate}
In general, the shrunken codimension one fibers form a subset of nodes of the affine Dynkin diagram of $G_i$.
In other words, blowing down $S_k$ (to either a curve or a point) will in general lead to a fibration of canonical singularities over $W_i$, whose ADE type $H_i$ correspond to the global symmetry $G_f$ of the 5d theory \cite{DelZotto:2017pti,Xie:2017pfl}.
If the fiber splittings are only of case 1., then the 5d theory only has abelian flavor symmetries.
In case 2., the weakly coupled gauge theory exhibits a non-abelian flavor symmetry.
As case 3.~illustrates, we generically expect a further non-abelian enhancement of $G_f$ from additional singularities over $W_i$ in the SCFT limit of the gauge theory.
We have summarized these possibilities in figure \ref{fig:non-flat_fiber_cartoon}.

Differently from the work of \cite{Xie:2017pfl}, our approach provides an explicit geometric link between 6d and 5d SCFTs and their flavor symmetries.
In particular, it is known that the only non-compact divisors in crepant resolutions of canonical Calabi--Yau threefold singularities are fibrations of resolved surface ADE singularities (see, e.g., footnote 5 in \cite{Jefferson:2018irk}).
Therefore it seems natural to identify these singularities as the fiber singularities of an non-flat elliptically fibered threefolds, which, as we argued, always has a 6d SCFT with prescribed flavor symmetries associated with it. In what follows we work under the assumption that the starting point of our analysis is the singular non-minimal elliptic fibration, which manifestly has the largest possible enhanced 6d flavor symmetry.
We believe that this is sufficient in order to see all the possible enhancements in 5d.
For example, we will consider the 6d E-string theory described by F-theory with an $E_8$ locus colliding an $I_1$, instead of two colliding $SO(8)$s.

Moreover, by combining the two conjectures that a), all 5d SCFTs arise from circle reductions of 6d SCFTs, and b), all canonical threefold singularities define a 5d SCFT, it seems natural to attempt a classification of threefold singularities via elliptically fibered Calabi--Yaus and F-theory.
Following our arguments, we believe that non-flat fibers would play a central role in this endeavor.
We leave a more thorough investigation of these speculations for future works.

For the remainder of this paper, we will instead provide examples of the interplay between non-flat fibers and 6d/5d SCFTs.
To have an easy method for efficiently producing different resolutions of non-minimal singularities, we will use toric methods, where some geometric aspects can be described using combinatorics.
Thus, before presenting the models, we will briefly review the necessary toric geometry tools.

\subsection{Non-flat resolutions from toric constructions}\label{sec:non_flat_resolution_toric}

The canonical singularities considered in \cite{Xie:2017pfl} allowed for a toric resolution.
That is, the compact divisors which are used to blow up the singularities were toric, and could be described via a 2d lattice polygon.
There, the authors make no reference to an elliptic fibration in which these surfaces are embedded in.

However, as we will describe now, such singularities and the corresponding resolutions appear naturally in toric constructions of elliptically fibered Calabi--Yau hypersurfaces.
In particular, a convenient way to resolve elliptic singularities torically is via so-called tops \cite{Candelas:1996su, Bouchard:2003bu}.
In the following we will give a qualitative description about these methods, focusing on the appearance of non-flat fibers and the combinatorics associated with different resolutions and flop transitions.

The construction is based on Batyrev's insight \cite{Batyrev:1994hm} that a pair of reflexive polytopes, $(\Diamond, \Diamond^*)$, define a Calabi--Yau hypersurface $\hat{Y} = \{P=0\}$ inside the toric variety ${\cal X}$ whose fan is given by a triangulation ${\cal T}$ of $\Diamond$.
The precise form of $P$ is determined by the dual $\Diamond^*$.
A suitable $\Diamond$ will give rise to an elliptically fibered hypersurface.\footnote{In fact, it has been recently shown that almost all such constructions of Calabi--Yau threefolds have multiple elliptic or genus-one fibrations \cite{Huang:2018esr}.}

Batyrev's construction can be further specialized in order to engineer elliptic fibrations with a desired codimension one Kodaira singularity.
The basic idea proposed in \cite{Candelas:1996su} is to ``chop up'' $\Diamond$ into two pieces, a ``top'' $\Delta$ and a ``bottom'' $\nabla$.
Then, following the classification in \cite{Bouchard:2003bu}, constructing the top specifies an elliptic fibration with a prescribed fiber degeneration over codimension non-compact one loci $W_i$.
Including the bottom can be then thought of as compactifying the base, i.e., embedding the local structures coming from the top into a global, compact elliptic Calabi--Yau.
Since we are interested in SCFTs, a local description of the singularity suffices, so the information contained in the top $\Delta$ and its dual $\Delta^*$ is all that we need.

Given a top $\Delta$, the polynomial cutting out the hypersurface $\hat{Y}$ takes the form
\begin{align}\label{eq:toric_hypersurface_general}
  P = \sum_k b_k \, \prod_l q_l^{n_l} \, .
\end{align}
Note that ${\cal X} \stackrel{\varphi}{\rightarrow} B$ is itself fibered over $B$, and the coefficients $b_k$ are sections of suitable line bundles on $B$.
The fibers of $\varphi$ are toric spaces with homogeneous coordinates $q_l$ corresponding to lattice vectors in $\Delta$, but $B$ can be generically non-toric.
The elliptic fibration can be viewed as elliptic curves defined by $\{P=0\} \equiv \{P\}$ inside the toric space $\varphi^{-1}(p)$, fibered over points $p \in B$ of the base.
The precise form of $P$ can be derived from the dual top $\Delta^*$, see \cite{Bouchard:2003bu}.

Irrespective of the expression of $P$, the geometry of $\hat{Y}$ has some generic features which we discuss now.
For that, let us denote the coordinates of the lattice vectors by $\vec{v} = ({\rm x}_1, {\rm x}_2, {\rm x}_3)$.
One 2d facet of $\Delta$ is one of the 16 reflexive 2d polygon $F_m$, which we can assume to sit at ${\rm x}_3=0$.
Vectors at vertices of $F_m$ define toric divisors of ${\cal X}$ which intersect the generic smooth elliptic fiber of $\hat{\pi}: \hat{Y} \rightarrow B$ and hence correspond to (multi-)sections of $\hat{Y}$ \cite{Rohsiepe:2005qg,Grassi:2012qw,Braun:2013nqa}.
Divisors with vectors interior to an edge $\cal E$ of $F_m$, on the other hand, do not intersect the generic fiber, but instead restrict to exceptional divisors which resolve an ADE singularity of type $G_{\cal E}$, fibered over a non-compact divisor $W_{\cal E} \subset B$.
The codimension one structures and the resulting F-theory physics of the elliptic fibrations associated with a choice of $F_m$ have been classified in \cite{Klevers:2014bqa}.
For an example, which we will analyze in detail later on, see figure \ref{fig:F10_with_blow-up}.

In addition to these structures coming from $F_m$, one can further engineer Kodaira singularities of ADE type $G$ over a codimension one locus $W$ in the base by including vectors $\vec{v}$ with ${\rm x}_3>0$, which fill out the top $\Delta$.
A full classification of possible tops for a given polygon $F_m$ at ${\rm x}_3=0$ has been worked out in \cite{Bouchard:2003bu}.
The toric divisors corresponding to vectors $\vec{v}$ with ${\rm x}_3>0$, which sit on edges (i.e., 1d facets) of $\Delta$, give rise to the exceptional divisors resolving the singularities over $W$.

However, the top can also have (2d) facets which contain interior vectors with ${\rm x}_3>0$.
In the ambient space ${\cal X}$, their corresponding divisors $D_\text{int}$ are also fibered over $W \subset B$, i.e., $\varphi(D_\text{int}) = W$.
However, they intersect the elliptically fibered hypersurface $\{P =0 \} \equiv \{P\}$ over a codimension two locus of the base \cite{Hu:2000pr,Braun:2011ux,Braun:2013nqa}.
For a fibration over a two dimensional base, this means that the four-cycle $D_\text{int} \cap \{P\}$ must be completely contained inside the fiber at that locus!
These four-cycles therefore give rise to non-flat fiber components of $\hat{Y}$.

In the top $\Delta$, each edge $\cal E$ of $F_m$ bounds a facet ${\cal F}_{\cal E}$.
Any internal vector of ${\cal F}_{\cal E}$ will then correspond to a non-flat fiber component over the point $W \cap W_{\cal E}$.
In M-theory, these compact surfaces support the 5d gauge sector, which arise from a circle reduction with non-trivial holonomies of a 6d $(G, G_{\cal E})$ conformal matter theory.

\begin{figure}[ht]
  \centering
  \subfigure[The $E_8$-top over $F_{10}$.]{
    \includegraphics[width=.47\hsize]{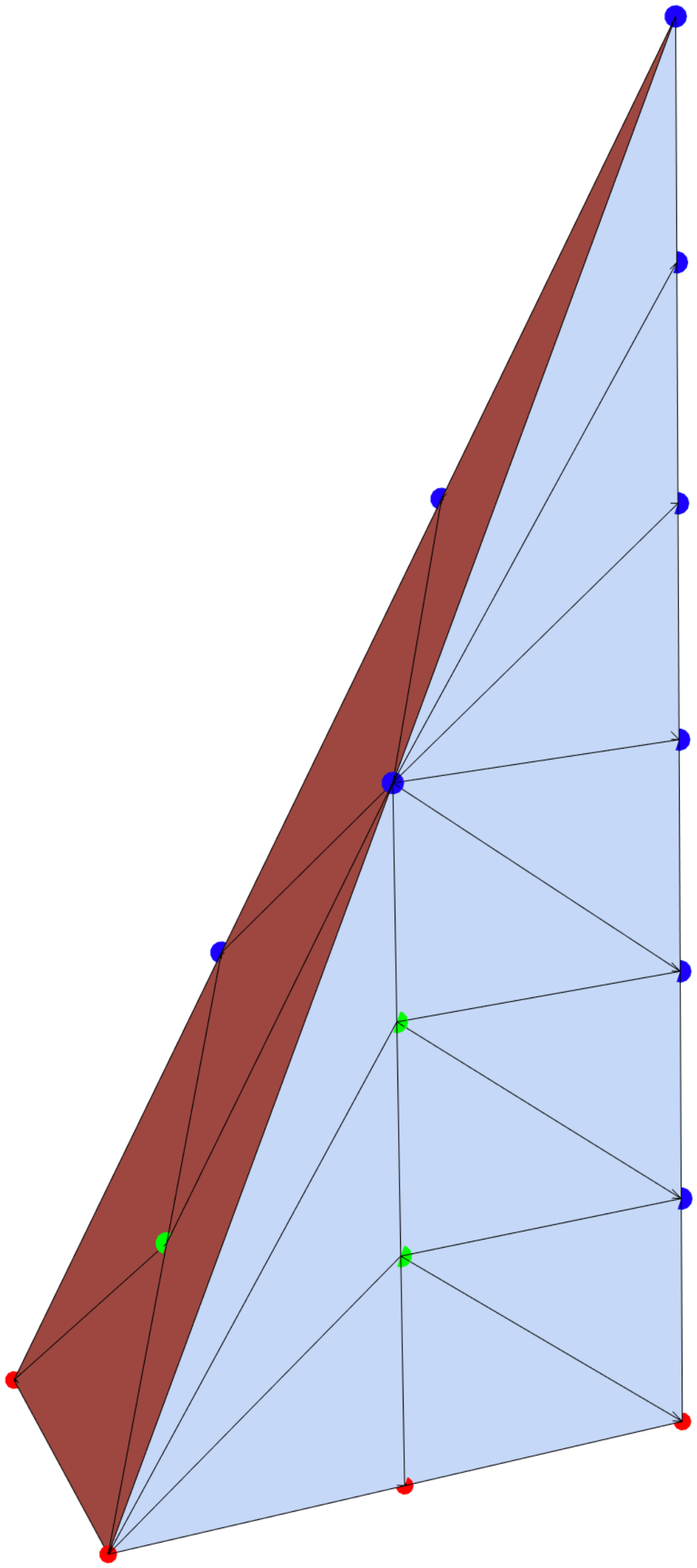}
  }
  \hfill
  \subfigure[Another perspective of the top.]{
    \includegraphics[width=.47\hsize]{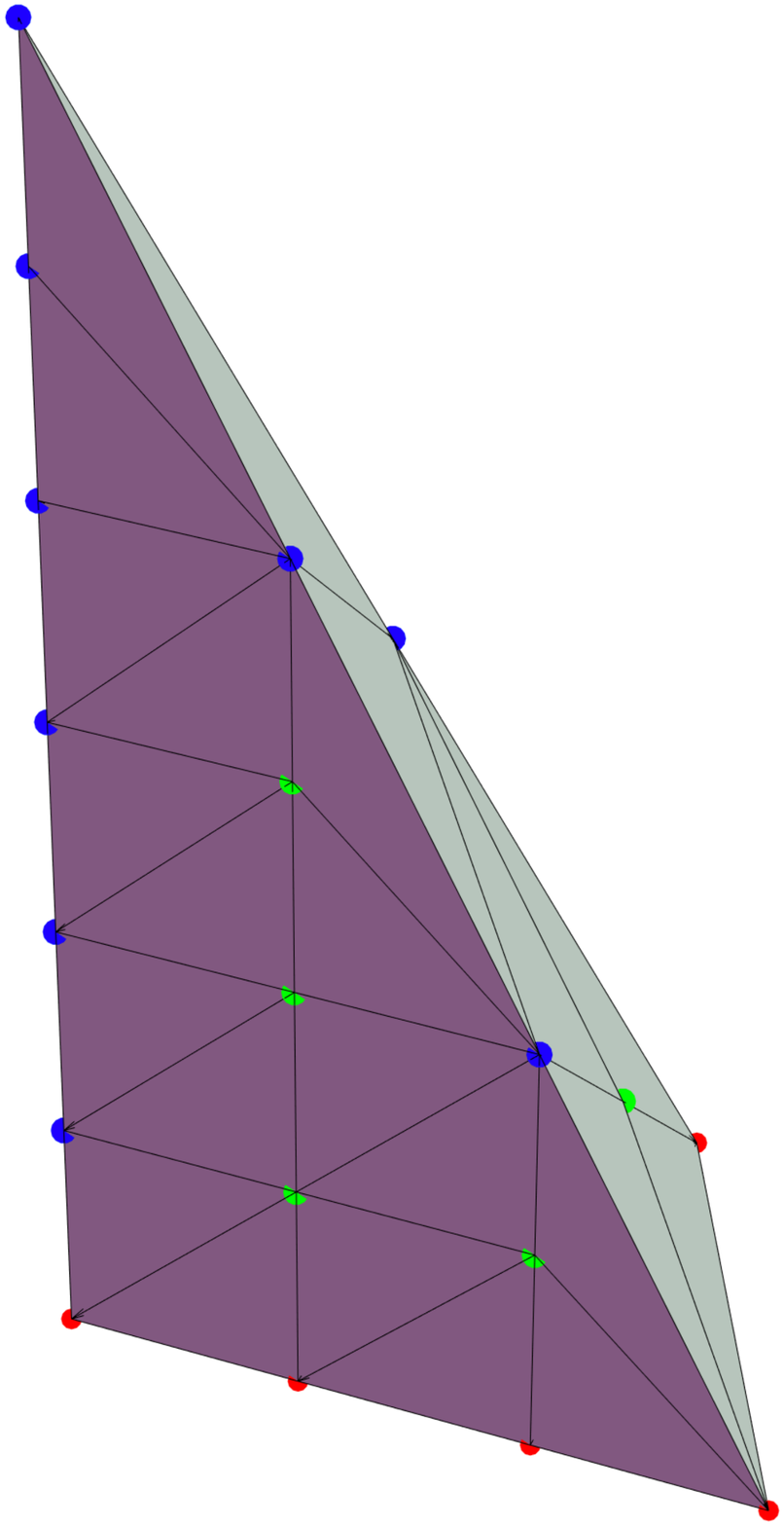}
  }
  \caption{A visualization of the $E_8$-top over $F_{10}$.
  The vertical direction is ${\rm x}_3$.
  At ${\rm x}_3 = 0$, the red points are the lattice vectors of $F_{10}$.
  For ${\rm x}_3 >0$, the blue points represent the vectors of the exceptional divisors of the $E_8$, as they sit on the edges of the top.
  The green points are facet interiors and correspond to non-flat surfaces.
  Different facets encode the local geometry at different non-flat fibers.
  The triangulation shown in the figures is arbitrary; for the rank one sector it defines the dP$_1$ phase.}\label{fig:top}
\end{figure}

As an illustration, we have displayed in figure \ref{fig:top} a graphical representation of the top that specifies the geometries we will consider in the next section.

\subsubsection{Resolution phases and flop transitions from triangulations}

To fully specify the toric geometry, one needs to define fan among the vectors of the toric divisors.
In the case of tops, such a fan is provided by a triangulation $\cal T$ of $\Delta$.\footnote{As usual, the triangulation has to be fine, star and regular.
In the case of a top, the star property is with respect to the single interior vector of the reflexive polygon $F_m$ at ${\rm x}_3=0$.}
For a three dimensional top, any triangulation $\cal T$ induces a triangulation ${\cal T}^{(\Sigma)}$ of its surface.
Geometrically, it encodes the intersection structure of the toric divisors:
Any line of ${\cal T}^{(\Sigma)}$ connecting two vectors $\vec{v}_i$ corresponds to a fibral curve $D_{v_1} \cap D_{v_2}$ of the fibration $\hat\pi: \hat{Y} \rightarrow B$.

A change of the triangulation now corresponds in general to a flop transition.
Indeed, one curve $C_{12}$ that is present in one resolution phase with ${\cal T}^{(\Sigma)}_1$ can be absent in a second phase, because in the second triangulation ${\cal T}^{(\Sigma)}_2$ the vectors $\vec{v}_1$ and $\vec{v}_2$ may not share a triangle.
Instead, there will be a different line connecting $\vec{v}_1$ with, say, $\vec{v}_3$, giving rise to a curve $C_{13}$ that was absent in ${\cal T}^{(\Sigma)}$.
Thus, changing the triangulation precisely corresponds to a transition that flops $C_{12}$ into $C_{13}$.

Since we are only interested in the local geometry of the non-flat fiber components, we actually do not need a full triangulation of the top, but only the triangulation ${\cal T}_{\cal F}$ it induces on a facet ${\cal F}$.
In particular, each non-flat fiber component $S_k$ corresponding to a toric divisor $D_{v_k}$ is a toric surface described itself by a 2d polygon ${\cal P}_{k}$ with a single interior vector.
This polygon is simply a sub-polygon of the facet ${\cal F}$ with $\vec{v}_k$ being the interior vector and the boundary vectors given by all vectors which are connected to $\vec{v}_k$ as dictated by the triangulation ${\cal T}_{\cal F}$.
Note that the resulting 2d polygon also appear in \cite{Xie:2017pfl} for toric resolutions of canonical threefold singularities.
Here, we see that these models can in principle arise as non-minimal elliptic singularities, which can be resolved in a non-flat manner via tops.

Before presenting the top we will be focusing on in the rest of this paper, we have to point out two main caveats of restricting to toric constructions.
First, even though there are many tops that give rise to a non-flat fibration (see, e.g., \cite{Dierigl:2018nlv}), we do not have a general argument that we can have a suitable toric construction that resolves any given 6d conformal matter model.
Second, not all resolution phases of a given elliptic fibration can be realized torically via a top (we will see this drawback explicitly later on).
We suspect that both of these restrictions can be overcome with non-toric resolution methods, such as in \cite{Lawrie:2012gg, Hayashi:2014kca,Braun:2014kla,Braun:2015hkv}.
However, we will defer an explicit analysis of this sort for future work \cite{work_in_progress_resolution}, and use toric constructions here for convenience and flexibility.

\subsection[Explicit example: \texorpdfstring{$E_8$}{E8}-top over \texorpdfstring{$F_{10}$}{F10}]{Explicit example: \boldmath{$E_8$}-top over \boldmath{$F_{10}$}}

To demonstrate our proposal of studying 5d SCFTs on non-flat fibrations, we will consider three explicit types of conformal matter theories in 6d, namely collisions of an $E_8$ singularity with
\begin{enumerate}
  \item an $I_1$ locus,
  \item an $SU(2)$ locus,
  \item an $SU(3)$ locus.
\end{enumerate}
In 6d, these theories are also known as the E-string theories of rank 1, 2 and the $(E_8, SU(3))$ conformal matter theory, whose 6d tensor branch in the notation of \cite{Heckman:2015bfa} reads
\begin{align}
\begin{split}
  &[E_8]-1 \, ,\\
  &[E_8]-1-2-[SU(2)] \, ,\\
  &[E_8]-1-\overset{\mathfrak{su}(1)}2-\overset{\mathfrak{su}(2)}2-[SU(3)] \, .
\end{split}
\end{align}
The numbers describe the negative self-intersection of the compact curves $\Sigma_i$ used to blow up the base.
The resulting threefold has only minimal singularities, with possible Kodaira fibers over the compact curves supporting the indicated gauge algebra.

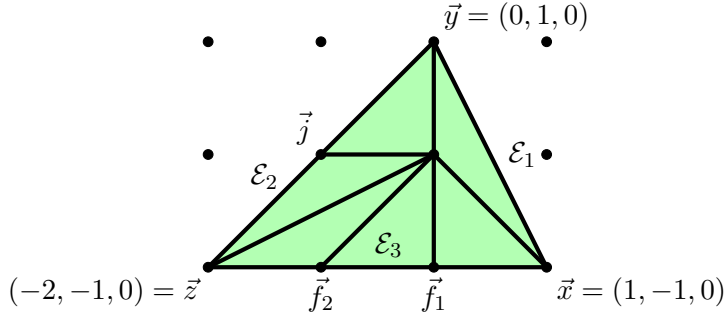
\begin{figure}[ht]
  \centering
  \begin{tikzpicture}[scale=1.5]
    \begin{scope}[xshift=0.33\textwidth]
      \filldraw [ultra thick, draw=black, fill=green!30!white] (-2,-1)--(0,1)--(1,-1)--cycle; 
      \filldraw [ultra thick, draw=black] (0,0)--(0,1); 
      \filldraw [ultra thick, draw=black] (0,0)--(1,-1); 
      \filldraw [ultra thick, draw=black] (0,0)--(0,-1); 
      \filldraw [ultra thick, draw=black] (0,0)--(-1,-1); 
      \filldraw [ultra thick, draw=black] (0,0)--(-1,0); 
      \filldraw [ultra thick, draw=black] (0,0)--(-2,-1); 
      \foreach \x in {-2,-1,...,1}{
        \foreach \y in {-1,0,1}{
          \node[draw,circle,inner sep=1.3pt,fill] at (\x,\y) {};
        }
        }
      \node [below left] at (-2,-1)  {$(-2,-1,0) = \vec{z}$};
      \node [above right] at (0,1) {$\vec{y} = (0,1,0) $};
      \node [below right] at (1,-1) {$\vec{x} = (1,-1,0)$};
      \node [below] at (-1,-1) {$\vec{f}_2$};
      \node [above left] at (-1,0) {$\vec{j}$};
      \node [below] at (0,-1) {$\vec{f}_1$};
      \node [left] at (1,0) {${\cal E}_1$};
      \node [above] at (-0.4,-1) {${\cal E}_3$};
      \node [below] at (-1.5,0) {${\cal E}_2$};
    \end{scope}
 \end{tikzpicture}
  \caption{Toric diagram of $F_{10}$ with blow-ups on the edges. 
  The vertices $(\vec{x}, \vec{y}, \vec{z})$ give rise to bi-, tri-, and rational sections, respectively, of the elliptic fibration.
  The $SU(2)$ singularity over $\{s_4 = 0\}$ is resolved by $j$, and the $SU(3)$ over $\{s_8 =0 \}$ is resolved by $f_1$ and $f_2$. The edges of length $l$ are labelled by ${\cal E}_l$.}\label{fig:F10_with_blow-up}
\end{figure}

These three different SCFT sectors can all be realized at different non-minimal singularities within the same elliptic fibration described by a single top, namely the only possible $E_8$ top over the polygon $F_{10}$ \cite{Bouchard:2003bu}.
This polygon has one edge of length 1, one of length 2 and one of length 3 (see figure \ref{fig:F10_with_blow-up}).

The elliptic Calabi--Yau threefold $\hat{Y}_3$ is the vanishing of the polynomial \cite{Klevers:2014bqa}
\begin{align}\label{eq:hypersurface_F10}
  \begin{split}
    P \equiv \, &b_1 \, j \, f_1\,f_2 \, x \, y \,z + b_2 \, j \, f_1^2\,f_2^2 \, x^2\,z^2 + b_3 \, j^2\,f_1\,f_2^2 \, y \,z^3 + b_4\, j^2\,f_1^2\,f_2^3 \, x\,z^4 + b_6\, j^3\,f_1^2\,f_2^4\,z^6 \\
    + & \, s_8\,j\,y^2 + s_4\,f_1^2\,f_2\,x^3 = 0 \, ,
  \end{split}
\end{align}
where the coefficients $b_i$ and $s_j$ are holomorphic sections of suitable line bundles over $B_2$.
Their vanishing loci define vertical divisors, i.e., pull-back of divisors on $B_2$.
All other variables that appear in \eqref{eq:hypersurface_F10} define toric divisors.

As shown in \cite{Klevers:2014bqa}, the resulting elliptic fibration has an $SU(2)$ and $SU(3)$ locus, located over $\{s_4 =0\} \equiv \{s_4\}$ and $\{s_8 =0\} \equiv \{s_8\}$, respectively (we will from now on use the notation $\{F\}$ to denote the vanishing locus of $F$).
To see this, we can set $s_4$ respectively $s_8$ to $0$ in \eqref{eq:hypersurface_F10}, in which case the hypersurface polynomial $P$ factorizes:
\begin{align}
  & P|_{s_4=0} = j \, (s_8\,y^2 + ...) \, , \label{eq:factorization_s4} \\
  & P|_{s_8=0} = f_1 \, f_2 \, (s_4\,f_1\,x^3 + ...) \label{eq:factorization_s8} \, .
\end{align}
From this, one can see that over $\{s_4\}$, the fiber splits into two components, and over $\{s_8\}$ into three.
A more close analysis (see appendix \ref{app:weierstrass_of_top}) reveals that the Kodaira types are $I_2$ and $I_3$, respectively, hence give rise to the promised $SU(2)$ and $SU(3)$ gauge groups in F-theory.
Finally, there is, as usual, also a divisor (the residual discriminant) in $B_2$ carrying $I_1$ fibers, see appendix \ref{app:weierstrass_of_top}.

Now, to incorporate an $E_8$ singularity along a divisor $W \subset B_2$, we construct the associated top over $F_{10}$.
The ruled divisors $A_i$, $i=0,...,8$, whose fibers give rise to the affine $E_8$ diagram in codimension one have associated polytope vectors $\vec{\alpha}_i$ given by
\begin{equation}\label{eq:top_vectors_E8_roots}
  \begin{aligned}
    & \vec{\alpha}_0 = (-2,-1,1) \, ,  \quad &\vec{\alpha}_1& = (-2,-1,2) \, , \quad & \vec{\alpha}_2 & = (-2,-1,3) \, , \\
    & \vec{\alpha}_3 = (-2,-1,4) \, , \quad &\vec{\alpha}_4 & = (-2,-1,5) \, , 
    \quad & \vec{\alpha}_5 &= (-2,-1,6) \, , \\
    & \vec{\alpha}_6 = (-1,-1,4) \, , \quad &\vec{\alpha}_7 & = (0,-1,2) \, , \quad & \vec{\alpha}_8 & = (-1,0,3) \, .
  \end{aligned}
\end{equation}
The convex hull of these lattice points and those in $F_{10}$ form a top $\Delta$ with three facets (not counting the one at height 0, which is just $F_{10}$ itself), each bounded by one of the edges ${\cal E}_l$ of $F_{10}$.
It can be easily checked that these facets contain the following interior vectors (cf.~fig.~\ref{fig:top}):
\begin{equation}\label{eq:face_interior_points}
  \begin{aligned}
    & \text{Face over } {\cal E}_1: \quad \vec{r} = (0,0,1) \, ,\\
    & \text{Face over } {\cal E}_2: \quad \vec{t}_1 = (-1,0,1) \, ,  &\vec{t}_2 &= (-1,0,2) \, , \\
    & \text{Face over } {\cal E}_3: \quad \vec{u}_1 = (-1,-1,1)\, ,  &\vec{u}_2 &= (0,-1,1) \, , & \vec{u}_3 & = (-1,-1,2) \, , & \vec{u}_4 & = (-1,-1,3) \, .
  \end{aligned}
\end{equation}

All vectors of the top above height ${\rm x}_3 = 0$ introduce additional variables, but no extra terms, into the hypersurface polynomial \eqref{eq:hypersurface_F10}.
The modified polynomial now reads
\begin{align}\label{eq:resolved_hypersurface}
  \begin{split}
    \hat{P} := &\, s_8\,\alpha_8\, r\, t_1\,t_2 \,j \, y^2 + s_4\,\alpha_6\, \alpha_7^2\, r\,u_1\,u_2^2\,u_3\,u_4\, f_1^2\,f_2 x^3  \\
    + & \, b_1\, \alpha_0\, \alpha_1\, \alpha_2\, \alpha_3\, \alpha_4\, \alpha_5\, \alpha_6\, \alpha_7\, \alpha_8\,r\,t_1\,t_2\,u_1\,u_2\,u_3\,u_4\,j\,f_1\,f_2\, x\, y\, z \\
    + & \, b_2\, \alpha_0^2\, \alpha_1^2\, \alpha_2^2\, \alpha_3^2\, \alpha_4^2\, \alpha_5^2\, \alpha_6^2\, \alpha_7^2\, \alpha_8\, r\, t_1\,t_2\,u_1^2\,u_2^2\,u_3^2\,u_4^2 \, j \, x^2\, z^2 \\
    + & \, b_3\, \alpha_0^3\, \alpha_1^3\, \alpha_2^3\, \alpha_3^3\, \alpha_4^3\, \alpha_5^3\, \alpha_6^2 \, \alpha_7\, \alpha_8^2\, r\, t_1^2\,t_2^2\, u_1^2\,u_2\,u_3^2\,u_4^2 \, j^2 \,f_1 \,f_2^2\, y\, z^3 \\
    + & \, b_4\, \alpha_0^4\, \alpha_1^4\, \alpha_2^4\, \alpha_3^4\, \alpha_4^4\, \alpha_5^4\, \alpha_6^3\, \alpha_7^2\, \alpha_8^2\, r\, t_1^2\,t_2^2 \, u_1^3\,u_2^2\,u_3^3\,u_4^3\, j^2\, f_1^2\,f_2^3\,  x\, z^4 \\
    + & \, b_6\, \alpha_0^5\, \alpha_1^4\, \alpha_2^3\, \alpha_3^2\, \alpha_4 \, t_1^2\,t_2 \, u_1^3\,u_2\,u_3^2\,u_4\, j^3 \,f_1^2\,f_2^4 \,z^6   \, .
  \end{split} 
\end{align}
The Weierstrass model of this elliptic fibration is presented in \eqref{eq:weierstrass_E8_top}.
From that, we see the non-minimal singularities at the collision of the $E_8$ with
\begin{enumerate}
  \item the $I_1$ at $\{b_6\} \cap W$ with ord$(f,g,\delta) = (4,6,12)$,
  \item the $SU(2)$ at $\{s_4\} \cap W$ with ord$(f,g,\delta) = (4,6,13)$,
  \item the $SU(3)$ at $\{s_8\} \cap W$ with ord$(f,g,\delta) = (4,6,14)$.
\end{enumerate}

Observe now the appearance of the non-flat fibers in the resolution \eqref{eq:resolved_hypersurface}.
By setting $b_6=0$, the remaining polynomial has an overall factor $r$.
This means that the surface $S \equiv \{b_6\} \cap \{r\}$ inside $\varphi: {\cal X}_4 \rightarrow B$ is completely contained in $\hat{Y}_3$.
Since $\varphi(S) = W \subset B$ in the ambient space, intersecting it further with $\{b_6\}$ which is a vertical divisor transverse to $W$ localizes to a point $W \cap \{b_6\}$ in the base, where the $I_1$ locus meets the $E_8$ locus.
Hence, the surface is a component of the fiber $\hat{\pi}^{-1} (W \cap \{b_6\})$ of the fibration $\hat{Y}$, which therefore is non-flat.

Likewise, setting $s_4$ to 0 leads to a factorization of the form
\begin{align}
  \hat{P}|_{s_4=0} = j\,t_1\,t_2\,(...) \, .
\end{align}
Because $\{t_{1,2}\}$ is again fibered over $W$, this factorization indicates the presence of now two non-flat surfaces, $T_i = \{s_4\} \cap \{t_i\}$, in the fiber of $\hat{Y}_3$ over $\{s_4\} \cap W$.
Note that the factor $j$, on the other hand, does not define a non-flat fiber, because the corresponding vector $\vec{j}$ is not a facet interior of the top $\Delta$.
Geometrically, the divisor $\{j\} \cap \hat{P} \subset \hat{Y}$ projects onto $\{s_4\}$, and is part of the resolved $I_2$ fiber of $\hat{Y}_3$ over that locus \cite{Klevers:2014bqa}.

Finally, setting $s_8$ to 0 in \eqref{eq:resolved_hypersurface} yields a factorization
\begin{align}
  \hat{P}_{s_8=0} = f_1\,f_2\,u_1\,u_2\,u_3\,u_4\,(...) \, .
\end{align}
Again, the interpretation for each of the $u_i$ factors is a non-flat fiber component of $\hat{Y}_3$ over $\{s_8\} \cap W$, while $\{f_{1,2}\}$ resolve the $I_3$ singularity of $\hat{Y}_3$ over $\{s_8\}$.

\section{5d Rank 1 Theories from 6d E-string}\label{sec:rank_one}

In this section, we turn to the familiar example of rank one theories in 5d, which are known to arise from circle reductions of the 6d E-string.
Our focus will be on demonstrating how the non-flat fiber captures the 5d physics, and how we can read off the global symmetries from the singularity enhancements over the codimension one fibers, which perfectly matches the already know flavor symmetry enhancements argued by other methods.

First, for the purpose of studying the fiber geometry around $\{b_6\} \cap W$, we can simplify the hypersurface polynomial \eqref{eq:resolved_hypersurface} by setting the coordinates $j, f_l, t_m, u_n$ to one, since they do not vanish over this point in the base.
Put differently, their associated divisors do not intersect the fiber over $\{b_6\} \cap W$.
The resulting polynomial defining the (local patch of the) Calabi--Yau threefold is then
\begin{align}\label{eq:resolved_hypersurface_rank_1}
  \begin{split}
    \hat{P}_1 := \, & s_8\, \alpha_8\, r\, y^2 + s_4\, \alpha_6\, \alpha_7^2\, r\, x^3  + b_1\, \alpha_0\, \alpha_1\, \alpha_2\, \alpha_3\, \alpha_4\, \alpha_5\, \alpha_6\, \alpha_7\, \alpha_8\, r\, x\, y\, z \\
    + &\, b_2\, \alpha_0^2\, \alpha_1^2\, \alpha_2^2\, \alpha_3^2\, \alpha_4^2\, \alpha_5^2\, \alpha_6^2\, \alpha_7^2\, \alpha_8\, r\, x^2\, z^2 +  b_3\, \alpha_0^3\, \alpha_1^3\, \alpha_2^3\, \alpha_3^3\, \alpha_4^3\, \alpha_5^3\, \alpha_6^2\, \alpha_7\, \alpha_8^2\, r\, y\, z^3 \\
    + & \,b_4\, \alpha_0^4\, \alpha_1^4\, \alpha_2^4\, \alpha_3^4\, \alpha_4^4\, \alpha_5^4\, \alpha_6^3\, \alpha_7^2\, \alpha_8^2\, r\, x\, 
z^4 + b_6\, \alpha_0^5\, \alpha_1^4\, \alpha_2^3\, \alpha_3^2\, \alpha_4\, z^6  \, .
  \end{split}
\end{align}
Over the generic point of the codimension one locus $W \subset B_2$, the elliptic fiber degenerates into the affine Dynkin diagram of $E_8$, whose fiber components are described by the equations $\{\alpha_i\} \cap \{\hat{P}_1 \}$ inside the fiber ambient space.

We focus on the non-flat fiber over $\{b_6\} \cap W \in B_2$.
The only compact surface $S$ in this fiber, given by $\{b_6\} \cap \{r\}$, supports a 5d rank one theory when we compactify M-theory on it.
By construction, this fiber component can be shrunk to a point (re-creating the singularity of the elliptic fibration).
In general, this can happen without shrinking all codimension one fiber components over $W$.
As we recall from section \ref{sec:M-/F-duality}, this implies a non-trivial $S^1$ holonomy of the $E_8$ in the reduction of F- to M-theory.
Thus, the 5d theory is expected to be a mass deformation of the 5d KK-theory, and hence should have a non-trivial SCFT limit.

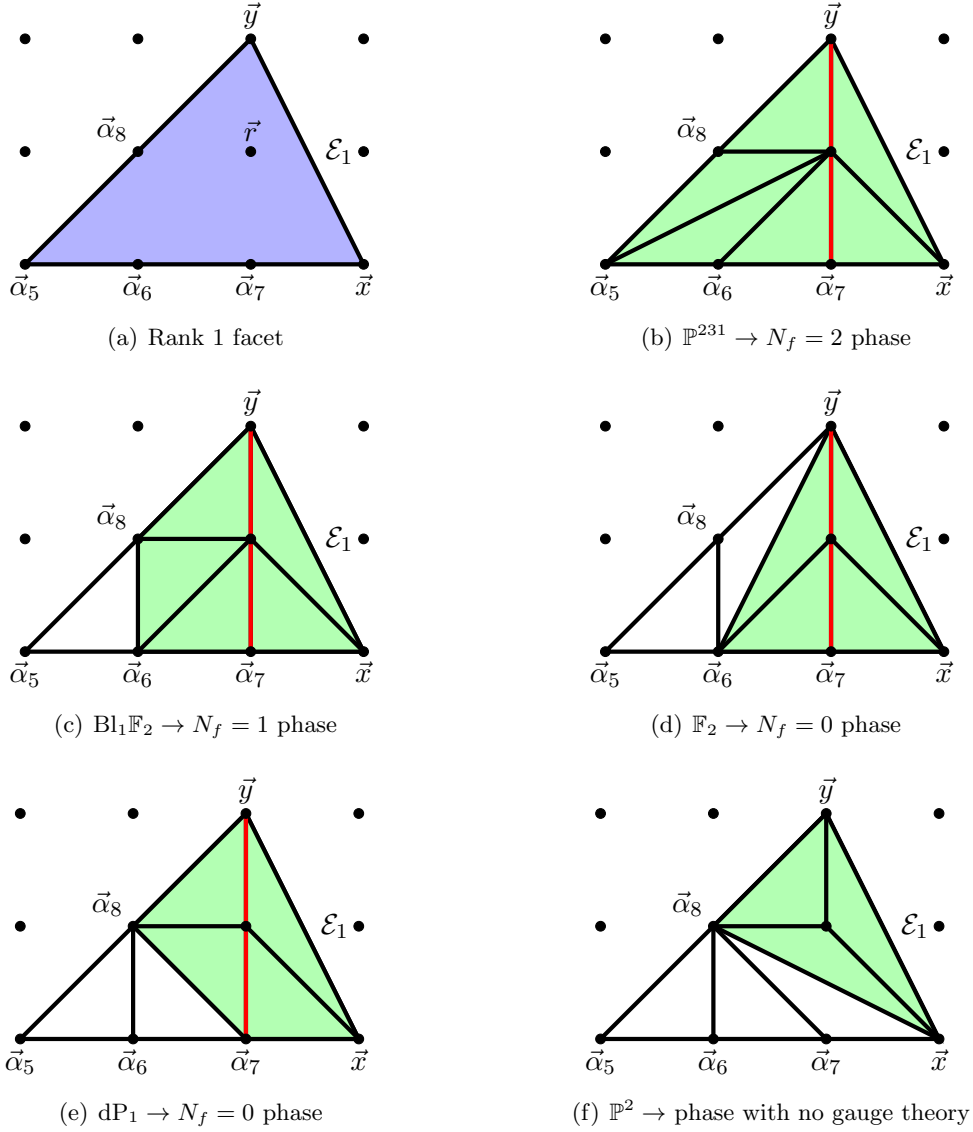
\begin{figure}[p]
  \centering
  \subfigure[Rank 1 facet]{\label{fig:rank_1_facette}
    	\begin{tikzpicture}[scale=1.5]
%
%
		\begin{scope}[xshift=0.33\textwidth]
			\filldraw [ultra thick, draw=black, fill=blue!30!white] (0,1)--(-2,-1)--(1,-1)--cycle;
			\foreach \x in {-2,-1,...,1}{
				\foreach \y in {-1,0,...,1}{
					\node[draw,circle,inner sep=1.3pt,fill] at (\x,\y) {};
				}
		    }
			\node [above] at (0,0)  {$\vec{r}$};
			\node [above] at (0,1) {$\vec{y}$};
			\node [below] at (1,-1) {$\vec{x}$};
			\node [below] at (0,-1) {$\vec{\alpha}_7$};
			\node [below] at (-1,-1) {$\vec{\alpha}_6$};
			\node [below] at (-2,-1) {$\vec{\alpha}_5$};
			\node [above left] at(-1,0)  {$\vec{\alpha}_8$};
			\node [left] at (1,0) {${\cal E}_1$};
		\end{scope}
	\end{tikzpicture}
  } 
  \hspace{2cm}
  \subfigure[$\bbP^{231} \rightarrow N_f = 2$ phase]{\label{fig:rank_1_dP3}
    	\begin{tikzpicture}[scale=1.5]
		\begin{scope}[xshift=0.33\textwidth]
			\filldraw [ultra thick, draw=black, fill=green!30!white] (-2,-1)--(1,-1)--(0,1)--cycle; 
			\filldraw [ultra thick, draw=black] (0,0)--(0,1); 
			\filldraw [ultra thick, draw=black] (0,0)--(1,-1); 
			\filldraw [ultra thick, draw=black] (0,0)--(0,-1); 
			\filldraw [ultra thick, draw=black] (0,0)--(-1,0); 
			\filldraw [ultra thick, draw=black] (0,0)--(-1,-1); 
			\filldraw [ultra thick, draw=black] (0,0)--(-2,-1); 
			\filldraw [ultra thick, draw=red] (0,1)--(0,-1);
			\node [above] at (0,1) {$\vec{y}$};
			\node [below] at (1,-1) {$\vec{x}$};
			\node [below] at (0,-1) {$\vec{\alpha}_7$};
			\node [below] at (-1,-1) {$\vec{\alpha}_6$};
			\node [below] at (-2,-1) {$\vec{\alpha}_5$};
			\node [above left] at(-1,0)  {$\vec{\alpha}_8$};
			\node [left] at (1,0) {${\cal E}_1$};
			\foreach \x in {-2,-1,...,1}{
				\foreach \y in {-1,0,...,1}{
					\node[draw,circle,inner sep=1.3pt,fill] at (\x,\y) {};
				}
		    }
		\end{scope}
	\end{tikzpicture}
  }
  \subfigure[$\text{Bl}_1 \bbF_2 \rightarrow N_f = 1$ phase]{\label{fig:rank_1_dP2}
    	\begin{tikzpicture}[scale=1.5]
		\begin{scope}[xshift=0.33\textwidth]
			\filldraw [ultra thick, draw=black, fill=white] (-2,-1)--(1,-1)--(0,1)--cycle; 
			\filldraw [ultra thick, draw=black, fill=green!30!white] (-1,-1)--(1,-1)--(0,1)--(-1,0)--cycle; 
			\filldraw [ultra thick, draw=black] (0,0)--(0,1); 
			\filldraw [ultra thick, draw=black] (0,0)--(1,-1); 
			\filldraw [ultra thick, draw=black] (0,0)--(0,-1); 
			\filldraw [ultra thick, draw=black] (0,0)--(-1,0); 
			\filldraw [ultra thick, draw=black] (0,0)--(-1,-1); 
			\filldraw [ultra thick, draw=red] (0,1)--(0,-1);
			\node [above] at (0,1) {$\vec{y}$};
			\node [below] at (1,-1) {$\vec{x}$};
			\node [below] at (0,-1) {$\vec{\alpha}_7$};
			\node [below] at (-1,-1) {$\vec{\alpha}_6$};
			\node [below] at (-2,-1) {$\vec{\alpha}_5$};
			\node [above left] at(-1,0)  {$\vec{\alpha}_8$};
			\node [left] at (1,0) {${\cal E}_1$};
			\foreach \x in {-2,-1,...,1}{
				\foreach \y in {-1,0,...,1}{
					\node[draw,circle,inner sep=1.3pt,fill] at (\x,\y) {};
				}
		    }
		\end{scope}
	\end{tikzpicture}
  }
  \hspace{2cm}
  \subfigure[$\bbF_2  \rightarrow N_f =0$ phase]{\label{fig:rank_1_P112}
    	\begin{tikzpicture}[scale=1.5]
		\begin{scope}[xshift=0.33\textwidth]
			\filldraw [ultra thick, draw=black, fill=white] (-2,-1)--(1,-1)--(0,1)--cycle; 
			\filldraw [ultra thick, draw=black, fill=green!30!white] (-1,-1)--(1,-1)--(0,1)--cycle; 
			\filldraw [ultra thick, draw=black] (0,0)--(0,1); 
			\filldraw [ultra thick, draw=black] (0,0)--(1,-1); 
			\filldraw [ultra thick, draw=black] (0,0)--(0,-1); 
			\filldraw [ultra thick, draw=black] (-1,-1)--(-1,0); 
			\filldraw [ultra thick, draw=black] (0,0)--(-1,-1); 
			\filldraw [ultra thick, draw=red] (0,1)--(0,-1);
			\foreach \x in {-2,-1,...,1}{
				\foreach \y in {-1,0,...,1}{
					\node[draw,circle,inner sep=1.3pt,fill] at (\x,\y) {};
				}
		    }
			\node [above] at (0,1) {$\vec{y}$};
			\node [below] at (1,-1) {$\vec{x}$};
			\node [below] at (0,-1) {$\vec{\alpha}_7$};
			\node [below] at (-1,-1) {$\vec{\alpha}_6$};
			\node [below] at (-2,-1) {$\vec{\alpha}_5$};
			\node [above left] at(-1,0)  {$\vec{\alpha}_8$};
			\node [left] at (1,0) {${\cal E}_1$};
		\end{scope}
	\end{tikzpicture}
  }
  \subfigure[$\text{dP}_1 \rightarrow N_f = 0$ phase]{\label{fig:rank_1_dP1}
    	\begin{tikzpicture}[scale=1.5]
		\begin{scope}[xshift=0.33\textwidth]
			\filldraw [ultra thick, draw=black, fill=white] (-2,-1)--(1,-1)--(0,1)--cycle; 
			\filldraw [ultra thick, draw=black, fill=green!30!white] (-1,0)--(0,-1)--(1,-1)--(0,1)--cycle; 
			\filldraw [ultra thick, draw=black] (0,0)--(0,1); 
			\filldraw [ultra thick, draw=black] (0,0)--(1,-1); 
			\filldraw [ultra thick, draw=black] (0,0)--(0,-1); 
			\filldraw [ultra thick, draw=black] (0,0)--(-1,0); 
			\filldraw [ultra thick, draw=black] (-1,-1)--(-1,0); 
			\filldraw [ultra thick, draw = red] (0,-1)--(0,1);
			\foreach \x in {-2,-1,...,1}{
				\foreach \y in {-1,0,...,1}{
					\node[draw,circle,inner sep=1.3pt,fill] at (\x,\y) {};
				}
		    }
			\node [above] at (0,1) {$\vec{y}$};
			\node [below] at (1,-1) {$\vec{x}$};
			\node [below] at (0,-1) {$\vec{\alpha}_7$};
			\node [below] at (-1,-1) {$\vec{\alpha}_6$};
			\node [below] at (-2,-1) {$\vec{\alpha}_5$};
			\node [above left] at(-1,0)  {$\vec{\alpha}_8$};
			\node [left] at (1,0) {${\cal E}_1$};
		\end{scope}
	\end{tikzpicture}
  }
  \hspace{2cm}
  \subfigure[$\bbP^2 \rightarrow$ phase with no gauge theory]{\label{fig:rank_1_P2}
    	\begin{tikzpicture}[scale=1.5]
		\begin{scope}[xshift=0.33\textwidth]
			\filldraw [ultra thick, draw=black, fill=white] (-2,-1)--(1,-1)--(0,1)--cycle; 
			\filldraw [ultra thick, draw=black, fill=green!30!white] (-1,0)--(1,-1)--(0,1)--cycle; 
			\filldraw [ultra thick, draw=black] (0,0)--(0,1); 
			\filldraw [ultra thick, draw=black] (0,0)--(1,-1); 
			\filldraw [ultra thick, draw=black] (0,-1)--(-1,0); 
			\filldraw [ultra thick, draw=black] (0,0)--(-1,0); 
			\filldraw [ultra thick, draw=black] (-1,-1)--(-1,0); 
			\foreach \x in {-2,-1,...,1}{
				\foreach \y in {-1,0,...,1}{
					\node[draw,circle,inner sep=1.3pt,fill] at (\x,\y) {};
				}
		    }
			\node [above] at (0,1) {$\vec{y}$};
			\node [below] at (1,-1) {$\vec{x}$};
			\node [below] at (0,-1) {$\vec{\alpha}_7$};
			\node [below] at (-1,-1) {$\vec{\alpha}_6$};
			\node [below] at (-2,-1) {$\vec{\alpha}_5$};
			\node [above left] at(-1,0)  {$\vec{\alpha}_8$};
			\node [left] at (1,0) {${\cal E}_1$};
		\end{scope}
	\end{tikzpicture}
  } 
  \caption{The full facet \subref{fig:rank_1_facette} of $\Diamond$ over the edge ${\cal E}_1$ (see figure \ref{fig:F10_with_blow-up}) containing one interior vector $\vec{r}$.
  Depending on the triangulation of top, which induces different triangulations \subref{fig:rank_1_dP3}--\subref{fig:rank_1_P2} of the facet, the non-flat geometry realizes a different weakly coupled phase of the rank one 5d SCFT.
  The similarity between this facet and the base polygon $F_{10}$ of $\Diamond$ is purely coincidental.
  The red lines indicate a projection of the corresponding toric diagram preserving the cones, and thus a ruling of the toric surface.}\label{fig:rank_1_phases}
\end{figure}

When the non-flat fiber is at finite size, it defines a weakly coupled phase of the SCFT.
For rank one, the possible phases are an $\mathfrak{su}(2)$ gauge theory with $0 \leq N_f \leq 7$ massless fundamental flavors, as well as a phase with no gauge group or matter, see section \ref{sec:review_su2_in_5d}.
In the toric set-up, we can only realize phases with $N_f \leq 2$ through different triangulations of the top, see below.
The associated induced triangulations of the corresponding facet of $\Delta$ are depicted in \ref{fig:rank_1_phases}.

As discussed in the previous section, the surface geometry of $S$ is determined by the triangulation induced on the facet ${\cal F}_{{\cal E}_1}$.
For this facet, the possibilities are rather limited and depicted in \ref{fig:rank_1_phases}.
In each, the green polygon represents the toric diagram $F_S$ of the surface $S$ in that phase.
The associated surfaces are then easily determined from basic toric geometry, and is included in the figure captions \ref{fig:rank_1_dP3} -- \ref{fig:rank_1_P2}.

For each such surface $S$, we can compute the associated number of flavors $N_f$ via \cite{Intriligator:1997pq}
\begin{align}
  8 - N_f = (S^3)|_{\hat{Y}_3} = (K_S^2)|_{S}\, ,
\end{align}
where the second equality follows from adjunction formula on an Calabi--Yau threefold $\hat{Y}_3$.
Obviously, this formula does not apply for the $\bbP^2$ case, cf.~figure \ref{fig:rank_1_P2}, because this phase has no gauge symmetry at weak coupling \cite{Morrison:1996xf}, and hence also no notion of flavors.

\subsection{Gauge enhancement at weak coupling}

Having a toric description of the surface $S$ immensely simplifies the analysis of the weakly coupled phase (see also \cite{Xie:2017pfl}).
First, to establish the notation, we will denote the toric curves $\{q\} \cap \{r\}$, for $\vec{q}$ a lattice point on $\Delta$, simply by $\{q\}$ whenever we are only focusing on the local geometry of $S$.
Furthermore, we will denote the divisor class of the curve $\{q\}$ in $S$ by $[q]$.
Finally, as usual, $K_S$ is the canonical divisor of $S$.
With that, one can immediately show with basic toric geometry that all toric curves appearing here are $\bbP^1$s

Now, to have a non-trivial gauge theory at weak coupling, the surface $S$ has to be ruled.
In the toric case, such rulings can be easily determined from the diagram $F_S$.
Concretely, any lattice projection $\psi_S : F_S \rightarrow L$ mapping the interior vector of $F_S$ onto the origin of a one-dimensional sub-lattice $L$, which collapses cones of $F_S$ onto cones of $L$, indicates a ruling of $S$.\footnote{Such a lattice projection gives rise to a so-called \textit{toric morphism}, i.e., a map between two toric spaces. Here, this map is simply the projection map of the ruling on $S$.}
Note that cones of $L$ are simply the ``left'' or ``right'' from the origin.
For the toric diagrams in \ref{fig:rank_1_phases}, one can easily spot such a projection $\psi$ for all phases except the $\bbP^2$:
It is giving by projecting along the vertical axis (i.e., it projects the red lines to the origin).
The sub-lattice $L$ in this case is the ``orthogonal'' sub-lattice spanned by $\pm (\vec{\alpha}_8 - \vec{r})$.
In all cases, the vectors $\vec{y}$ and $\vec{\alpha}_7$ project onto the origin.
This signals that in $S$, $\{y\}$ and $\{\alpha_7\}$ define sections of the ruling.
The other curves correspond to fibers of the ruling.

Furthermore, a simple toric computation reveals that $K_S \cdot [x] = -2$ for all relevant phases, which means that $\{x\}$ is generic fiber of $\psi_S$.\footnote{Because $\{x\}$ is toric, it is a $\bbP^1$, so we know that $[x] \cdot_S ([x] + K_S) = -2$.
But since $[x] \cdot_S K_S = -2$, it implies $[x]^2 = 0$, which by adjunction is the degree of the normal bundle of $[x]$ in $S$. Hence, $[x]$ moves in a family.}
This also immediately implies that by collapsing the generic fiber of $S$ (except when $S \cong \bbP^2$) which shrinks $S$ to a curve, we obtain an $\mathfrak{su}(2)$ gauge symmetry, whose W-bosons come from M2-branes wrapping curves in $[x]$.

Lastly, $S$ also contains fibral curves with intersection numbers $-1$ and $0$ with $K_S$.
For $S \cong \bbP^{231}$, these are $\{\alpha_{5,6,8}\}$ with $[\alpha_5] \cdot K_S = -1$ and $[\alpha_{6,8}] \cdot K_S = 0$ (cf.~figure \ref{fig:rank_1_dP3}).
These now give rise to the $N_f = 2$ fundamental hypermultiplets as follows:
There are exactly four linear combinations of these curve classes for which the (arithmetic) genus is 0 and the intersection with $K_S$ is $-1$:
$[\alpha_5], [\alpha_5] + [\alpha_8], [\alpha_5]+[\alpha_6]$ and $[\alpha_5] + [\alpha_6] + [\alpha_8]$.
Each such class gives rise to one hypermultiplet with $\mathfrak{su}(2)$ Cartan-charge $-1$.
Because the representations of $\mathfrak{su}(2)$ are pseudo-real, these four states can actually assemble into two full hypermultiplets of the ${\bf 2}$-representation.
Since all these curves are fibral, blowing down the generic fiber will also force these curves to shrink.
Said differently, the curve classes satisfy $[x] = [\alpha_6] + 2\,[\alpha_5] + [\alpha_8]$, and since $S$ is K\"ahler, sending the volume of $[x]$ to zero means that all classes on the right-hand side also shrink.
Thus, when we enhance the gauge symmetry to an $\mathfrak{su}(2)$, the two fundamental hypermultiplets also become massless.

Field theoretically, one can transition from $N_f = 2$ to $N_f = 1$ via a mass deformation.
Geometrically, the corresponding flop transition must hence eliminate curves which carry fundamental charges under the $\mathfrak{su}(2)$.
Indeed, we see from the toric diagrams in figures \ref{fig:rank_1_dP3} and \ref{fig:rank_1_dP2}, that the transitions flops out $\{\alpha_5\} \cap \{r\}$ from the geometry and blows up the curve $\{\alpha_6\} \cap \{\alpha_8\}$, where the latter no longer lies inside the compact surface $S \cong \text{Bl}_1 \bbF_2$.
In this phase, we only have the fibral curves $\{\alpha_{6,8}\} \subset S$ with intersection numbers $K_S \cdot [\alpha_{6,8}] = -1$.
M2-branes on these two curves then give rise to one fundamental hypermultiplet, which are massless in the shrinking fiber limit, because now $[x] = [\alpha_6] + [\alpha_8]$.

Further mass deforming the $N_f=1$ phase yields a pure $\mathfrak{su}(2)$ gauge theory with no flavors.
Physically, however, there are two distinct $N_f=0$ theories with different $\theta$-angle.
Geometrically, we can see these two phases arising by flopping different curves in the $S \cong \text{Bl}_1 \bbF_2$ surface, namely either $\{r\} \cap \{\alpha_8\}$ or $\{r\} \cap \{\alpha_6\}$, cf.~figure \ref{fig:rank_1_dP2}.
The first flop yields the $S \cong \bbF_2$ phase, corresponding to the $\theta = 0$ theory, while the second flop produces the $\theta = \pi$ theory on $S \cong \text{dP}_1$.
Lastly, the $\theta= \pi$ theory allows for another transition, which geometrically is described by the flop transition from dP$_1$ to $\bbP^2$.
Note that the two $N_f =0$ phases are not related by a simple flop transition, but rather by an extremal transition, in which the volume of the surface has to pass through zero \cite{Morrison:1996xf}.
In the toric diagrams, we can see that, by starting from $\bbF_2$ in figure \ref{fig:rank_1_P112}, one would have to first blow down the curve $\{\alpha_6\}$, which corresponds to a generic fiber of $\bbF_2$.
Hence, this blow-down would contract $\bbF_2$ to a curve.

\subsection{Global symmetries at weak and strong coupling}

So far, we have simply reproduced the known results on the gauge dynamics by considering M-theory on the local surface geometry, similarly to \cite{Morrison:1996xf,Intriligator:1997pq,Jefferson:2017ahm,Jefferson:2018irk}.
In the following, we will discuss the flavor symmetries $G_f$ of these theories, both at weak and strong coupling.
The idea is analogous to that presented in \cite{Xie:2017pfl}, namely that the toric diagram of the compact surface $S$ indicates the existence of ADE singularities along a non-compact curve in the Calabi--Yau threefold $\hat{Y}$, once $S$ is partially or complete blown down.
The novelty here will be to identify these singularities explicitly as codimension one singularities of $\hat\pi: \hat{Y} \rightarrow B$ which corresponds to the remnant flavor symmetries of the 6d conformal matter theory from F-theory on $\hat{Y}$.
We will match these exactly with the known field theoretic results presented in section \ref{sec:review_su2_in_5d}, which we summarize here again for convenience:
\begin{equation}
	\begin{array}{c|c|c|c}
		\text{surface geometry} & N_f & G_f \, \text{ (weak coupling)} & G_f \, \text{ (SCFT)} \\ [.6ex] \hline \rule{0pt}{2.5ex} 
		\bbP^{231} & 2 & SO(4) \times U(1)_T \cong SU(2)^2 \times U(1)_T & SU(3) \times SU(2) \\
		\text{Bl}_1 \bbF_2 & 1 & SO(2) \times U(1)_T \cong U(1)^2 & SU(2) \times U(1) \\
		\bbF_2 & 0 & U(1)_T & SU(2) \\
		\text{dP}_1 & 0 & U(1)_T & U(1) \\
		\bbP^2 & - & \emptyset & \emptyset
	\end{array}
\end{equation}

To begin, let us first observe that the toric phases of the top (see figure \ref{fig:rank_1_phases}) only allow non-trivial intersections between the non-flat surface $S$ and the fibers of the exceptional divisors $A_i = \{\alpha_i\}$ with $i = 5,...,8$.
These fibral curves, which at generic points over $W$ will be an irreducible rational curve, will split at the non-flat point $\{b_6\} \cap W$ according to the factorization of the equations $\{\hat{P}_1\} \cap \{\alpha_i\} \cap \{b_6\}$:
\begin{align}
  & A_5: \qquad \alpha_5 = 0 \quad \text{and} \quad r \, (s_4\,\alpha_6 \, \alpha_7^2 \, x^3 + s_8\, \alpha_8 \, y^2) = 0\, , \label{eq:rank_1_e5} \\
  & A_6: \qquad \alpha_6 = 0 \quad \text{and} \quad  s_8\,\alpha_8 \, r \, y^2 = 0 \, , \label{eq:rank_1_e6} \\
  & A_7: \qquad \alpha_7 = 0 \quad \text{and} \quad s_8\,\alpha_8 \, r \, y^2 = 0\, , \label{eq:rank_1_e7} \\
  & A_8: \qquad \alpha_8 =0 \quad \text{and} \quad s_4\,\alpha_6 \, \alpha_7^2 \, r\, x^3 = 0 \, . \label{eq:rank_1_e8}
\end{align}
First, note that for generic choice of coefficients $s_i$, they cannot vanish at the non-flat point $\{b_6\} \cap W$, hence we can formally set them to 1 (or any other non-zero constant).
Furthermore, these equations are subject to the restrictions coming from the triangulation.
More precisely, even though the polynomial $\hat{P}_1$ factorizes for $\alpha_j = 0$, a component defined by the vanishing of a single toric coordinate $q_i = 0$ could actually be absent in a specific resolution phase, if $\vec{\alpha}_j$ and $\vec{q}_i$ do not share a cone in the corresponding triangulation of $\Delta$.
As we will see now, all the relevant information is encoded within the induced triangulation on the facet.

\subsubsection{\boldmath{$N_f = 2$}}\label{sec:rank_1_Nf=2}

Given that the gauge dynamics arises entirely from the compact surface $S$, all states that can become massless must arise from M2-branes wrapping curves classes in $S$.
Since $S$ is toric, all such classes can be generated by the toric ones, i.e., $\Gamma = \{\alpha_i\}$ with $i=5,...,8$, $\{x\}$ and $\{y\}$.


The global symmetry $G_f$ of this theory is expected to be a subgroup of the global $E_8$ symmetry of the 6d SCFT affinized by the Kaluza--Klein $U(1)$ from the circle reduction.
As explained in \cite{Xie:2017pfl}, the rank of the flavor symmetry $G_f$ is $\#(F_S) - 3$, where $\#(F_S)$ is the number of lattice points on the boundary of the toric diagram of $S$.

We remark that this number agrees in all cases with the rank of the intersection matrix of all toric curves in $S$ with all non-compact divisors of $\hat{Y}$, which is spanned by the exceptional divisors $A_i$ and a (multi-)section of the fibration.
Physically, this is expected if the 5d gauge sector indeed arises from an F-theory conformal matter theory on $\hat{Y}$, because the full 5d flavor symmetry must be generated by the 6d flavor group realized via $A_i$ and the KK-$U(1)$ dual to the (multi-)section.\footnote{Here, we are ignoring possible abelian factors of the 6d flavor symmetry, which could arise from a non-trivial Mordell--Weil group of $\hat{Y}$, see \cite{Lee:2018ihr}.
In principle, these can also be captured in toric models via the ``base'' polygon $F_m$ of the top, though they are trivial for $F_{10}$.
We leave a more thorough analysis of other examples for the future.}

In the $N_f = 2$ phase, we see that the toric diagram of $S \cong \bbP^{231}$ has 6 boundary vectors (cf.~figure \ref{fig:rank_1_dP3}), hence $\text{rank}(G_f) = 3$.
To go beyond rank counting, we will identify how the codimension one fibers of $\hat{Y}$ split at the non-flat fiber.
This requires us to re-evaluate the equations \eqref{eq:rank_1_e5} -- \eqref{eq:rank_1_e8} using the triangulation depicted in figure \ref{fig:rank_1_dP3} for the polygon $F_S$.
There we see that the vector $\vec{\alpha}_8$ is not connected by a line with $\{\vec{\alpha}_{6,7}\}$ and $\vec{x}$.
This implies that despite the factorization \eqref{eq:rank_1_e8}, only $\{\alpha_8\} \cap \{b_6\} \cap \{r\}$ corresponds to an actual curve in the resolved geometry.
Using the same logic, we find that, geometrically, $\{\alpha_i\} \cap \{b_6\} \cap \{\hat{P}_1\} = \{\alpha_i\} \cap \{b_6\} \cap \{r\}$ for $i = 6,7$ as well.
For the fiber of $A_5$ however, the second factor in \eqref{eq:rank_1_e5} is a sum, which can vanish even if the summands are non-zero.
Therefore, the triangulation does not forbid the factorization of $\{\alpha_5\} \cap \{b_6\} \cap \{\hat{P}_1\} = \{\alpha_5\} \cap \{b_6\} \cap ( \{r\} \cup \{\alpha_6\,\alpha_7^2\,x^3 + \alpha_8\,y^2\} )$.

From this analysis, we see that the fibers of the exceptional divisors $A_6$, $A_7$ and $A_8$ do not split, and are in fact entirely contained in the surface $S$.
Furthermore, recall from above that the corresponding fibers of $\{\alpha_6\}$ and $\{\alpha_8\}$ are also fibral curves with respect to the ruling on $S$.
Following our general discussion in \ref{sec:global_symmetry_general}, we therefore conclude that, when the fiber of $S$ is blown down, the codimension one fibers of $A_6$ and $A_8$ are also shrunk.
Being two disconnected nodes of the $E_8$'s Dynkin diagram, the threefold $\hat{Y}_3$ therefore develops two $SU(2)$ singularities in the generic fiber over $W$.

Physically, one thus sees an enhancement of an $SU(2)^2 \cong SO(4) = SO(2 N_f)$ flavor subgroup of the $E_8$, when the surface $S$ shrinks to a curve, i.e., when the gauge sector is a weakly coupled $\mathfrak{su}(2)$ theory with $N_f$ flavors.
Since this non-abelian flavor symmetry has rank 2, there must be a remaining $U(1)$ part.
Indeed, this $U(1)$ must be the topological $U(1)_T$ inherent to any 5d gauge theory!
There is some redundancies in identifying the $U(1)$s in terms of the non-compact divisors.
One natural choice for the Cartans of the $SO(4)$ flavor group is simply $A_6$ and $A_8$.
The topological $U(1)_T$ is determined by requiring all massless states of the weakly coupled gauge theory to be uncharged under it.
One such divisor is $D_T = A_5 + A_6 + A_7 - Y$, which makes apparent the involvements of the KK-$U(1)$ in form of $Y$ and the Cartan generator $A_7$ of $E_8$.

Finally, we can also see how the flavor symmetry enhances when we pass to strong coupling.
To do that, we have to shrink the surface $S \cong P^{231}$ completely to a point.
This amounts for not only to blow down the fiber, but also the base of the ruling on $S$.
Recall from earlier that $\{\alpha_7\}$ is a section, i.e., a copy of the base of $S$.
But in this resolution phase, it is also homologically equivalent to the generic fiber of the exceptional divisor $A_7$.
Thus, blowing down $S$ to a point will inevitable force the fibers of $A_6$, $A_7$ and $A_8$ to shrink everywhere over $W$.

We note that this is in accord with the statement in \cite{Xie:2017pfl} that the intersection pattern of vectors on the \textit{interior} of edges of $F_S$ corresponds to the Dynkin diagram of the non-abelian part of $G_f$.
In this case, we see from figure \ref{fig:rank_1_dP3} that there are two edges with interior vectors, one having only a single vector ($\vec\alpha_8$) and one with two connected vectors ($\vec\alpha_6$ and $\vec\alpha_7$).
The first corresponds to the Dynkin diagram of $SU(2)$, and the second is that of $SU(3)$, compatible with the embedding of $G_f$ into the $E_8$ that we saw above.

\subsubsection{\boldmath{$N_f=1$} phase}

We can repeat the analysis in analogous fashion for the $N_f=1$ phase, where the surface geometry is $S \cong \text{Bl}_1 \bbF_2$.
Since the corresponding toric diagram $F_S$ has 5 boundary vectors, the rank of the global flavor symmetry is 2.

Utilizing the facet triangulation depicted in figure \ref{fig:rank_1_dP2}, we see that the relevant splittings of the codimension one fibers are only for $A_{6,7,8}$.
In particular, we find that the fiber of $A_6$ in \eqref{eq:rank_1_e6} splits into two components, $\{\alpha_6\} \cap \{\alpha_8\}$ and $\{\alpha_6\} \cap \{r\}$, the latter of the two being the toric curve $\{\alpha_6\}$ inside the surface $S$.
Similarly, \eqref{eq:rank_1_e8} reveals that also the fiber of $A_8$ splits into two, one being $\{\alpha_8 \} \cap \{\alpha_6\}$ sitting outside $S$, and one being $\{\alpha_8\} \cap \{r\}$ which is the toric curve $\{\alpha_8\} \subset S$.
On the other hand, the fiber of $A_7$ in \eqref{eq:rank_1_e7} does not split and simply becomes the toric curve $\{\alpha_7\}$ inside the non-flat surface.

Now, to identify possible non-abelian parts of $G_f$ at weak coupling, we ask if we create any singularities over $W$ when we blow down the fiber in the ruling of $S$.
As discussed in the previous section, this results in shrinking the curves $\{\alpha_6\}$, $\{\alpha_8\}$ and $\{x\}$ on $S$.
However, because the fibers of $A_{6,8}$ each split into two components, one of which lies outside of $S$, the volume of these divisors' generic fibers over $W$ remains finite size, even if we blow down $S$.
Thus, there is no singularity enhancement in the fiber over $W$.
Hence, the rank two flavor symmetry is abelian at weak coupling, which of course agrees with the known result $G_f = SO(2) \times U(1)_T \cong U(1)^2$.

To go to strong coupling, we now have to also shrink the base of the ruling on $S$, which again is given by $\{\alpha_7\}$.
Like before, this curve is homologous to the generic fiber of $A_7$, because it did not split at the non-flat point.
This means that at strong coupling, we find a single node of the $E_8$ Dynkin diagram that is blown down.
Therefore, we confirm that in the $N_f= 1$ phase, the global symmetry enhances from $SO(2) \times U(1)$ at weak coupling to $SU(2) \times U(1) \cong E_2$ at strong coupling.
This is also compatible with \cite{Xie:2017pfl}, since the polygon $F_S$ in this case has only one vector ($\vec\alpha_7$) interior to edges.

\subsubsection{Different global symmetries for \boldmath{$N_f = 0$}}

Proceeding analogously, we can verify easily that the two $N_f = 0$ phases differ by their global symmetry enhancement at strong coupling.
First, since both diagrams in figures \ref{fig:rank_1_P112} and \ref{fig:rank_1_dP1} have 4 boundary vectors, we verify the existence of a non-trivial global symmetry to begin with.

Furthermore, we can verify that in the $S \cong \bbF_2$ phase, the fiber of $S$ contains only one of the two components, into which the fiber of $A_6$ splits, see figure \ref{fig:rank_1_P112} and equation \eqref{eq:rank_1_e6}.
Similarly, the fiber of $S \cong \text{dP}_1$ only contains a split component of $A_8$, cf.~figure \ref{fig:rank_1_dP1} and equation \eqref{eq:rank_1_e8}.
Thus blowing down $S$ to a curve along its ruling does not produce any singularities over $W$, thus the global symmetry is just the topological $U(1)$ in both cases.

However, in the $S \cong \bbF_2$ phase, the surface contains the full fiber of $A_7$, which does not split.
Thus, when we shrink $S$ to a point, we do observe an enhancement of the codimension one singularity.
This indicates the enhancement of the global symmetry, $U(1)_T \rightarrow SU(2) \cong E_1$.
Both cases are of course in agreement with \cite{Xie:2017pfl}.

On the other hand, when $S \cong \text{dP}_1$, the fiber of $A_7$ also splits, see figure \ref{fig:rank_1_dP1} and equation \eqref{eq:rank_1_e8}, such that the base of $S$ is identified with only one of the two components.
Thus, even when the surface is blown down to a point, the generic fiber of $A_7$ is still at finite size.
Hence, the symmetry at strong coupling remains a $U(1)$, which has been labelled $\tilde{E}_1$ in \cite{Morrison:1996xf}.

Finally, let us briefly comment on the phase $S \cong \bbP^2$.
Since the corresponding toric diagram only has 3 boundary vectors, there is no flavor symmetry, see figure \ref{fig:rank_1_P2}.
Furthermore, the fiber of $A_8$---the only exceptional divisor intersecting the non-flat component---splits into several components, only one of which is contained in $S$.
Thus, consistent with the familiar field theory result, there is no singularity enhancement over $W$ at strong coupling.

Before we proceed to the higher rank theories, let us briefly mention that the triangulations of the facet in figure \eqref{fig:rank_1_phases} allow us to determine the full structure of the non-flat fibers in each case, which we have collected in appendix \ref{app:fiber_structure}.
From the hopefully intuitive notation there, one can very quickly read off the non-abelian flavor symmetry enhancements of the SCFT.

\section{Higher Rank Theories from Non-flat Fibers}\label{sec:rank_two}

In this section, we provide further examples of non-flat fibers which realize higher rank 5d SCFTs.
These theories come from the other two facets of the $E_8$ top and can be analyzed similarly as the rank one facet.

\subsection[5d Rank 2 Theories from 6d \texorpdfstring{$(E_8, SU(2))$}{(E8, SU2)} Conformal Matter]{5d Rank 2 Theories from 6d \boldmath{$(E_8, SU(2))$} Conformal Matter}

Let us turn to examples of 5d rank two theories obtained from circle reductions of the 6d $(E_8 , SU(2))$ conformal matter theory, also known as the 6d rank 2 E-string.
In the singular elliptic fibration $Y_3$, this theory is supported at the non-minimal singularity over $W \cap \{s_4\}$, where the $E_8$ and $SU(2)$ loci collide.
It is known that the circle reduction to 5d (without automorphism twists) will yield a rank two theory.
And indeed, this is realized by the top.

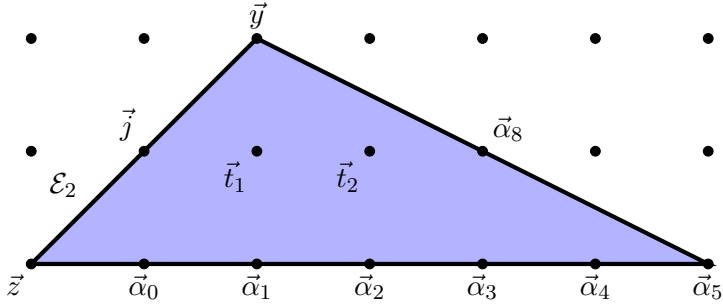
\begin{figure}[ht]
  \centering
  	\begin{tikzpicture}[scale=1.5]
		\begin{scope}[xshift=0.33\textwidth]
			\filldraw [ultra thick, draw=black, fill=blue!30!white] (3,-1)--(-3,-1)--(-1,1)--cycle; 
			\foreach \x in {-3,-2,...,3}{
				\foreach \y in {-1,...,1}{
					\node[draw,circle,inner sep=1.3pt,fill] at (\x,\y) {};
				}
		    }
			\node [above] at (-1,1)  {$\vec{y}$};
			\node [above left] at (-2,0) {$\vec{j}$};
			\node [below left] at (-3,-1) {$\vec{z}$};
			\node [below] at (-2,-1) {$\vec{\alpha}_0$};
			\node [below] at (-1,-1) {$\vec{\alpha}_1$};
			\node [below] at (0,-1) {$\vec{\alpha}_2$};
			\node [below] at (1,-1) {$\vec{\alpha}_3$};
			\node [below] at (2,-1) {$\vec{\alpha}_4$};
			\node [below] at (3,-1) {$\vec{\alpha}_5$};
			\node [above right] at (1,0)  {$\vec{\alpha}_8$};
			\node [below left] at (-1,0) {$\vec{t}_1$};
			\node [below left] at (0,0) {$\vec{t}_2$};
			\node [above left] at (-2.5, -0.5) {${\cal E}_2$};
		\end{scope}
	\end{tikzpicture}
  \caption{The facet of the $E_8$ top over the edge ${\cal E}_2$, containing the non-flat surfaces $\{t_1\}$ and $\{t_2\}$ over $\{s_4\} \cap W$.}\label{fig:rank_2_facette}
\end{figure}

Inside its facet over the edge ${\cal E}_2$ of the $F_{10}$ polygon (see figure \ref{fig:F10_with_blow-up}), the top now contains two internal vectors, $\vec{t}_1$ and $\vec{t}_2$, as shown in figure \ref{fig:rank_2_facette}.
Including the coordinates $t_1$, $t_2$ as well as the resolution divisor $\{j\}$ of the $SU(2)$ over $\{s_4\}$ into the blow-up of the hypersurface, the resulting equation (neglecting the $r$ blow-up at $\{b_6 = w=0\}$ now) becomes
\begin{align}\label{eq:resolved_hypersurface_rank_2}
  \begin{split}
    \hat{P}_2 := &\, s_8\,\alpha_8\, t_1\,t_2 \,j \, y^2 + s_4\,\alpha_6\, \alpha_7^2\, x^3  + b_1\, \alpha_0\, \alpha_1\, \alpha_2\, \alpha_3\, \alpha_4\, \alpha_5\, \alpha_6\, \alpha_7\, \alpha_8\,t_1\,t_2\,j\, x\, y\, z \\
    + & \, b_2\, \alpha_0^2\, \alpha_1^2\, \alpha_2^2\, \alpha_3^2\, \alpha_4^2\, \alpha_5^2\, \alpha_6^2\, \alpha_7^2\, \alpha_8\, t_1\,t_2\, j \, x^2\, z^2 +b_3\, \alpha_0^3\, \alpha_1^3\, \alpha_2^3\, \alpha_3^3\, \alpha_4^3\, \alpha_5^3\, \alpha_6^2 \, \alpha_7\, \alpha_8^2\, t_1^2\,t_2^2\, j^2 \, y\, z^3 \\
      + & \,  b_4\, \alpha_0^4\, \alpha_1^4\, \alpha_2^4\, \alpha_3^4\, \alpha_4^4\, \alpha_5^4\, \alpha_6^3\, \alpha_7^2\, \alpha_8^2\, t_1^2\,t_2^2 \, j^2\, x\, z^4 + b_6\, \alpha_0^5\, \alpha_1^4\, \alpha_2^3\, \alpha_3^2\, \alpha_4\, \, t_1^2\,t_2 \, j^3 \,z^6  \, .
  \end{split} 
\end{align}
All terms but $s_4...$ contain the factors $t_1\,t_2$, so that $S_1 = \{s_4 = t_1 =0\}$ and $S_2 = \{s_4 = t_2 =0\}$ indeed define two surfaces in the ambient space that lie on the hypersurface, which are non-flat fiber components over the point $\{s_4\} \cap W$.

As previously, the physics of the 5d theory will depend on the resolution phase of the geometry.
A subset of these phases is encoded via the triangulation of the top, which induces a triangulation of the facet in figure \ref{fig:rank_2_facette}.
While for the rank one case, the combinatorics only allows for five different triangulations (see figure \ref{fig:rank_1_facette}), the number for the rank two facet is 156.

Of these, 131 lead to geometries where the two non-flat fibers intersect; the other 25 will give two independent $\mathfrak{su}(2)$ theories.
Those configuration where the surfaces intersect can realize $\mathfrak{su}(3)$ or $\mathfrak{su}(2) \times \mathfrak{su}(2)$ gauge theories, but never $\mathfrak{sp}(2)$.
To realize the latter, the two surfaces $S_1 \equiv \{t_1\}$ and $S_2 \equiv \{t_2\}$ would have to intersect along a curve which is a bisection on one of the two \cite{Intriligator:1997pq}.
However, in the present toric realization, the intersection curve $\{t_1\} \cap \{t_2\}$ is always a toric curve, and thus cannot be a bisection of either.\footnote{More precisely, this holds for toric rulings, i.e., when the generic fiber corresponds to a vector of the toric diagram.
In this case, intersections of two distinct toric divisors is either 0 or 1 because the surfaces are smooth.}
Still, it would be interesting to identify the torically realized theories here with those appearing in the classification \cite{Jefferson:2018irk}.
We will leave this for future works.

\subsubsection{Non-flat fiber with \boldmath{$\mathfrak{su}(3) + 6 {\bf F}$} theory}

\begin{figure}[ht]
\centering
	\begin{tikzpicture}[scale=1.5]
		\begin{scope}[xshift=0.33\textwidth]
			\filldraw [ultra thick, draw=black, fill=blue!30!white] (3,-1)--(-1,-1)--(0,0)--(-1,1)--cycle; 
			\filldraw [ultra thick, draw=black, fill=green!30!white] (-3,-1)--(-1,-1)--(-1,1)--cycle; 
			\filldraw [ thick, draw=black, fill=green!30!blue!30!white] 
			(-1,-1)--(0,0)--(-1,1)--cycle; 
		    \filldraw [ thick, draw=black] (-1,0)--(0,0); 
			\filldraw [ thick, draw=black] (-1,0)--(-2,0); 
			\filldraw [ thick, draw=black] (-1,0)--(-3,-1); 
			\filldraw [ thick, draw=black] (-1,0)--(-2,-1); 
			\filldraw [ thick, draw=black] (-1,0)--(-1,1); 
			\filldraw [ thick, draw=black] (-1,0)--(-1,-1); 
			\filldraw [ thick, draw=black] (0,0)--(1,0); 
			\filldraw [ thick, draw=black] (0,0)--(0,-1); 
			\filldraw [ thick, draw=black] (0,0)--(1,-1); 
			\filldraw [ thick, draw=black] (0,0)--(2,-1); 
			\filldraw [ thick, draw=black] (0,0)--(3,-1); 
			\node [above] at (-1,1)  {$\vec{y}$};
			\node [above left] at (-2,0) {$\vec{j}$};
			\node [below left] at (-3,-1) {$\vec{z}$};
			\node [below] at (-2,-1) {$\vec{\alpha}_0$};
			\node [below] at (-1,-1) {$\vec{\alpha}_1$};
			\node [below] at (0,-1) {$\vec{\alpha}_2$};
			\node [below] at (1,-1) {$\vec{\alpha}_3$};
			\node [below] at (2,-1) {$\vec{\alpha}_4$};
			\node [below] at (3,-1) {$\vec{\alpha}_5$};
			\node [above right] at (1,0)  {$\vec{\alpha}_8$};
			\node [above left] at (-1,0) {$\vec{t}_1$};
			\node [above right] at (0,0) {$\vec{t}_2$};
			\filldraw [ultra thick, draw=red] (-2,0)--(1,0);
			\foreach \x in {-3,-2,...,3}{
				\foreach \y in {-1,...,1}{
					\node[draw,circle,inner sep=1.3pt,fill] at (\x,\y) {};
				}
		    };
		\end{scope}
	\end{tikzpicture}
\caption{A triangulation of the rank two facet ``involving'' all vectors.
The red line indicates the ruling of the two surfaces, along which a blow-down produces an $\mathfrak{su}(3)$ gauge theory.}\label{fig:rank_2_example_1}
\end{figure}
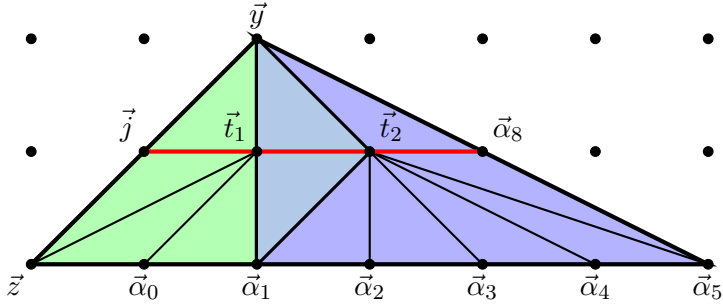

The first example we examine comes a triangulation which involves the all vectors of the facet, i.e., all vectors in figure \ref{fig:rank_2_facette} give rise to a curve on one of the two surfaces.
The explicit triangulation is depicted in figure \ref{fig:rank_2_example_1}.
The topology of the surface $\{t_1\}$ is that of a dP$_2$, blown up further at a non-generic point (represented by the vector $\vec{z}$).
The surface $\{t_2\}$ can be viewed as a four-fold blow-up of the Hirzebruch surface $\bbF_2$---spanned by vectors $(\vec{y}, \vec{t}_1, \vec{\alpha}_5, \vec{\alpha}_8)$---with three being non-generic ($\vec{\alpha}_{1,2,3}$).
The triple intersection numbers of these surfaces in the threefold $\hat{Y}$ are
\begin{align}\label{eq:triple_intersections_rank2}
  [t_1]^3 = 6 \, , \qquad [t_1]^2 \cdot [t_2] = -2 \, , \qquad [t_1]\cdot [t_2]^2 = 0 \, , \qquad [t_2]^3 = 4 \, .
\end{align}

Right away, we can see that both surfaces are ruled, and glued along a curve which is a section of both rulings.
Indeed, the diagram admits a projection along the horizontal axis which maps 2d-cones (triangles in figure \ref{fig:rank_2_example_1}) onto 1d-cones (vertical lines starting at origin, which is the projection of $\vec{t}_i$).
Therefore, the weakly coupled gauge theory is has an $\mathfrak{su}(3)$ algebra \cite{Intriligator:1997pq}.
One can easily verify that the curves $\{y\} \cap \{t_1\}$ and $\{y\} \cap \{t_2\}$ have the appropriate intersection numbers to support the simple roots of $\mathfrak{su}(3)$.
This is of course no surprise, since they are also the generic fibers of the rulings on either surface.

Charged matter arise all from the curves on the bottom line of vectors in figure \ref{fig:rank_2_example_1}. Their intersection numbers with $([t_1], [t_2])$ are:
\begin{align}\label{eq:matter_curves_rank2_example1}
  \begin{split}
    \{t_1\} \cap \left\{ 
    \begin{array}{lr}
      \{z\} : & (-1,0) \\
      \{\alpha_0\} : & (0,0) \\
      \{\alpha_1\} : & (-1,1) \\
    \end{array}
    \right. \qquad
    \{t_2\} \cap \left\{ 
    \begin{array}{lr}
      \{\alpha_1\} : & (1,-1) \\
      \{\alpha_2\} : & (0,0) \\
      \{\alpha_3\} : & (0,0) \\
      \{\alpha_4\} : & (0,0) \\
      \{\alpha_5\} : & (0,-1) \\
    \end{array}\right.
  \end{split}
\end{align}
Based on some combinatorial methods \cite{unpublished}, from this collection of curves one can form connected linear combinations with genus 0, on which the M2-brane states would fill out $N_f = 6$ fundamental hypermultiplets of $\mathfrak{su}(3)$.

Using the intuition about the global symmetries, we can also see this arising as follows:
With the triangulation structure on the facet as in figure \ref{fig:rank_2_example_1}, we can straightforwardly check that out of the $E_8$ exceptional divisors $A_i$, $i=0,...,8$, only $A_1$ and $A_5$ factorize into two.
While the fiber splitting of $A_5$ yields a curve ($\{\alpha_5\} \cap \{\alpha_8\}$) not inside the non-flat surfaces, the factors of $A_1$'s fiber are each contained inside $\{t_1\}$ and $\{t_2\}$, respectively:
\begin{align}
  \{\hat{P}_2|_{s_4 =0} \} \cap \{\alpha_1\} \longrightarrow (\{t_1\} \cap \{\alpha_1\}) \, \cup (\{t_2\} \cap \{\alpha_1\} )\,.
\end{align}
When we now blow down the fibers of $\{t_1\}$ and $\{t_2\}$, thus shrinking all the curves in \eqref{eq:matter_curves_rank2_example1}, we are hence also shrinking the generic fibers of $A_i$, $i=0,...,4$.
The resulting singularity---collapsing the first six nodes of the affine $E_8$ Dynkin diagram---corresponds to that of ADE type $SU(6)$, which can only be part of the global symmetry $G_f$ if $N_f = 6$.

Furthermore, because the toric diagram has 10 vectors on the boundary, the rank of the flavor  group must be 7 \cite{Xie:2017pfl}.
Thus, the full flavor symmetry at weak coupling must be $SU(6) \times U(1) \times U(1)' \cong U(6) \times U(1)_T$, where $U(1)_T$ is the universal topological $U(1)$.

Lastly, we observe that in order to reach strong coupling, we have to also blow down the base of the fibrations, which are the curves $\{t_1\} \cap \{j\}$ and $\{t_2\} \cap \{\alpha_8\}$.
Again, it is easily verified that, the triangulation (figure \ref{fig:rank_2_example_1}) for this resolution phase will not split the fibers of the corresponding exceptional divisors $J$ and $A_8$ at the non-minimal point.
Hence, in the SCFT limit, we further enhance the singularities of $\hat{Y}_3$ in codimension one by shrinking two disjoint $\bbP^1$ fibers (one over $\{s_4\}$ and one over $W$), which furthermore are also not connected to the $SU(6)$ singularity.
In total, we therefore find an $SU(6) \times SU(2) \times SU(2)'$ flavor symmetry at the SCFT point, consistent with the structure of the vectors on edge interiors.

This flavor symmetry enhancement is only compatible with the $\mathfrak{su}(3) + 6{\bf F}$ theory if the Chern--Simons level is $\kappa = 0$ (see, e.g., \cite{Tachikawa:2015mha}). In fact, the prepotential of an $SU(3)$ gauge theory with $N_f$ hypermultiplets reads
\begin{align} \label{eq:gensu3prep} 
\begin{split}
 6 \mathcal  F & = 3 m_0 \left( \begin{bmatrix}\phi^1 \\\phi^2 \end{bmatrix}^T \begin{bmatrix} 2,-1 \\
 -1,2 \end{bmatrix} \begin{bmatrix} \phi^1  \\ \phi^2\end{bmatrix} \right)+ 3 \kappa((\phi^1)^2 \phi^2 - (\phi^2)^2 \phi^1)\\
 & +\left|\begin{bmatrix}2\\-1 \end{bmatrix}^T \begin{bmatrix} \phi^1  \\ \phi^2\end{bmatrix} \right|^3 + \left|\begin{bmatrix}1\\1 \end{bmatrix}^T \begin{bmatrix} \phi^1  \\ \phi^2\end{bmatrix} \right|^3+  \left|\begin{bmatrix}-1\\2 \end{bmatrix}^T \begin{bmatrix} \phi^1   \\ 
 \phi^2\end{bmatrix} \right|^3 \\ 
& - \frac{N_f-a}{2} \left( \left|\begin{bmatrix}1\\0 \end{bmatrix}^T \begin{bmatrix} \phi^1  \\ \phi^2\end{bmatrix}+m_f \right|^3+\left|\begin{bmatrix}-1\\1 \end{bmatrix}^T \begin{bmatrix} \phi^1  \\ \phi^2\end{bmatrix} + m_f \right|^3+\left|\begin{bmatrix}0\\-1 \end{bmatrix}^T \begin{bmatrix} \phi^1  \\ \phi^2\end{bmatrix}+ m_f \right|^3 \right)  \\
& - \frac{a}{2} \left( \left|\begin{bmatrix}1\\0 \end{bmatrix}^T \begin{bmatrix} \phi^1  \\ \phi^2\end{bmatrix}+m_a \right|^3+\left|\begin{bmatrix}-1\\1 \end{bmatrix}^T \begin{bmatrix} \phi^1  \\ \phi^2\end{bmatrix} + m_a \right|^3+\left|\begin{bmatrix}0\\-1 \end{bmatrix}^T \begin{bmatrix} \phi^1  \\ \phi^2\end{bmatrix}+ m_a \right|^3 \right) 
\end{split}
\end{align}
where the product between matrices is the standard Cartesian one, and we have 
\begin{equation}
2\phi^1 - \phi^2\geq 0, \qquad \phi^1 + \phi^2\geq 0, \qquad \qquad- \phi^1 + 2\phi^2\geq 0
\end{equation}
The contributions from the hypermultiplets splits accordingly to the sign of the middle terms in the third and fourth row. 
$(N_f-a)$ hypers have $\phi^1 >0$, $\phi^2>0$, $\phi^2-\phi^1+m_f\geq0$ and $a$ hypers have $\phi^1 >0$, $\phi^2>0$, $-\phi^2+\phi^1+m_a\geq0$. 
For $N_f=6$ the match between \eqref{eq:gensu3prep} and the geometric prepotential computed with the intersection numbers \eqref{eq:triple_intersections_rank2} fixes $a=2$ and Chern--Simons level $\kappa=0$.

\subsubsection{Physically equivalent flop transitions and \boldmath{$\mathfrak{su}(2)$} quiver phases}

\begin{figure}[p]
  \centering
  \subfigure[Triangulation 1. The blue line indicates a different ruling of the surface $\{t_2\}$, under which the curve $\{t_1\} \cap \{t_2\}$ is a fiber.]{\label{fig:rank_2_tri1}
    	\begin{tikzpicture}[scale=1.5]
		\begin{scope}[xshift=0.33\textwidth]
			\filldraw [ultra thick, draw=black, fill=blue!30!white] (3,-1)--(-3,-1)--(0,0)--(-1,1)--cycle; 
			\filldraw [ultra thick, draw=black, fill=green!30!white] (-3,-1)--(-1,0)--(-1,1)--cycle; 
			\filldraw [ thick, draw=black, fill=green!30!blue!30!white] 
			(-3,-1)--(0,0)--(-1,1)--(-1,0)--cycle; 
		    \filldraw [ thick, draw=black] (-1,0)--(0,0); 
			\filldraw [ thick, draw=black] (-1,0)--(-2,0); 
			\filldraw [ thick, draw=black] (-1,0)--(-1,1); 
			\filldraw [ thick, draw=black] (0,0)--(1,0); 
			\filldraw [ thick, draw=black] (0,0)--(0,-1); 
			\filldraw [ thick, draw=black] (0,0)--(1,-1); 
			\filldraw [ thick, draw=black] (0,0)--(2,-1); 
			\filldraw [ thick, draw=black] (0,0)--(3,-1); 
			\filldraw [ thick, draw=black] (0,0)--(-2,-1); 
			\filldraw [ thick, draw=black] (0,0)--(-1,-1); 
			\node [above] at (-1,1)  {$\vec{y}$};
			\node [below left] at (-3,-1) {$\vec{z}$};
			\node [above right] at (1,0)  {$\vec{\alpha}_8$};
			\node [above left] at (-1,0) {$\vec{t}_1$};
			\node [above right] at (0,0) {$\vec{t}_2$};
			\filldraw [ultra thick, draw=red] (-2,0)--(1,0);
			\filldraw [ultra thick, draw=blue] (-1,1)--(1,-1);
			\foreach \x in {-3,-2,...,3}{
				\foreach \y in {-1,...,1}{
					\node[draw,circle,inner sep=1.3pt,fill] at (\x,\y) {};
				}
		    };
		\end{scope}
	\end{tikzpicture}
  }

  \vspace{.5cm}
  \subfigure[Triangulation 2. The surface $\{t_1\}$ still has only one ruling (red line).]{\label{fig:rank_2_tri2}
    	\begin{tikzpicture}[scale=1.5]
		\begin{scope}[xshift=0.33\textwidth]
			\filldraw [ultra thick, draw=black, fill=blue!30!white] (3,-1)--(-2,-1)--(0,0)--(-1,1)--cycle; 
			\filldraw [ultra thick, draw=black, fill=green!30!white] (-3,-1)--(-2,-1)--(-1,0)--(-1,1)--cycle; 
			\filldraw [ thick, draw=black, fill=green!30!blue!30!white] 
			(-2,-1)--(0,0)--(-1,1)--(-1,0)--cycle; 
		    \filldraw [ thick, draw=black] (-1,0)--(0,0); 
			\filldraw [ thick, draw=black] (-1,0)--(-2,0); 
			\filldraw [ thick, draw=black] (-1,0)--(-1,1); 
			\filldraw [ thick, draw=black] (-1,0)--(-3,-1); 
			\filldraw [ thick, draw=black] (0,0)--(1,0); 
			\filldraw [ thick, draw=black] (0,0)--(0,-1); 
			\filldraw [ thick, draw=black] (0,0)--(1,-1); 
			\filldraw [ thick, draw=black] (0,0)--(2,-1); 
			\filldraw [ thick, draw=black] (0,0)--(3,-1); 
			\filldraw [ thick, draw=black] (0,0)--(-2,-1); 
			\filldraw [ thick, draw=black] (0,0)--(-1,-1); 
			\node [above left] at (-1,0) {$\vec{t}_1$};
			\node [above right] at (0,0) {$\vec{t}_2$};
			\filldraw [ultra thick, draw=red] (-2,0)--(1,0);
			\filldraw [ultra thick, draw=blue] (-1,1)--(1,-1);
			\foreach \x in {-3,-2,...,3}{
				\foreach \y in {-1,...,1}{
					\node[draw,circle,inner sep=1.3pt,fill] at (\x,\y) {};
				}
		    };
		\end{scope}
	\end{tikzpicture}
  }

  \vspace{.5cm}
  \subfigure[Triangulation 3. Each surface has a second ruling (blue lines).]{\label{fig:rank_2_tri3}
    	\begin{tikzpicture}[scale=1.5]
		\begin{scope}[xshift=0.33\textwidth]
			\filldraw [ultra thick, draw=black, fill=blue!30!white] (3,-1)--(-1,-1)--(0,0)--(-1,1)--cycle; 
			\filldraw [ultra thick, draw=black, fill=green!30!white] (-3,-1)--(-1,-1)--(-1,1)--cycle; 
			\filldraw [ thick, draw=black, fill=green!30!blue!30!white] 
			(-1,-1)--(0,0)--(-1,1)--cycle; 
		    \filldraw [ thick, draw=black] (-1,0)--(0,0); 
			\filldraw [ thick, draw=black] (-1,0)--(-2,0); 
			\filldraw [ thick, draw=black] (-1,0)--(-3,-1); 
			\filldraw [ thick, draw=black] (-1,0)--(-2,-1); 
			\filldraw [ thick, draw=black] (-1,0)--(-1,1); 
			\filldraw [ thick, draw=black] (-1,0)--(-1,-1); 
			\filldraw [ thick, draw=black] (0,0)--(1,0); 
			\filldraw [ thick, draw=black] (0,0)--(0,-1); 
			\filldraw [ thick, draw=black] (0,0)--(1,-1); 
			\filldraw [ thick, draw=black] (0,0)--(2,-1); 
			\filldraw [ thick, draw=black] (0,0)--(3,-1); 
			\node [above left] at (-1,0) {$\vec{t}_1$};
			\node [above right] at (0,0) {$\vec{t}_2$};
			\filldraw [ultra thick, draw=red] (-2,0)--(1,0);
			\filldraw [ultra thick, draw=blue] (-1,1)--(-1,-1);
			\filldraw [ultra thick, draw=blue] (-1,1)--(1,-1);
			\foreach \x in {-3,-2,...,3}{
				\foreach \y in {-1,...,1}{
					\node[draw,circle,inner sep=1.3pt,fill] at (\x,\y) {};
				}
		    };
		\end{scope}
	\end{tikzpicture}
  }

  \vspace{.4cm}
  \subfigure[Triangulation 4. Each surface has a second ruling (blue lines).]{\label{fig:rank_2_tri4}
    	\begin{tikzpicture}[scale=1.5]
		\begin{scope}[xshift=0.33\textwidth]
			\filldraw [ultra thick, draw=black, fill=blue!30!white] (3,-1)--(0,-1)--(0,0)--(-1,1)--cycle; 
			\filldraw [ultra thick, draw=black, fill=green!30!white] (-3,-1)--(0,-1)--(-1,0)--(-1,1)--cycle; 
			\filldraw [ thick, draw=black, fill=green!30!blue!30!white] 
			(-1,0)--(0,-1)--(0,0)--(-1,1)--cycle; 
		    \filldraw [ thick, draw=black] (-1,0)--(0,0); 
			\filldraw [ thick, draw=black] (-1,0)--(-2,0); 
			\filldraw [ thick, draw=black] (-1,0)--(-3,-1); 
			\filldraw [ thick, draw=black] (-1,0)--(-2,-1); 
			\filldraw [ thick, draw=black] (-1,0)--(-1,1); 
			\filldraw [ thick, draw=black] (-1,0)--(-1,-1); 
			\filldraw [ thick, draw=black] (0,0)--(1,0); 
			\filldraw [ thick, draw=black] (0,0)--(0,-1); 
			\filldraw [ thick, draw=black] (0,0)--(1,-1); 
			\filldraw [ thick, draw=black] (0,0)--(2,-1); 
			\filldraw [ thick, draw=black] (0,0)--(3,-1); 
			\node [above left] at (-1,0) {$\vec{t}_1$};
			\node [above right] at (0,0) {$\vec{t}_2$};
			\filldraw [ultra thick, draw=red] (-2,0)--(1,0);
			\filldraw [ultra thick, draw=blue] (-1,1)--(-1,-1);
			\filldraw [ultra thick, draw=blue] (-1,1)--(1,-1);
			\foreach \x in {-3,-2,...,3}{
				\foreach \y in {-1,...,1}{
					\node[draw,circle,inner sep=1.3pt,fill] at (\x,\y) {};
				}
		    };
		\end{scope}
	\end{tikzpicture}
  }
  \caption{The four inequivalent triangulations of the full facet of the top over ${\cal E}_2$. Depending on the triangulation, each of the two surfaces $\{t_i\}$ can have a different ruling in addition to the one which glues them together along sections.
  In the case where both surfaces are ruled separately (blue lines), the corresponding blow downs produces an $\mathfrak{su}(2) \times \mathfrak{su}(2)$ quiver.
  }\label{fig:rank_2_flops}
\end{figure}
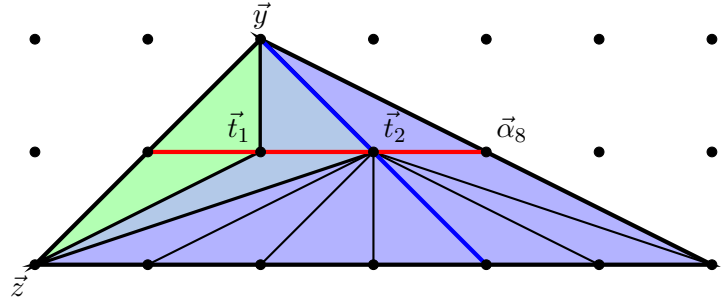
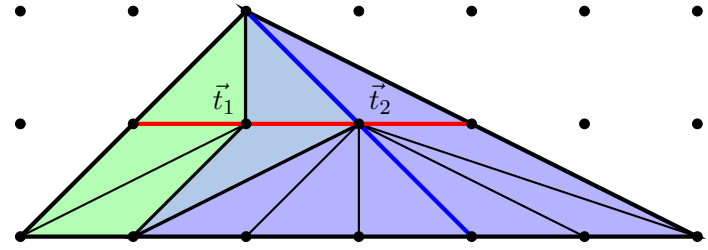
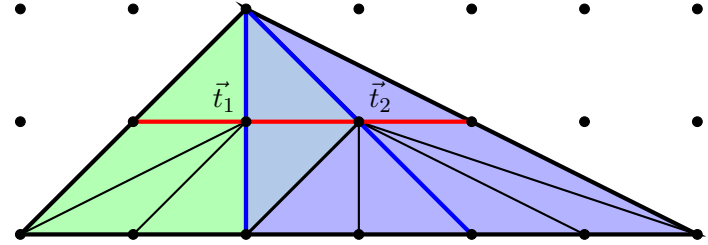
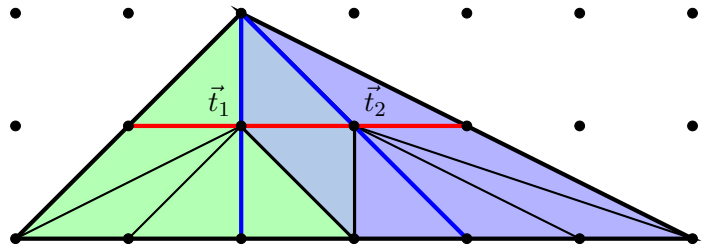

Note that the facet in figure \ref{fig:rank_2_facette} possesses four inequivalent triangulations, in which $\vec{t}_1$ and $\vec{t}_2$ are connected, i.e., $\{t_1\} \cap \{t_2\} \neq \emptyset$, and where every vector on the boundary is connected to either of the two interior vectors, see \ref{fig:rank_2_flops}.
These four triangulations are related via transitions which flop a curve ``out'' from one surface ``into'' the other.\footnote{Strictly speaking, there are seven triangulations, as evident from the figure.
However, physically, the other three triangulations are equivalent upon exchanging $t_1 \leftrightarrow t_2$.}
However, these flops clearly does not change the ``combined'' polygon, which is simply the full facet.
Thus, according to \cite{Xie:2017pfl}, the singular limit where both surfaces shrink to a point is the same.
Hence the 5d SCFT should also be the same.

However, depending on the triangulation, the geometry can also realize an $\mathfrak{su}(2) \times \mathfrak{su}(2)$ quiver theory.
Indeed, the phases with corresponding diagrams \ref{fig:rank_2_tri3} and \ref{fig:rank_2_tri4} admit a separate ruling for each surface $\{t_i\}$, indicated by the blue lines.
Note that diagram \ref{fig:rank_2_tri3} is precisely the phase we considered previously.
One can quickly convince oneself that the corresponding gauge theory includes, in addition to the bifundamental hypermultiplet, two flavors for each of the two $\mathfrak{su}(2)$ factors. 
That is, we have in both cases an $[\mathfrak{su}(2) + 2{\bf F} ] \times [\mathfrak{su}(2) + 2{\bf F} ]$ theory. 
In this case the geometric prepotential \eqref{eq:geomprep} computed by plugging in the intersection numbers \eqref{eq:matter_curves_rank2_example1} (with $S_i\equiv[t_i]$) matches directly with the field theory one, which for general $[\mathfrak{su}(2) +  N_{f_1}{\bf F} ] \times [\mathfrak{su}(2) + N_{f_2}{\bf F} ]$ theories reads
\begin{align}
	\begin{split}
		6\mathcal{F}&=3 m^{(1)}_0 (\phi^1)^2 + 3 m^{(2)}_0 (\phi^2)^2  +(8-N_{f_1}(\phi^1+m_{f_1})^3 + (8-N_{f_2})(\phi^2+m_{f_2})^3 \\
		&- (\phi^1 + \phi^2)^3 - (\phi^2 - \phi^1)^3.
	\end{split}
\end{align}
The apparent paradox with the quiver having an SCFT limit \cite{Intriligator:1997pq} is resolved by the obvious duality to the $\mathfrak{su}(3)$ phase, which is provided by the alternative ruling of both surfaces (red lines in figure \ref{fig:rank_2_flops}).
Meanwhile, in phases corresponding to triangulations \ref{fig:rank_2_tri1} and \ref{fig:rank_2_tri2}, only the surface $\{t_2\}$ admits a separate ruling.
Blowing down $\{t_2\}$ along this ruling would degenerate $\{t_1\}$ to a (singular) curve, thus introducing additional massless instantonic states which indicate the break-down of a potential $\mathfrak{su}(2)$ gauge theory interpretation. 
This is also confirmed by the prepotential analysis, and for this the hypermultiplets masses can be ignored.
In fact, similarly to \cite{Jefferson:2017ahm}, in the wedge $\phi_2>\phi_1>0$ (where for this analysis we set the hypermultiplets masses to zero) the tension $T_1 = \partial {\cal F} / \partial \phi^1$ of monopole strings changes sign at $\phi_1=2\phi_2$.
This signals the appearance of extra light degrees of freedom, and the breakdown of the quiver gauge theory, which is valid only in the subwedge $\phi_2>\phi_1>\phi_2/2>0$.
Hence, the gauge theory which is allowed by positivity of the metric of the Coulomb branch and string tensions in the entire chamber is $\mathfrak{su}(3) + 6 {\bf F}$. This gauge theory interpretation given by the ruling of these surfaces is always present in these triangulations. 
In particular, the triple intersection numbers of phase \ref{fig:rank_2_tri1},
\begin{align}
  [t_1]^3 = 8 \, , \qquad [t_1]^2 \cdot [t_2] = -4 \, , \qquad [t_1] \cdot [t_2]^2 =2 \, , \qquad [t_2]^3 = 2 \, ,
\end{align}
match perfectly with the prepotential, \eqref{eq:gensu3prep}, of $\mathfrak{su}(3) + 6{\bf F}$ with $a= \kappa = 0$.

Since all these geometries are related by flops, the corresponding 5d UV fixed points must be equivalent.
Concretely, we believe that the four triangulations of figure \ref{fig:rank_2_example_1} are physically equivalent to the four geometries appearing in the classification \cite{Jefferson:2018irk} which realize $\mathfrak{su}(3)$ with six flavors at CS-level 0.\footnote{Such a theory has seven mass deformations, see figure 16 of \cite{Jefferson:2017ahm}.}
More specifically, triangulation 1 in figure \ref{fig:rank_2_tri1} should be physically equivalent to the geometry $\text{Bl}_6 \bbF_4 \stackrel{F+E}{\cup} \bbF_0$.
This is supported by the fact that the blue polygon indeed corresponds to a six-fold blow-up (at non-generic points) of an $\bbF_4$ (spanned by the vectors $(\vec{y},\vec{t}_1,\vec{z}, \vec{\alpha}_8)$, cf.~figure \ref{fig:rank_2_tri1}), and $\bbF_0$ is believed to be physically equivalent to $\bbF_2$ which is the green polygon.
Continuing along the chains of flops, we can then identify the other three geometries with the triangulations presented here, with agreement on the existence of the $\mathfrak{su}(2)$ quiver phase.

Interestingly, this example seem to suggest that, in order to embed the rank two geometries presented in the classification \cite{Jefferson:2018irk} into a non-flat elliptic fibration, one would in general need to find physically equivalent surfaces that are not generic blow-ups of Hirzebruch or del Pezzo surfaces.
A better understanding of the possibilities here, and also for higher rank, would be essential for an attempt to classify all possible circle reductions of 6d SCFTs via M-/F-theory duality.

\subsection[5d Rank 4 Theories from 6d \texorpdfstring{$(E_8, SU(3))$}{(E8, SU3)} Conformal Matter]{5d Rank 4 Theories from 6d \boldmath{$(E_8, SU(3))$} Conformal Matter}

\begin{figure}[ht]
\centering
	\begin{tikzpicture}[scale=1.5]
		\begin{scope}[xshift=0.33\textwidth]
			\filldraw [ultra thick, draw=black, fill=blue!30!white] (3,-1)--(-3,-1)--(0,2)--cycle; 
			\foreach \x in {-3,-2,...,3}{
				\foreach \y in {-1,...,2}{
					\node[draw,circle,inner sep=1.3pt,fill] at (\x,\y) {};
				}
		    }
			\node [above] at (0,2)  {$\vec{x}$};
			\node [above left] at (-2,0) {$\vec{f}_2$};
			\node [above left] at (-1,1) {$\vec{f}_1$};
			\node [below left] at (-3,-1) {$\vec{z}$};
			\node [below] at (-2,-1) {$\vec{\alpha}_0$};
			\node [below] at (-1,-1) {$\vec{\alpha}_1$};
			\node [below] at (0,-1) {$\vec{\alpha}_2$};
			\node [below] at (1,-1) {$\vec{\alpha}_3$};
			\node [below] at (2,-1) {$\vec{\alpha}_4$};
			\node [below] at (3,-1) {$\vec{\alpha}_5$};
			\node [above right] at (2,0)  {$\vec{\alpha}_6$};
			\node [above right] at (1,1) {$\vec{\alpha}_7$};
			\node [below left] at (-1,0) {$\vec{u}_1$};
			\node [below ] at (0,0) {$\vec{u}_3$};
			\node [below right] at (1,0) {$\vec{u}_4$};
			\node [below] at (0,1) {$\vec{u}_2$};
			\node [above left] at (-2.5, -0.5) {${\cal E}_3$};
		\end{scope}
	\end{tikzpicture}
\caption{The facet of the $E_8$-top over the edge ${\cal E}_3$. The four vectors in the interior give rise to a rank four theory in 5d.
The divisors corresponding to $\vec{f}_i$ resolve the $SU(3)$ fiber singularities of the elliptic fibration over $\{s_8\}$.}\label{fig:rank_4_facet}
\end{figure}
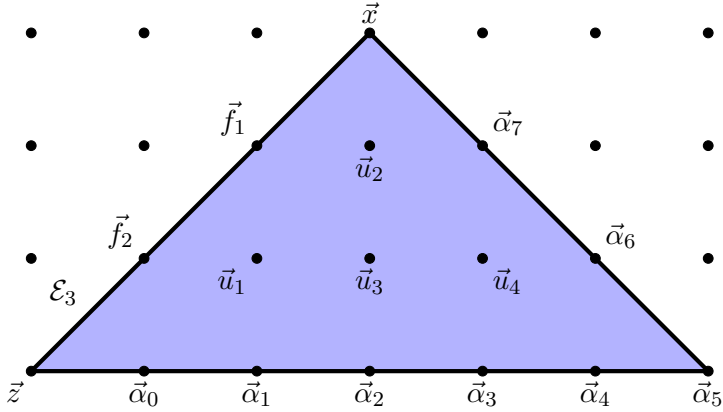

In this part, we will take a look at the rank four facet of the $E_8$ top.
This facet, sitting over the edge ${\cal E}_3$ of the base polygon, is depicted in figure \ref{fig:rank_4_facet}.
This facet describes the non-flat fiber over the collision locus $W \cap \{s_8\}$ of the $E_8$ and $SU(3)$ divisors.
In F-theory, such collision realizes an $(E_8 , SU(3))$ conformal matter system.

The fiber resolution of the singularities is described by the vanishing locus of the following polynomial:
\begin{align}\label{eq:rank_4_polynomial}
  \begin{split}
    \hat{P}_3 = \,& s_4\,\alpha_6\,\alpha_7^2\,f_1^2\,f_2\,u_1\,u_2^2\,u_3\,u_4\,x^3 + s_8\,\alpha_8\,y^2 \\
    + \,& b_1\,\alpha_0\,\alpha_1\,\alpha_2\,\alpha_3\,\alpha_4\,\alpha_5\,\alpha_6\,\alpha_7\,\alpha_8\,f_1\,f_2\,u_1\,u_2\,u_3\,u_4\,x\,y\,z \\
    + \,& b_2\,\alpha_0^2\,\alpha_1^2\,\alpha_2^2\,\alpha_3^2\,\alpha_4^2\,\alpha_5^2\,\alpha_6^2\,\alpha_7^2\,\alpha_8\,f_1^2\,f_2^2\,u_1^2\,u_2^2\,u_3^2\,u_4^2\,x^2\,z^2 \\ 
    + \,& b_3\,\alpha_0^3\,\alpha_1^3\,\alpha_2^3\,\alpha_3^3\,\alpha_4^3\,\alpha_5^3\,\alpha_6^2\,\alpha_7\,\alpha_8^2\,f_1\,f_2^2\,u_1^2\,u_2\,u_3^2\,u_4^2\,y\,z^3 \\
    + \,& b_4\,\alpha_0^4\,\alpha_1^4\,\alpha_2^4\,\alpha_3^4\,\alpha_4^4\,\alpha_5^4\,\alpha_6^3\,\alpha_7^2\,\alpha_8^2\,f_1^2\,f_2^3\,u_1^3\,u_2^2\,u_3^3\,u_4^3\,x\,z^4 \\
    + \,& b_6\,\alpha_0^5\,\alpha_1^4\,\alpha_2^3\,\alpha_3^2\,\alpha_4\,f_1^2\,f_2^4\,u_1^3\,u_2\,u_3^2\,u_4\,z^6 \, .
  \end{split}
\end{align}
We observe the appearance of the $SU(3)$ divisors $\{f_i\}$ over $\{s_8\}$, as well as the four non-flat fiber components $S_j \equiv \{u_j\}$ at $W \cap \{s_8\}$ from the factorization $\hat{P}_3|_{s_8=0} = f_1\,f_2\,u_1\,u_2\,u_3\,u_4 \, (...)$.
The precise fiber structure depends on the triangulation of the facet.
Again, there is an enormous number of possible triangulations ($\sim 14000$), which we do not attempt to classify here.
However, there are some universal properties which can be inferred from the toric realization right away.

First, as we already noted in the rank two case, the toric setting does not admit gauge groups of non-simply laced ADE type, because the gluing curve dictated by triangulations is always a $\bbP^1$ in every surface.
Second, for the facet in figure \ref{fig:rank_4_facet}, we can only have quiver gauge theories at weak coupling, but no single gauge factor theories, i.e., no $\mathfrak{su}(5)$ or $\mathfrak{so}(8)$.
For the first, one would need the four vectors $\vec{u}_i$ to be aligned along a straight line, such that each surface intersects its next two neighbors (one for the surfaces at the end) along sections of the same ruling \cite{Intriligator:1997pq}; this is clearly not possible with the position of $\vec{u}_2$.
One the other hand, $\mathfrak{so}(8)$ requires one surface $S$ which intersects the other three along sections of the same ruling.
However, at most two of them can be toric curves inside $S$, because any line subdividing the cones of $F_S$ (the toric diagram of $S$) passes only through two vectors on the boundary of $F_S$.

Again, we expect these restrictions to be lifted in non-toric resolutions of the non-flat fiber.
Some of these phases should also be only one flop away from a toric resolution.
For now, we will content ourselves with the torically available phases and leave detailed studies of such flops for the future \cite{work_in_progress_resolution}.

\subsubsection{Rank 4 quiver theories}\label{sec:rank_4_example}

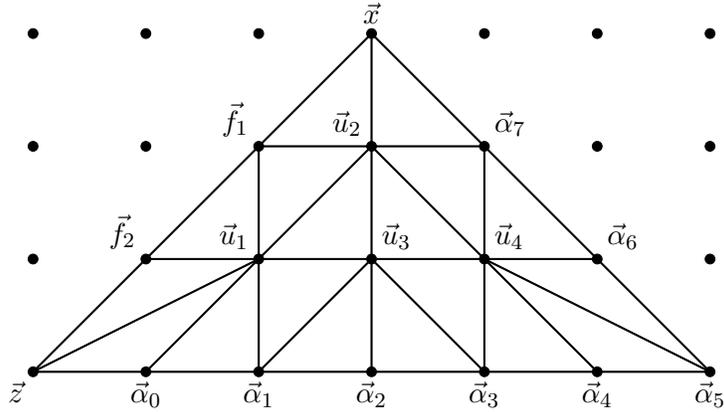
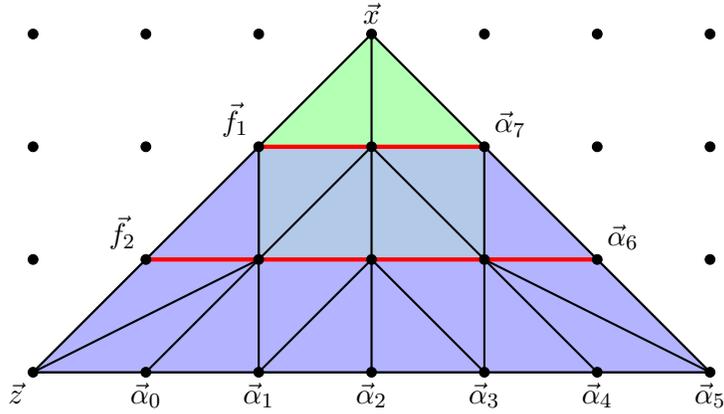
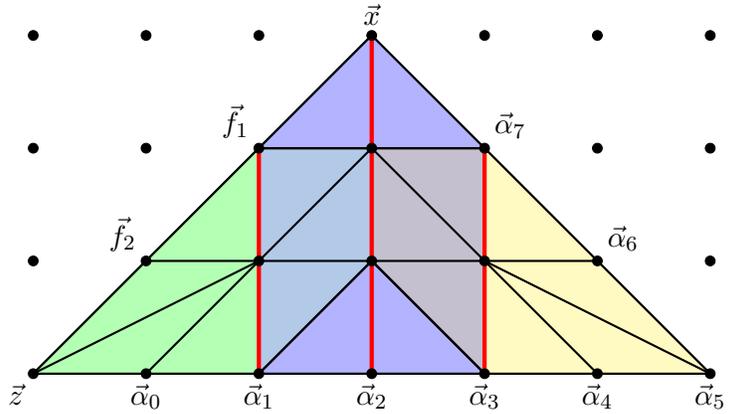
\begin{figure}[p]
  \centering
  \subfigure[Triangulation of the rank four facet.]{\label{fig:rank_4_triangulation}
    	\begin{tikzpicture}[scale=1.5]
		\begin{scope}[xshift=0.33\textwidth]
			\filldraw [ thick, draw=black, 
				fill = white] (3,-1)--(-3,-1)--(0,2)--cycle; 
			\filldraw [thick, draw=black] (0,1) -- (0,2);
			\filldraw [thick, draw=black] (0,1) -- (1,1);
			\filldraw [thick, draw=black] (0,1) -- (-1,1);
			\filldraw [thick, draw=black] (0,1) -- (0,0);
			\filldraw [thick, draw=black] (0,1) -- (-1,0);
			\filldraw [thick, draw=black] (0,1) -- (1,0);
			\filldraw [thick, draw=black] (-1,0) -- (0,0);
			\filldraw [thick, draw=black] (-1,0) -- (-1,1);
			\filldraw [thick, draw=black] (-1,0) -- (-2,0);
			\filldraw [thick, draw=black] (-1,0) -- (-3,-1);
			\filldraw [thick, draw=black] (-1,0) -- (-2,-1);
			\filldraw [thick, draw=black] (-1,0) -- (-1,-1);
			\filldraw [thick, draw=black] (0,0) -- (-1,-1);
			\filldraw [thick, draw=black] (0,0) -- (0,-1);
			\filldraw [thick, draw=black] (0,0) -- (1,-1);
			\filldraw [thick, draw=black] (0,0) -- (1,-0);
			\filldraw [thick, draw=black] (1,0) -- (1,1);
			\filldraw [thick, draw=black] (1,0) -- (1,-1);
			\filldraw [thick, draw=black] (1,0) -- (2,-1);
			\filldraw [thick, draw=black] (1,0) -- (3,-1);
			\filldraw [thick, draw=black] (1,0) -- (2,0);
			\foreach \x in {-3,-2,...,3}{
				\foreach \y in {-1,...,2}{
					\node[draw,circle,inner sep=1.3pt,fill] at (\x,\y) {};
				}
		    }
			\node [above] at (0,2)  {$\vec{x}$};
			\node [above left] at (-2,0) {$\vec{f}_2$};
			\node [above left] at (-1,1) {$\vec{f}_1$};
			\node [below left] at (-3,-1) {$\vec{z}$};
			\node [below] at (-2,-1) {$\vec{\alpha}_0$};
			\node [below] at (-1,-1) {$\vec{\alpha}_1$};
			\node [below] at (0,-1) {$\vec{\alpha}_2$};
			\node [below] at (1,-1) {$\vec{\alpha}_3$};
			\node [below] at (2,-1) {$\vec{\alpha}_4$};
			\node [below] at (3,-1) {$\vec{\alpha}_5$};
			\node [above right] at (2,0)  {$\vec{\alpha}_6$};
			\node [above right] at (1,1) {$\vec{\alpha}_7$};
			\node [above left] at (-1,0) {$\vec{u}_1$};
			\node [above right ] at (0,0) {$\vec{u}_3$};
			\node [above right] at (1,0) {$\vec{u}_4$};
			\node [above left] at (0,1) {$\vec{u}_2$};
		\end{scope}
	\end{tikzpicture}
  }
  %
  %
  \subfigure[One choice of ruling which gives rise to an $\mathfrak{su}(2) \times \mathfrak{su}(4)$ quiver.]{\label{fig:rank_4_phase_1}
    	\begin{tikzpicture}[scale=1.5]
		\begin{scope}[xshift=0.33\textwidth]
			\filldraw [ thick, draw=black, 
				fill = white] (3,-1)--(-3,-1)--(0,2)--cycle; 
			\filldraw [ thick, draw=black, fill=blue!30!white]
			(-3,-1)--(3,-1)--(1,1)--(1,0)--(-1,0)--(-1,1)--cycle;
			\filldraw [ thick, draw=black, fill=green!30!white]
			(1,1)--(0,2)--(-1,1)--cycle;
			\filldraw [ thick, draw=black, fill=green!30!blue!30!white]
			(-1,0)--(1,0)--(1,1)--(-1,1)--cycle;
			\filldraw [thick, draw=black] (0,1) -- (0,2);
			\filldraw [thick, draw=black] (0,1) -- (1,1);
			\filldraw [thick, draw=black] (0,1) -- (-1,1);
			\filldraw [thick, draw=black] (0,1) -- (0,0);
			\filldraw [thick, draw=black] (0,1) -- (-1,0);
			\filldraw [thick, draw=black] (0,1) -- (1,0);
			\filldraw [thick, draw=black] (-1,0) -- (0,0);
			\filldraw [thick, draw=black] (-1,0) -- (-1,1);
			\filldraw [thick, draw=black] (-1,0) -- (-2,0);
			\filldraw [thick, draw=black] (-1,0) -- (-3,-1);
			\filldraw [thick, draw=black] (-1,0) -- (-2,-1);
			\filldraw [thick, draw=black] (-1,0) -- (-1,-1);
			\filldraw [thick, draw=black] (0,0) -- (-1,-1);
			\filldraw [thick, draw=black] (0,0) -- (0,-1);
			\filldraw [thick, draw=black] (0,0) -- (1,-1);
			\filldraw [thick, draw=black] (0,0) -- (1,-0);
			\filldraw [thick, draw=black] (1,0) -- (1,1);
			\filldraw [thick, draw=black] (1,0) -- (1,-1);
			\filldraw [thick, draw=black] (1,0) -- (2,-1);
			\filldraw [thick, draw=black] (1,0) -- (3,-1);
			\filldraw [thick, draw=black] (1,0) -- (2,0);
			\filldraw [ultra thick, draw=red] (-2,0)--(2,0);
			\filldraw [ultra thick, draw=red] (-1,1)--(1,1);
			\foreach \x in {-3,-2,...,3}{
				\foreach \y in {-1,...,2}{
					\node[draw,circle,inner sep=1.3pt,fill] at (\x,\y) {};
				}
		    }
			\node [above] at (0,2)  {$\vec{x}$};
			\node [above left] at (-2,0) {$\vec{f}_2$};
			\node [above left] at (-1,1) {$\vec{f}_1$};
			\node [below left] at (-3,-1) {$\vec{z}$};
			\node [below] at (-2,-1) {$\vec{\alpha}_0$};
			\node [below] at (-1,-1) {$\vec{\alpha}_1$};
			\node [below] at (0,-1) {$\vec{\alpha}_2$};
			\node [below] at (1,-1) {$\vec{\alpha}_3$};
			\node [below] at (2,-1) {$\vec{\alpha}_4$};
			\node [below] at (3,-1) {$\vec{\alpha}_5$};
			\node [above right] at (2,0)  {$\vec{\alpha}_6$};
			\node [above right] at (1,1) {$\vec{\alpha}_7$};
		\end{scope}
	\end{tikzpicture}
  }
  %
  %
    \subfigure[A different ruling giving rise to an $\mathfrak{su}(2) \times \mathfrak{su}(3) \times \mathfrak{su}(2)$ quiver.]{\label{fig:rank_4_phase_2}
    	\begin{tikzpicture}[scale=1.5]
		\begin{scope}[xshift=0.33\textwidth]
			\filldraw [ thick, draw=black, 
				fill = white] (3,-1)--(-3,-1)--(0,2)--cycle; 
			\filldraw [thick, draw=black, fill=green!30!white]
			(-3,-1)--(-1,-1)--(-1,1)--cycle;
			\filldraw [thick, draw=black, fill=green!30!blue!30!white]
			(-1,-1)--(0,0)--(0,1)--(-1,1)--cycle;
			\filldraw [thick, draw=black, fill=yellow!30!white]
			(1,-1)--(3,-1)--(1,1)--cycle;
			\filldraw [thick, draw=black, fill=yellow!30!blue!30!white]
			(0,0)--(1,-1)--(1,1)--(0,1)--cycle;
			\filldraw [thick, draw=black, fill=blue!30!white]
			(-1,1)--(1,1)--(0,2)--cycle;
			\filldraw [thick, draw=black, fill=blue!30!white]
			(-1,-1)--(1,-1)--(0,0)--cycle;
			\filldraw [thick, draw=black] (0,1) -- (0,2);
			\filldraw [thick, draw=black] (0,1) -- (1,1);
			\filldraw [thick, draw=black] (0,1) -- (-1,1);
			\filldraw [thick, draw=black] (0,1) -- (0,0);
			\filldraw [thick, draw=black] (0,1) -- (-1,0);
			\filldraw [thick, draw=black] (0,1) -- (1,0);
			\filldraw [thick, draw=black] (-1,0) -- (0,0);
			\filldraw [thick, draw=black] (-1,0) -- (-1,1);
			\filldraw [thick, draw=black] (-1,0) -- (-2,0);
			\filldraw [thick, draw=black] (-1,0) -- (-3,-1);
			\filldraw [thick, draw=black] (-1,0) -- (-2,-1);
			\filldraw [thick, draw=black] (-1,0) -- (-1,-1);
			\filldraw [thick, draw=black] (0,0) -- (-1,-1);
			\filldraw [thick, draw=black] (0,0) -- (0,-1);
			\filldraw [thick, draw=black] (0,0) -- (1,-1);
			\filldraw [thick, draw=black] (0,0) -- (1,-0);
			\filldraw [thick, draw=black] (1,0) -- (1,1);
			\filldraw [thick, draw=black] (1,0) -- (1,-1);
			\filldraw [thick, draw=black] (1,0) -- (2,-1);
			\filldraw [thick, draw=black] (1,0) -- (3,-1);
			\filldraw [thick, draw=black] (1,0) -- (2,0);
			\filldraw [ultra thick, draw=red] (-1,-1) -- (-1,1);
			\filldraw [ultra thick, draw=red] (1,-1) -- (1,1);
			\filldraw [ultra thick, draw=red] (0,-1) -- (0,2);
			\foreach \x in {-3,-2,...,3}{
				\foreach \y in {-1,...,2}{
					\node[draw,circle,inner sep=1.3pt,fill] at (\x,\y) {};
				}
		    }
			\node [above] at (0,2)  {$\vec{x}$};
			\node [above left] at (-2,0) {$\vec{f}_2$};
			\node [above left] at (-1,1) {$\vec{f}_1$};
			\node [below left] at (-3,-1) {$\vec{z}$};
			\node [below] at (-2,-1) {$\vec{\alpha}_0$};
			\node [below] at (-1,-1) {$\vec{\alpha}_1$};
			\node [below] at (0,-1) {$\vec{\alpha}_2$};
			\node [below] at (1,-1) {$\vec{\alpha}_3$};
			\node [below] at (2,-1) {$\vec{\alpha}_4$};
			\node [below] at (3,-1) {$\vec{\alpha}_5$};
			\node [above right] at (2,0)  {$\vec{\alpha}_6$};
			\node [above right] at (1,1) {$\vec{\alpha}_7$};
		\end{scope}
	\end{tikzpicture}
  }
  \caption{In this triangulation, the surface $\{u_1\}$ has polygon bounded by $\vec{f}_1-\vec{u}_2-\vec{u}_3-\vec{\alpha}_1-\vec{\alpha}_0-\vec{z}-\vec{f}_2$, that of $\{u_2\}$ by $\vec{x}-\vec{\alpha}_7-\vec{u}_4-\vec{u}_3-\vec{u}_1-\vec{f}_1$, that of $\{u_3\}$ by $\vec{u}_2-\vec{u}_4-\vec{\alpha}_3-\vec{\alpha}_3-\vec{\alpha}_1-\vec{u}_1$, and that of $\{u_4\}$ by $\vec{\alpha}_7-\vec{\alpha}_6-\vec{\alpha}_5-\vec{\alpha}_4-\vec{\alpha}_3-\vec{u}_3-\vec{u}_2$.
  Depending on the rulings, along which the fibers are shrunk, one obtains a different gauge theory at weak coupling.}\label{fig:rank_4_interpretations}
\end{figure}

\noindent The resolution phase we will consider in the following is given by the triangulation depicted in figure \ref{fig:rank_4_triangulation}.
This triangulation allows for two dual descriptions at weak coupling, which can be reached by blowing down the surfaces along different rulings.

The first choice, corresponding to the rulings marked by red lines in figure \ref{fig:rank_4_phase_1}, realizes an $\mathfrak{su}(2)$ gauge sector via the green polygon and an $\mathfrak{su}(4)$ sector via the blue polygon.
The green polygon by itself has the geometry of $\text{Bl}_2 \bbF_2$, with $\{x\} \cap \{u_2\}$ being the generic fiber, and thus supports two flavors.
However, both flavors are actually charged under the $\mathfrak{su}(4)$ (they are realized on the curves $\{u_2\} \cap \{u_1\}$ and $\{u_2\} \cap \{u_4\}$), and are completed by additional curves inside the blue polygon into a full bifundamental hypermultiplet.
There are additional curves, represented by the black lines below the lower red line in figure \ref{fig:rank_4_phase_1}, which give rise to additional matter charged only under the $\mathfrak{su}(4)$.
With some combinatorics, we find six fundamental hypermultiplets.
Thus, the weakly coupled theory obtained via this ruling is $\mathfrak{su}(2) \times [\mathfrak{su}(4) + 6{\bf F}]$.

By computing the rank of the intersection matrix between all non-compact divisors with vectors on the boundary and curves inside the surfaces,\footnote{Recall that these correspond to edges of the triangulation in the facet interior.} we find a rank 9 flavor symmetry group $G_f$.
Using the same method as in the lower rank examples, we can easily verify that by blowing down along the ruling indicated by the red lines in figure \ref{fig:rank_4_phase_1}, the codimension one fibers of $A_i$ with $i=0,...,4$ shrink.
Thus, $G_f = SU(6) \times U(1)^4$ at weak coupling.
The six fundamental flavors of the $\mathfrak{su}(4)$ transform under the $SU(6) \times U(1)_{B_1} \cong U(6)$ part of $G_f$, where $U(1)_{B_1}$ is the baryonic symmetry of the $\mathfrak{su}(4)$.
The remaining three abelian factors are $U(1)_{B_2} \times U(1)_{T_1} \times U(1)_{T_2}$, where $U(1)_{B_2}$ is the baryonic $U(1)$ of $\mathfrak{su}(2)$, and the $U(1)_{T_i}$ are the topological instanton $U(1)$s of each factor.

The second choice of ruling is shown in figure \ref{fig:rank_4_phase_2}.
Here, we blow down along the vertical direction of the toric diagram.
The green and yellow polygons are each dP$_3$ blown up at one non-generic points, hence each by itself would realize an $\mathfrak{su}(2)$ with three flavors.
Two flavors of each $\mathfrak{su}(2)$ are realized on the curves around the edges ($\{z\}, \{\alpha_2\}, \{f_2\}$ in green and $\{\alpha_4\},\{\alpha_5\}, \{\alpha_6\}$ in yellow polygon), similarly to the situation of the $N_f=2$ phase in the rank one example, see section \ref{sec:rank_1_Nf=2} and figure \ref{fig:rank_1_dP3}.
Clearly, these curves do not intersect any of the other non-flat surfaces.
The remaining flavor of each $\mathfrak{su}(2)$ factor is charged under the $\mathfrak{su}(3)$ realized on the blue polygon and are completed into a bifundamental hypermultiplet.
Thus, the weakly coupled theory is $[\mathfrak{su}(2)+ 2{\bf F}] \times \mathfrak{su}(3) \times [\mathfrak{su}(2)+ 2{\bf F}]$.
This is consistent with the flavor symmetry, which is again of rank 9.
Here, the non-abelian part is $SU(2)^4 \cong SO(4)_L \times SO(4)_R$, coming from the shrinking of the fibers of $\{f_2\}$ and $A_0$ on the one side and $A_4$ and $A_6$ and the other side.
The rank 5 abelian part is composed out of three topological and two baryonic $U(1)$s.\footnote{Naively, one would expect three baryonic $U(1)$s for each gauge factor. However, because the two $\mathfrak{su}(2)$ sectors are coupled to the same $\mathfrak{su}(3)$, one linear combination actually decouples.}

By construction, these two seemingly different gauge theories must have the same SCFT limit, which is reached by blowing down the surfaces to a point.
We can read off the flavor symmetry enhancement at this limit to be $SU(6) \times SU(3) \times SU(3)'$. 
The $SU(6) \times SU(3)$ is a subgroup of the $\hat{E}_8$ (the affinized version of the $E_8$ over $W$) coming from shrinking the fibers of $A_{0,...,4}$ and $A_{6,7}$, while the $SU(3)'$ factor is the flavor symmetry group over $\{s_8\}$ with fibers of $\{f_{1,2}\}$ shrunk.

Finally, we note that the enhanced flavor symmetry matches the field theory analysis in \cite{Yonekura:2015ksa, Tachikawa:2015mha, Zafrir:2014ywa}, so based on this we claim that the Chern--Simons levels and $\mathfrak{su}(2)$ theta angle are $[\mathfrak{su}(2)_{\theta=0}]\times [\mathfrak{su}(4)_{\kappa=0}+ 6 \mathbf F]$, and $[\mathfrak{su}(2)_{\theta=0} + 2 \mathbf F] \times [\mathfrak{su}(3)_{\kappa=-1}] \times [\mathfrak{su}(2)_{\theta=0} + 2 \mathbf F]$. 
We can also confirm these Chern--Simons levels geometrically from the intersection numbers.
The explicit computation is detailed in appendix \ref{app:prepot}.

\section{Summary and Outlook}\label{sec:summary}

In this paper we have motivated a systematic study of elliptic Calabi--Yau threefolds $\hat{Y}$ with non-flat fibers.
These geometries connect 6d conformal matter theories defined via F-theory on the singular limit $Y$ with their circle reduced 5d SCFTs from M-theory on degenerations of $\hat{Y}$.
Compared to previous approaches \cite{DelZotto:2017pti,Jefferson:2017ahm,Jefferson:2018irk,Bhardwaj:2018yhy}, the non-flat geometries have both the 6d and the 5d SCFT manifest.

To see the difference, recall that in those references, the non-minimal singularities of $Y$ are resolved by first blowing up the base, which pushes the 6d theory onto the tensor branch.
A subsequent circle reduction is a 5d KK-theory, which does not have an honest strong coupling limit in 5d.
To obtain 5d SCFTs, one has to further mass deform, which is equivalent to turning on holonomies in the circle reduction.
Our approach is based on the intuition that in the M-/F-theory duality, resolving the fiber corresponds precisely to such non-trivial holonomies.

The proposal is confirmed by the observation that any crepant fiber resolution of non-minimal elliptic singularities introduces surface components in the fiber, hence being non-flat.
Physically, these surfaces are precisely the compact divisors of a (non-compact) Calabi--Yau threefold which support 5d gauge theories \cite{Intriligator:1997pq}.
By construction, they can be collapsed to a point without shrinking the generic elliptic fiber, hence they have a well-defined SCFT limit in 5d.

The non-flat fibers have another feature, namely they allow us to identify the global flavor symmetry of the 5d system, both at weak and at strong coupling, as a subgroup of the flavor symmetry of the 6d conformal matter theory times the KK-$U(1)$.
Physically, this is clearly expected if the 5d theory ought to arise from a reduction of a 6d SCFT.
What makes this identification possible is a clean way to track how the codimension one fibers of the resolved elliptic fibration $\hat{Y}$ split and become parts of the non-flat fiber components.
Blowing down these components then leads to curves of ADE singularities in $\hat{Y}$, which are the Kodaira singularities of $\hat{Y}$ over codimension one loci.
While it has been known that such singularities are the source of 5d flavor symmetries \cite{DelZotto:2017pti,Xie:2017pfl}, the non-flat fibration provides an explicit embedding of these into the 6d theory.\footnote{More precisely, this holds for 5d theories arising from 6d conformal matter theories.}

We have tested our proposal using 5d theories of rank 1, 2 and 4, which arise from the 6d E-string theories of rank 1, 2 and the $(E_8, SU(3))$ conformal matter. 
For simplicity, we have restricted ourselves to elliptic fibrations constructed via so-called tops which provide resolutions in terms of toric geometry.
The advantage of these constructions is that they incorporate many different resolution phases, and thus different 5d theories, in terms of combinatorial data.
For these, we showed that the ``usual'' fiber analysis in spirit of \cite{Lawrie:2012gg, Hayashi:2014kca, Braun:2014kla,Braun:2015hkv} matches previous toric constructions of 5d SCFTs \cite{Xie:2017pfl}.

However, we also saw clear restrictions of such toric models.
For example, it is impossible to obtain non-simply laced gauge algebras, or non-$SU(N)$ types of flavor symmetries.
On the other hand, we also know \cite{DelZotto:2017pti,Jefferson:2017ahm,Jefferson:2018irk} that such geometric phases must exist and be related to toric resolutions via flops.
We believe that an explicit construction and analysis of these phases will provide further evidence for the efficacy of our proposal.

Of course, the ideal scenario would be to have a classification of all non-flat fiber resolutions of non-minimal singularities, similar to the Kodaira classification of minimal ones.
Such a classification could provide a systematic approach, orthogonal to that of \cite{Bhardwaj:2018yhy}, to find all circle reductions of 6d SCFTs, which hopefully brings us closer to a classification of all 5d SCFTs.
Note that there is also some evidence that fiber resolutions sometimes goes hand in hand with twisting by discrete automorphisms (see \cite{Apruzzi:2017iqe} for the classification of 6D automorphisms compatible with the string lattice), which would reduce the rank of 5d theory \cite{Braun:2014oya,Baume:2017hxm}.
A precise understanding of such phenomena \cite{bisection_to_appear} could vastly increase the predictive power of our proposal.

Mathematically, our proposal might also open up a new perspective on the classification of canonical Calabi--Yau threefold singularities.
Indeed, combining the two conjectures---1), all 5d SCFTs arise from circle reductions \cite{Jefferson:2018irk}, and 2), any canonical singularity gives rise to a 5d SCFT \cite{Xie:2017pfl}---seems to suggest the classification \cite{Heckman:2015bfa,Bhardwaj:2015xxa} of 6d SCFTs via elliptic fibrations also encodes one for canonical threefold singularities.
If true, then our analysis suggests that non-flat fibers will play a central role to decipher this code.

Additionally, it is worth pointing out that the phenomenon of non-flat fibers has haunted the F-theory model building community for some time.
However, in these constructions one encounters them as surface components in codimension three fibers of elliptic Calabi--Yau fourfolds \cite{Candelas:2000nc,Denef:2005mm,Braun:2011ux,Mayrhofer:2012zy,Lawrie:2012gg,Braun:2013nqa,Cvetic:2013uta,Borchmann:2013hta,Lin:2014qga}.\footnote{Similar degenerations also occur in codimension four on Calabi--Yau fivefolds, which define 2d $(0,2)$ theories \cite{Schafer-Nameki:2016cfr,Apruzzi:2016iac}.}
Hence, though geometrically they are similar to what we have studied in this work, the physics is qualitatively different as surfaces on a fourfold do not give rise to any gauge sectors in M-theory.
Though there had been some recent progress in this direction \cite{Achmed-Zade:2018idx}, a more thorough investigation is needed to fully understand the physical ramifications in 4d/3d.

As a last comment, we highlight that our construction can be easily embedded inside a compact Calabi--Yau threefold.
In the top construction, this simply amounts to include a ``bottom'', see section \ref{sec:non_flat_resolution_toric}.
In some limits of the moduli space of M-theory on this compact Calabi--Yau, a complicated quiver description of the effective theory is likely to exist and probably easy to determine from our constructions.
In the putative quiver all flavor symmetries determined in the non-compact setting will be gauged.
Computing BPS invariants on compact Calabi--Yau threefolds is in general a difficult task, and the computation of these invariants involves evaluating at all loops the topological string partition function \cite{Gopakumar:1998ii, Gopakumar:1998jq}.
A description of the compact threefold in terms of a quiver with 5d gauge theories/SCFTs building blocks \cite{Hayashi:2019fsa} might, in combination with the correspondences between the partition function (or index) of gauge/superconformal theories and topological string partition function \cite{Nekrasov:2003rj, Aganagic:2004js, Lockhart:2012vp, Hayashi:2017jze}, be important for the computation of these invariants.

\subsection*{Acknowledgments}

We thank C.~Closset, M.~del Zotto, S.~Giacomelli, M.~Fazzi, J.~Heckman, P.~Jefferson, C.~Lawrie, P.~Oehlmann, S.~Sch\"afer-Nameki, C.~Uhlemann, L.~Tizzano and G.~Zoccarato for stimulating communications. 
L.L.~and C.M.~are particularly grateful to E.~Palti and T.~Weigand for valuable discussions and collaboration at early stages of the project.
The work of F.A.~is supported by the ERC Consolidator Grant 682608 ``Higgs bundles: Supersymmetric Gauge Theories and Geometry'' (HIGGSBNDL). F.A.~thanks the 2018 Summer Workshop at the Simons Center for Geometry and Physics and the Aspen Center for Physics (2018 summer program: Superconformal Field Theories and Geometry), supported by National Science Foundation grant PHY-1607611, for hospitality during part of this work. 
The work of L.L.~was supported by DOE Award DE-SC0013528.

\newpage

\appendix

\section[Weierstrass Model of the \texorpdfstring{\boldmath{$E_8$}}{E8} top on \texorpdfstring{\boldmath{$F_{10}$}}{F10}]{Weierstrass Model of the \boldmath{$E_8$} top on \boldmath{$F_{10}$}}\label{app:weierstrass_of_top}

For completeness, we present the Weierstrass model of the elliptic fibration $\hat{Y}$ which we constructed through the $E_8$ top on the polygon $F_{10}$.
This is obtained first by mapping the generic fibration in $F_{10}$, without the specialization from the top, into Weierstrass form.
Recall that the hypersurface polynomial in this case is
\begin{align}
  P_{F_{10}} \equiv s_8\,y^2 + s_4\,x^3 + b_1 \, x \, y \,z + b_2 \, x^2\,z^2 + b_3 \, y \,z^3 + b_4 \, x\,z^4 + b_6\,z^6 \, ,
\end{align}
 where we have omitted the blow-up coordinates $j$ and $f_i$ compared to \eqref{eq:hypersurface_F10}.
The careful reader might notice the resemblance with the ordinary Tate-form of an elliptic fibration, where $b_i$ (often denoted $a_i$) are sections of powers $K_B^{\otimes (-i)}$ of the canonical bundle $K_B$ of the base.

Including non-trivial sections $s_{4,8}$ as coefficients of the $y^2$ and $x^3$ term introduces $I_2$ and $I_3$ singularities in codimension one.
The corresponding Weierstrass model can be found in \cite{Klevers:2014bqa} with a modified notation\footnote{The sections $b_i$ here appear as $s_j$ in the reference. Explicitly: $b_1 \leftrightarrow s_6, b_2 \leftrightarrow s_3, b_3 \leftrightarrow s_5, b_4 \leftrightarrow s_2, b_6 \leftrightarrow s_1$.}
Furthermore, when we include the top, then at the level of the Weierstrass functions, it factorizes the coefficients $b_i \rightarrow b_i \, w^{n_i}$ such that the elliptic fibration further degenerates over $W = \{w\}$.
For the $E_8$ top, the factorization corresponds to the well-known result of Tate's algorithm to engineer an $E_8$ singularity, namely $b_1 \rightarrow b_1 w , b_2 \rightarrow b_2 w^2, b_3 \rightarrow b_3 w^3, b_4 \rightarrow b_4 w^4 , b_6 \rightarrow b_6 w^5$.
All in all, the Weierstrass functions of the elliptic fibration \eqref{eq:resolved_hypersurface} are
\begin{align}\label{eq:weierstrass_E8_top}
  \begin{split}
    f & \,= w^4 \left(\frac{b_1^2\,b_2\,s_8}{6} -\frac{b_1^4}{48}  -\frac{b_1\,b_3\,s_4\,s_8}{2} - \frac{b_2^2\,s_8^2}{3} + b_4\,s_4\,s_8^2 \right) \, , \\
    g & \,= w^5 \left( b_6\,s_4^2\,s_8^3 - \frac{b_1^6\,w}{864} + \frac{b_1^4\,b_2\,s_8\,w}{72} - \frac{b_1^3\,b_3\,s_4\,s_8\,w}{24} - \frac{b_1^2\,b_2^2\,s_8^2\,w}{18} \right. \\
       & \qquad \qquad \left. + \frac{b_1\,b_2\,b_3\,s_4\,s_8^2\,w}{6} + \frac{b_1^2\,b_4\,s_4\,s_8^2\,w}{12} - \frac{b_3^2\,s_4^2\,s_8^2\,w}{4} + \frac{2\,b_2^3\,s_8^3\,w}{27} - \frac{b_2\,b_4\,s_4\,s_8^3\,w}{3} \right) , \\
    \delta & \, = s_4^2\,s_8^3\,w^{10} \left( 
    27\,b_6^2\,s_4^2\,s_8^3 - \frac{b_1^6\,b_6\,w}{16} + \frac{3\,b_1^4\,b_2\,b_6\,s_8\,w}{4} - \frac{9\,b_1^3\,b_3\,b_6\,s_4\,s_8\,w}{4} \right. \\
    & \qquad \qquad - 3\,b_1^2\,b_2^2\,b_6\,s_8^2\,w + 9\,b_1\,b_2\,b_3\,b_6\,s_4\,s_8^2\,w + \frac{9\,b_1^2\,b_4\,b_6\,s_4\,s_8^2\,w}{2} - \frac{27\,b_3^2\,b_6\,s_4^2\,s_8^2\,w}{2} \\
    & \qquad \qquad + 4\,b_2^3\,b_6\,s_8^3\,w - 18\,b_2\,b_4\,b_6\,s_4\,s_8^3\,w - \frac{b_1^4\,b_2\,b_3^2\,w^2}{16} + \frac{b_1^5\,b_3\,b_4\,w^2}{16} + \frac{b_1^3\,b_3^3\,s_4\,w^2}{16} \\
    &\qquad \qquad + \frac{b_1^2\,b_2^2\,b_3^2\,s_8\,w^2}{2} - \frac{b_1^3\,b_2\,b_3\,b_4\,s_8\,w^2}{2} - \frac{b_1^4\,b_4^2\,s_8\,w^2}{16} - \frac{9\,b_1\,b_2\,b_3^3\,s_4\,s_8\,w^2}{4} \\
    &\qquad \qquad + \frac{15\,b_1^2\,b_3^2\,b_4\,s_4\,s_8\,w^2}{8} + \frac{27\,b_3^4\,s_4^2\,s_8\,w^2}{16} - b_2^3\,b_3^2\,s_8^2\,w^2 + b_1\,b_2^2\,b_3\,b_4\,s_8^2\,w^2 \\
    &\left. \qquad \qquad + \frac{b_1^2\,b_2\,b_4^2\,s_8^2\,w^2}{2} + \frac{9\,b_2\,b_3^2\,b_4\,s_4\,s_8^2\,w^2}{2} - 6\,b_1\,b_3\,b_4^2\,s_4\,s_8^2\,w^2 - b_2^2\,b_4^2\,s_8^3\,w^2 + 4\,b_4^3\,s_4\,s_8^3\,w^2 
    \right) \, .
  \end{split}
\end{align}
With these, one can easily verify the codimension one singularities to be $I_2$ over $\{s_4\}$, $I_3$ over $\{s_8\}$, and $E_8$ over $\{w\} = W$.

\section{Geometric Computation of Chern--Simons Levels for Rank Four Example}\label{app:prepot}

In this appendix, we present the details of the prepotential computation for the rank four example discussed in section \ref{sec:rank_4_example}.
The intersection numbers for this geometry can be read off from the toric diagram \ref{fig:rank_4_triangulation}:
\begin{align} \label{eq:intnumbrank4}
  \begin{split}
    & [u_1]^3 = [u_4]^3 = 5 \, , \quad [u_2]^3 = [u_3]^3 = 6 \, , \quad [u_1]\cdot[u_2]\cdot[u_3] = [u_2] \cdot [u_3]\cdot [u_4] = 1 \, , \\
    & [u_1]^2 \cdot [u_2] = [u_1]^2 \cdot[u_3] = [u_2]^2\cdot [u_1] = [u_2]^2 \cdot [u_4] = -1 \, ,\\
    & [u_3]^2 \cdot [u_1] = [u_3]^2 \cdot [u_4] = [u_4]^2 \cdot [u_2] = [u_4]^2 \cdot [u_3] = -1 \, , \quad [u_3]^2\cdot [u_2] = -2 \, , \\
    & [u_1]^2 \cdot [u_4] = [u_2]^2 \cdot [u_3] = [u_4]^2 \cdot [u_1] = 0 = [u_1]\cdot [u_2] \cdot [u_4] = [u_1]\cdot [u_3] \cdot [u_4] = [u_2] \, ,
  \end{split}
\end{align}
where $\{u_i\}$ are the four non-flat surfaces.
By expanding the K\"ahler form as $J=\phi^i [u_i]$, we recall from section \ref{sec:5d_gauge_theories} that these intersection numbers determine the prepotential via
\begin{equation} \label{eq:geomprep2}
\mathcal F = \frac{1}{6}\int_Y J \wedge J \wedge J = \frac{1}{6} \sum_{i,j,k} [u_i] \cdot [u_j] \cdot [u_k] \, \phi^i \phi^j \phi^k. 
\end{equation}

We wish to match this with the field theoretic expression of the prepotential, which for $[\mathfrak{su}(2) + N_{f_1} \mathbf F] \times [\mathfrak{su}(3)_{\kappa}] \times [\mathfrak{su}(2) + N_{f_2}  \mathbf F]$ is
\begin{align} \label{eq:gensu232prep}
 & 6 \mathcal  F \nonumber\\
 & = 3 m^{(1)}_0 (\phi^1)^2 + 3 m^{(3)}_0 (\phi^4)^2+3 m^{(2)}_0  \left( \begin{bmatrix}\phi^3 \\\phi^2 \end{bmatrix}^T \begin{bmatrix} 2,-1 \\
 -1,2 \end{bmatrix} \begin{bmatrix} \phi^3  \\ \phi^2\end{bmatrix} \right)+ 3 \kappa((\phi^3)^2 \phi^2 - (\phi^2)^2 \phi^3) \\
 & +(2\phi^1)^3 +\left(\begin{bmatrix}2\\-1 \end{bmatrix}^T \begin{bmatrix} \phi^3  \\ \phi^2\end{bmatrix} \right)^3 + \left(\begin{bmatrix}1\\1 \end{bmatrix}^T \begin{bmatrix} \phi^3 \\ \phi^2\end{bmatrix} \right)^3+  \left(\begin{bmatrix}-1\\2 \end{bmatrix}^T \begin{bmatrix} \phi^3   \\ \phi^2\end{bmatrix} \right)^3 +(2\phi^4 )^3 \nonumber \\
 &- N_{f_1} (\phi^1+ m_{f_1})^3- N_{f_4} (\phi^4 + m_{f_2})^3  \nonumber \\ 
& - \frac{1}{2} \left( \left(\begin{bmatrix}1\\0 \end{bmatrix}^T \begin{bmatrix} \phi^3  \\ \phi^2\end{bmatrix}+\phi^1 \right)^3- \left(\begin{bmatrix}1\\0 \end{bmatrix}^T \begin{bmatrix} \phi^3  \\ \phi^2\end{bmatrix}-\phi^1 \right)^3+\left(\begin{bmatrix}1\\-1 \end{bmatrix}^T \begin{bmatrix} \phi^3  \\ \phi^2\end{bmatrix} + \phi^1 \right)^3-\left(\begin{bmatrix}1\\-1 \end{bmatrix}^T \begin{bmatrix} \phi^3  \\ \phi^2\end{bmatrix} - \phi^1 \right)^3 \right. \nonumber \\
& \left. -\left(\begin{bmatrix}0\\-1 \end{bmatrix}^T \begin{bmatrix} \phi^3  \\ \phi^2\end{bmatrix}+ \phi^1 \right)^3 -\left(\begin{bmatrix}0\\-1 \end{bmatrix}^T \begin{bmatrix} \phi^3  \\ \phi^2\end{bmatrix}-  \phi^1 \right)^3 + \left(\begin{bmatrix}1\\0 \end{bmatrix}^T \begin{bmatrix} \phi^3  \\ \phi^2\end{bmatrix}+\phi^4 \right)^3+ \left(\begin{bmatrix}1\\0 \end{bmatrix}^T \begin{bmatrix} \phi^3  \\ \phi^2\end{bmatrix}-\phi^4 \right)^3 \right. \nonumber\\
& \left.+\left(\begin{bmatrix}1\\-1 \end{bmatrix}^T \begin{bmatrix} \phi^3  \\ \phi^2\end{bmatrix} + \phi^4 \right)^3-\left(\begin{bmatrix}1\\-1 \end{bmatrix}^T \begin{bmatrix} \phi^3  \\ \phi^2\end{bmatrix} - \phi^4 \right)^3 -\left(\begin{bmatrix}0\\-1 \end{bmatrix}^T \begin{bmatrix} \phi^3  \\ \phi^2\end{bmatrix}+ \phi^4 \right)^3 -\left(\begin{bmatrix}0\\-1 \end{bmatrix}^T \begin{bmatrix} \phi^3  \\ \phi^2\end{bmatrix}-  \phi^4 \right)^3 \right) \nonumber 
\end{align} 
where we have already fixed the signs of the absolute values in \eqref{eq:prepq}.
We can now see that the geometric prepotential computed with the intersection numbers \eqref{eq:intnumbrank4} matches the field theory one \eqref{eq:gensu232prep} with $N_{f_1}=N_{f_2}=2$, and $\kappa=-1$.\\

Likewise, the field theory prepotential for $\mathfrak{su}(2)\times [\mathfrak{su}(4)_{\kappa}+ N_f \mathbf F]$ is
\begin{footnotesize}
\begin{align} \label{eq:gensu24prep}
 & 6 \mathcal  F \nonumber \\
 & = 3 m^{(1)}_0 (\phi^2)^2 +3 m^{(2)}_0  \left( \begin{bmatrix}\phi^1 \\\phi^3 \\  \phi^4 \end{bmatrix}^T \begin{bmatrix} 2,-1,0 \\
 -1,2,-1\\ 0,-1,2 \end{bmatrix} \begin{bmatrix} \phi^1  \\ \phi^3 \\ \phi^4 \end{bmatrix} \right)+ 3 \kappa((\phi^1)^2 \phi^3 - (\phi^3)^2 \phi^1+(\phi^3)^2 \phi^4 - (\phi^4)^2 \phi^3) \nonumber \\
 & +(2\phi^1)^3 +\left(\begin{bmatrix}2\\-1\\0 \end{bmatrix}^T \begin{bmatrix} \phi^1  \\ \phi^3 \\ \phi^4\end{bmatrix} \right)^3 + \left(\begin{bmatrix}0\\-1\\2 \end{bmatrix}^T \begin{bmatrix} \phi^1  \\ \phi^3 \\ \phi^4\end{bmatrix} \right)^3 + \left(\begin{bmatrix}-1\\2\\-1 \end{bmatrix}^T \begin{bmatrix} \phi^1  \\ \phi^3 \\ \phi^4\end{bmatrix} \right)^3  \\ 
& + \left(\begin{bmatrix}1\\0\\1 \end{bmatrix}^T \begin{bmatrix} \phi^1  \\ \phi^3 \\ \phi^4\end{bmatrix} \right)^3 + \left(\begin{bmatrix}-1\\1\\1 \end{bmatrix}^T \begin{bmatrix} \phi^1  \\ \phi^3 \\ \phi^4\end{bmatrix} \right)^3 + \left(\begin{bmatrix}1\\1\\-1 \end{bmatrix}^T \begin{bmatrix} \phi^1  \\ \phi^3 \\ \phi^4\end{bmatrix} \right)^3  - \frac{1}{2} \left( \left(\begin{bmatrix}1\\0\\0 \end{bmatrix}^T \begin{bmatrix} \phi^1  \\ \phi^3 \\ \phi^4\end{bmatrix} + \phi^2 \right)^3 \right. \nonumber \\
&\left. + \left(\begin{bmatrix}-1\\1\\0 \end{bmatrix}^T \begin{bmatrix} \phi^1  \\ \phi^3 \\ \phi^4\end{bmatrix} + \phi^2 \right)^3- \left(\begin{bmatrix}-1\\1\\0 \end{bmatrix}^T \begin{bmatrix} \phi^1  \\ \phi^3 \\ \phi^4\end{bmatrix} - \phi^2 \right)^3+ \left(\begin{bmatrix}0\\-1\\1 \end{bmatrix}^T \begin{bmatrix} \phi^1  \\ \phi^3 \\ \phi^4\end{bmatrix} + \phi^2 \right)^3- \left(\begin{bmatrix}0\\-1\\1 \end{bmatrix}^T \begin{bmatrix} \phi^1  \\ \phi^3 \\ \phi^4\end{bmatrix} - \phi^2 \right)^3 \right. \nonumber \\
& \left. + \left(\begin{bmatrix}1\\0\\0 \end{bmatrix}^T \begin{bmatrix} \phi^1  \\ \phi^3 \\ \phi^4\end{bmatrix} - \phi^2 \right)^3 - \left(\begin{bmatrix}0\\0\\-1 \end{bmatrix}^T \begin{bmatrix} \phi^1  \\ \phi^3 \\ \phi^4\end{bmatrix} + \phi^2 \right)^3- \left(\begin{bmatrix}0\\0\\-1 \end{bmatrix}^T \begin{bmatrix} \phi^1  \\ \phi^3 \\ \phi^4\end{bmatrix} - \phi^2 \right)^3 \right) - \frac{N_f-a_1-a_2}{2}\nonumber \\
& \left(  \left(\begin{bmatrix}1\\0\\0 \end{bmatrix}^T \begin{bmatrix} \phi^1  \\ \phi^3 \\ \phi^4\end{bmatrix} + m_f \right)^3 + \left(\begin{bmatrix}-1\\1\\0 \end{bmatrix}^T \begin{bmatrix} \phi^1  \\ \phi^3 \\ \phi^4\end{bmatrix} + m_f \right)^3+ \left(\begin{bmatrix}0\\-1\\1 \end{bmatrix}^T \begin{bmatrix} \phi^1  \\ \phi^3 \\ \phi^4\end{bmatrix} + m_f \right)^3 - \left(\begin{bmatrix}0\\0\\-1 \end{bmatrix}^T \begin{bmatrix} \phi^1  \\ \phi^3 \\ \phi^4\end{bmatrix} + m_f \right)^3 \right) \nonumber \\
& - \frac{a_1}{2}\left(  \left(\begin{bmatrix}1\\0\\0 \end{bmatrix}^T \begin{bmatrix} \phi^1  \\ \phi^3 \\ \phi^4\end{bmatrix} + m_{a_1} \right)^3 + \left(\begin{bmatrix}-1\\1\\0 \end{bmatrix}^T \begin{bmatrix} \phi^1  \\ \phi^3 \\ \phi^4\end{bmatrix} + m_{a_1} \right)^3- \left(\begin{bmatrix}0\\-1\\1 \end{bmatrix}^T \begin{bmatrix} \phi^1  \\ \phi^3 \\ \phi^4\end{bmatrix} + m_{a_1}  \right)^3- \left(\begin{bmatrix}0\\0\\-1 \end{bmatrix}^T \begin{bmatrix} \phi^1  \\ \phi^3 \\ \phi^4\end{bmatrix} + m_{a_1}  \right)^3 \right) \nonumber \\
& - \frac{a_2}{2}\left(  \left(\begin{bmatrix}1\\0\\0 \end{bmatrix}^T \begin{bmatrix} \phi^1  \\ \phi^3 \\ \phi^4\end{bmatrix} + m_{a_2}  \right)^3 - \left(\begin{bmatrix}-1\\1\\0 \end{bmatrix}^T \begin{bmatrix} \phi^1  \\ \phi^3 \\ \phi^4\end{bmatrix} + m_{a_2} \right)^3+ \left(\begin{bmatrix}0\\-1\\1 \end{bmatrix}^T \begin{bmatrix} \phi^1  \\ \phi^3 \\ \phi^4\end{bmatrix} + m_{a_2} \right)^3 - \left(\begin{bmatrix}0\\0\\-1 \end{bmatrix}^T \begin{bmatrix} \phi^1  \\ \phi^3 \\ \phi^4\end{bmatrix} + m_{a_2} \right)^3 \right)  \nonumber
\end{align} 
\end{footnotesize} 
where we have already fixed the signs of the absolute values in \eqref{eq:prepq}.
The geometric prepotential \eqref{eq:geomprep2} with the intersection numbers \eqref{eq:intnumbrank4} matches \eqref{eq:gensu24prep} with $N_f=6$ and $a_1=a_2=2$ if and only if $\kappa=0$, as expected.

\section{Fiber Structure in Non-flat Resolutions of the E-String}\label{app:fiber_structure}

\begin{figure}[p]
  \centering
  \includegraphics[width=.9\hsize]{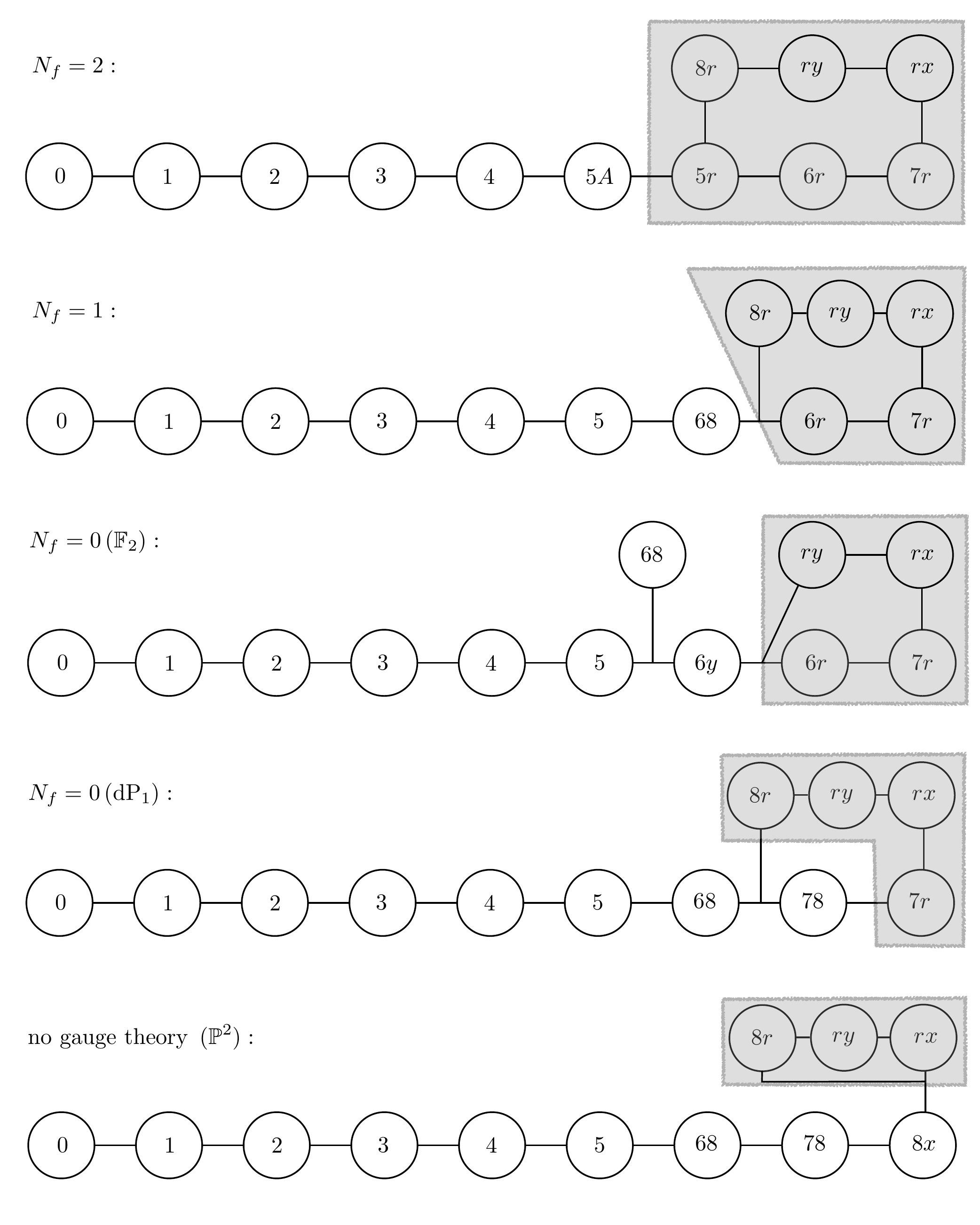}
  \caption{The resolved non-flat fibers realizing the rank one theories.
  The $\bbP^1$s (represented by circles) carrying a single number $n$ inside are the non-splitting fibers of the exceptional $E_8$ divisors $A_n$.
  Labels $q_1\,q_2$ indicate a curve defined by $\{q_1\} \cap \{q_2\}$, where $q_1 \cong n$ stands for $\alpha_n$.
  The only exception is the curve $5A$, which is the vanishing of $\alpha_5$ and a two-term polynomial (see equation \eqref{eq:rank_1_e5}).
  Lines indicate pairwise intersections; $\perp$ signals that three $\bbP^1$s intersect at one point.
  The gray box with enclosed circles represents the non-flat surface with curves contained inside of it.
  }\label{fig:resolved_fiber_structure}
\end{figure}

Here, we present the structures of the non-flat fiber after resolving the $E_8 - I_1$ collision of the Weierstrass model \eqref{eq:weierstrass_E8_top} at $W \cap \{b_6\}$.
Depending on the triangulation of the top (see figure \ref{fig:rank_1_phases}) and the corresponding 5d gauge theory phase, the resolved fiber has different curve splittings, intersection patterns, and positioning of the surface component.
All cases are displayed in figure \ref{fig:resolved_fiber_structure}.

\newpage

\bibliography{FTheory}{}
\bibliographystyle{JHEP} 

\end{document}